\documentclass[aps,prx,twocolumn,groupedaddress,showpacs,amsmath,amssymb,longbibliography, amsfonts,10pt]{revtex4-2}

\usepackage{amsmath}
\usepackage{amsfonts}
\usepackage{amssymb}
\usepackage{graphicx}
\usepackage{color}
\usepackage{xcolor}
\usepackage{bbold}
\usepackage{appendix}
\usepackage[citecolor=blue,urlcolor=blue]{hyperref}
\hypersetup{colorlinks}
\usepackage[caption=false]{subfig}
\usepackage{bookmark}
\usepackage{makecell}
\usepackage{mathtools}
\usepackage{comment}
\usepackage{tikz}
\usetikzlibrary{arrows.meta}
\usepackage{array}
\usepackage{multirow}
\usepackage[nolist,nohyperlinks]{acronym}
\usepackage[export]{adjustbox}
\newacro{RG}[RG]{renomalization group}
\newacro{fdr}[FDR]{fluctuation-dissipation relation}
\newacro{EP}[EP]{exceptional point}
\usepackage[normalem]{ulem}

\newacro{CEP}[CEP]{critical exceptional point}
\newacro{frg}[FRG]{functional RG}
\newacro{DSE}[DSE]{Dyson-Schwinger equation}
\newacro{MSRJD}[MSRJD]{Martin-Siggia-Rose-Janssen-DeDominicis}
\newacro{BKT}[BKT]{Berezinskii–Kosterlitz–Thouless}
\newacro{KPZ}[KPZ]{Kardar–Parisi–Zhang}
\newacro{EFT}[EFT]{effective field theory}
\newacro{Nlsm}[NL$\sigma$M]{non-linear $\sigma$ model}
\usepackage{lipsum}

\newcommand{\Svec}{\mathbf{S}}

\newcommand{\vecxi}{\boldsymbol\xi}
\newcommand{\vecphi}{\boldsymbol{\phi}}

\newcommand{\vecx}{\mathbf{x}}
\newcommand{\vecq}{\mathbf{q}}
\newcommand{\vecp}{\mathbf{p}}

\newcommand{\cCZ}[1]{\textbf{\textcolor{blue}{#1}}}
\newcommand{\cRD}[1]{\textbf{\emph{\textcolor{brown}{#1}}}}

\definecolor{rotatingColor}{rgb}{0.741176, 0.811765, 1}
\definecolor{xyColor}{rgb}{0.611765, 0.905882, 0.658824}
\definecolor{paraColor}{rgb}{0.952941, 0.952941, 0.913725}
\definecolor{cepColor}{rgb}{0.858824, 0, 0}

\begin{document}
\title{Universal Phenomenology at Critical Exceptional Points of Nonequilibrium \texorpdfstring{$O(N)$}{O(N)} Models}

\author{Carl Philipp Zelle}
\thanks{These authors contributed equally to this work.\\  Emails: zelle@thp.uni-koeln.de  daviet@thp.uni-koeln.de}

\author{Romain Daviet} 
\thanks{These authors contributed equally to this work.\\  Emails: zelle@thp.uni-koeln.de  daviet@thp.uni-koeln.de}

\author{Achim Rosch} 
\author{Sebastian Diehl}

\affiliation{Institut f\"ur Theoretische Physik, Universit\"at zu K\"oln, 50937 Cologne, Germany}

\begin{abstract}
In thermal equilibrium the dynamics of phase transitions is largely controlled by  fluctuation-dissipation relations: On the one hand, friction suppresses fluctuations, while on the other hand the thermal noise is proportional to friction constants. Out of equilibrium, this balance dissolves and one can  have situations where friction vanishes due to antidamping in the presence of a finite noise level. We study a wide class of $O(N)$ field theories where this situation is realized at a phase transition, which we identify as a critical exceptional point. In the ordered phase, antidamping induces a continuous limit cycle rotation of the order parameter with an enhanced number of $2N-3$ Goldstone modes. Close to the critical exceptional point, however, fluctuations diverge so strongly 
due to the suppression of friction that in dimensions $d<4$ they universally either destroy a preexisting static order, or give rise to a fluctuation-induced first order transition. This is demonstrated within a full resummation of loop corrections via Dyson-Schwinger equations for $N=2$, and a generalization for arbitrary $N$, which can be 
solved in the long wavelength limit. We show that in order to realize this physics it is not necessary to drive a system far out of equilibrium: Using the peculiar protection of Goldstone modes, the transition from an $xy$ magnet to a ferrimagnet is governed by an exceptional critical point once weakly perturbed away from thermal equilibrium.

\end{abstract}

\date{\today}

\maketitle
\section{Introduction}

The quest for universal structure in phases and phase transitions far from equilibrium is a long standing challenge and has acquired a lot of attention in the recent years. Paradigmatic examples, which manifestly go beyond thermal equilibrium \cite{Goldenfeld1992a,Hohenberg1977}, are provided by problems such as interface growth, membering the Kardar-Parisi-Zhang universality class \cite{Kardar1986,Krug1997,Takeuchi2018}, wetting transitions in the directed percolation class \cite{Hinrichsen2000,OdorRev}, or self-organized criticality \cite{Bak1987,pruessner2012self}. In these systems, detailed balance is violated on the microscopic scale, and this manifests itself in the macroscopic observables. An important arena where these phenomena have been identified recently are instances of driven and open quantum matter, including systems like Rydberg gases in the dissipative regime~\cite{Gutierrez2017absorbing,HelmrichSOC} or exciton-polariton systems~\cite{Fontaine2021}. The rapid experimental developments in these directions in turn inspires theory to identify novel forms of non-equilibrium universality without equilibrium counterparts \cite{marino2016quantum,young2020nonequilibrium,Hanai2020,Scarlatella2019,Soriente_2018,Soriente_2021}.  Beyond such  condensed matter platforms, non-equilibrium dynamics occurs rather as a rule than an exception in biological, economic, and even social systems, which are only at the verge of being studied from the stance of universality \cite{Bowick2022,Vicsek95,Toner1995,vicsek2012,Ballerini2008, Helbing2000,Bain2019,Shankar2022}. 

These systems share in common that their generators of dynamics -- be it quantum or classical -- generically consist of reversible and irreversible terms, which occur on equal footing. This circumstance makes the generator non-Hermitian, and in turn, enables the existence of \textit{\ac{EP}} -- points in the space of tuneable parameters which show degeneracies in the excitation and damping spectra \cite{Berry2004,Heiss2004}. These \ac{EP}s have recently fueled an active stream of research in condensed matter, atomic condensates, and optics: On the one hand, they hold promises for applications, such as sensing due to an enhanced response to external perturbations in their vicinity~\cite{Hodaei17,Chen17,Poli15,AshidaGongUeda}. At the same time, they host conceptually new topological phenomena, such as nodal topological phases with open Fermi surfaces, or an anomalous bulk-boundary correspondence \cite{Zeuner15,Zhou18,Weimann17,Xiao20,Helbig2020,cerjan19,BergholtzBudichKunstRMP}. 

All these intriguing phenomena appear on an effective single particle level, describing, for example, the linear excitations above a more complex underlying non-linear dynamics. Taking such non-linearity into account is then an important step towards a more comprehensive many-body theory of non-Hermitian systems: It offers the possibility to describe qualitatively distinct stable phases in systems with many degrees of freedom. It also paves the way to describe hallmark non-equilibrium phenomena such as pattern formation \cite{Cross1993}. Dynamical limit cycles provide one prominent incarnation of this general phenomenology, which have recently  met great interest~\cite{Walter_2014,Walter_2014b,Iemini_2018,Dutta_2019,Buca_2019,Buca_2019b,Tindall_2020,BucaGoold2020,Ben_Arosh_2021,Buca_2022,BucaDonner2022,Mivehvar_2021,Scarlatella2023}. Experimental examples range from the paradigmatic van der Pol oscillator~\cite{vanderPol1920} over recent realizations of driven collective spin ensembles in quantum cavities \cite{Dogra_2019,Dreon_2022} and driven magnonics \cite{wang2020,Yuan2023} to active matter systems~\cite{Fruchart2021,Weis2022,suchanek2023}.

These developments spark a fundamental question: Which novel universal behaviors emerge when an exceptional point coincides with a critical point at a \ac{CEP}? Universal critical behavior, in and as well as out of equilibrium, is closely linked to stochastic fluctuations that become macroscopically large at a critical point, "washing out" all details of the microscopic scales and only depend on symmetries and conservation laws.  It is thus paramount to include these fluctuations on a nonlinear level to determine the fate of exceptional point criticality in terms of universality, a task not undertaken up to now in
 $d<4$. This final layer of complexity beyond deterministic or linear approximations is characteristic for a many body problem. 
\begin{figure*}
    \centering
    \subfloat[\label{fig:PhaseDiagram_a}]{
    \begin{tikzpicture}

    \fill[paraColor, opacity=1] (0,0) -- (4,0) -- (4,3) -- (0,3) -- (0,0);
    \fill[xyColor, opacity=1] (0,0) -- (-4,0) -- (-4,3) -- (0,3) -- (0,0);
    \fill[xyColor, opacity=1] (0,0) -- (-4,0) -- (-4,-3) -- (0,0);
    \fill[rotatingColor, opacity=1] (0,0) -- (-4,-3) -- (0,-3) -- (0,0);
    \fill[rotatingColor, opacity=1] (0,0) -- (0,-3) -- (4,-3) -- (4,0)-- (0,0);
    
          \draw[->,black] (-4,0)--(4,0);
    \draw[->,black] (0,-3)--(0,3);
    \draw[black] (0.1,2.8) node[right]{\large $ \gamma$};
    \draw[black] (3.8,-0.2) node[left]{\large $r$};
    
    \draw[black] (2,1.75) node{$O(N)$ symmetric};
    \draw[black] (2,1.25) node{$\boldsymbol{\phi}_{s}=0$};
    \draw[black] (-2,2) node{Static phase};
    \draw[black] (-2,1.5) node{$O(N)\to O(N-1)$};
    \draw[black] (-2,1.) node{$\rho_{0}\neq 0$, $E=0$};
    
    \draw[black] (2,-1) node{Rotating phase};
    \draw[black] (2,-1.5) node{$O(N)\to O(N-2)$};
    \draw[black] (2,-2) node{$\rho_{0}\neq 0$, $E\neq 0$};

\draw[black] (-2.5,-0.5) node{$N-1$ Goldstones};
\draw[black] (-1.75,-2.7) node{$2N-3$ Goldstones};

\draw[orange,very thick] (0,0) --(0,3) node [pos=0.5, left](TextNode) {A};
\draw[brown,very thick] (0,0) --(4,0) node [pos=0.5, below](TextNode) {B};

\draw[very thick,cepColor] (0,0)--(-4,-3) node [pos=0.6, above, sloped] (TextNode) {C=CEP};
    \end{tikzpicture}
    }
    \hfill
        \subfloat[\label{fig:PhaseDiagram_b}]{
    \begin{tikzpicture}
    
\coordinate (A) at (-1,-1);
\coordinate (B) at (-0.3,3);
\coordinate (C) at (4,-0.3);

\fill[paraColor, opacity=1,smooth]  (0,3) .. controls (-0.3,0) .. (A) .. controls (0,-0.4) .. (4,0) -- (4,3) -- (B);
    \fill[xyColor, opacity=1] (0,3) .. controls (-0.3,0) .. (A) -- (-4,-3) -- (-4,3) -- (B);
    \fill[rotatingColor, opacity=1] (A) .. controls (0,-0.4) .. (4,0) --(4,-3) -- (-4,-3) --(A);
\draw[very thick,cepColor] (-4,-3)--(A) node [pos=0.6, above, sloped](TextNode) {1st-order};

\draw[brown,very thick,dashed]  (A) .. controls (0,-0.4) .. (4,0);
\draw[orange,very thick] (0,3) .. controls (-0.3,0) .. (A);
            \draw[->,black] (-4,0)--(4,0);
    \draw[->,black] (0,-3)--(0,3);
    \draw[black] (0.1,2.8) node[right]{\large $\gamma$};
    \draw[black] (3.8,-0.2) node[left]{\large $r$};
    
    \draw[black] (2,1.75) node{$O(N)$ symmetric};
    \draw[black] (2,1.25) node{$\boldsymbol{\phi}_{s}=0$};

    \draw[black] (-2,2) node{Static phase};
    \draw[black] (-2,1.5) node{$O(N)\to O(N-1)$};
    \draw[black] (-2,1.) node{$\rho_{0}\neq 0$, $E=0$};

    \draw[black] (2,-1) node{Rotating phase};
    \draw[black] (2,-1.5) node{$O(N)\to O(N-2)$};
    \draw[black] (2,-2) node{$\rho_{0}\neq 0$, $E\neq 0$};

    \end{tikzpicture}
    }
\caption{(a) mean-field phase diagram of the nonconservative $O(N)$ model. For positive $r$ and $\gamma$, the symmetric, disordered phase is stable (light yellow). Upon tuning the gap $r$ through zero, the model A transition (orange line A) into a statically ordered phase (green) that breaks $O(N)$ spontaneously to $O(N-1)$ occurs. A nonequilibrium dynamically ordered phase with a rotating order parameter (blue) emerges for negative dampings. The phase transition between statically ordered and rotating phase (red line C) is described by a critical exceptional point. The point in the middle of the phase diagram, at which all phase transition lines meet, forms a multicritical exceptional point. (b) Schematic phase diagram beyond mean-field ($d<4$). The CEP line is replaced by a first-order phase transition where the angular velocity $E$ jumps from zero in the ordered phase to a finite value in the rotating phase. For initial value of $\delta$ and $\gamma$ closer to zero, the enhanced fluctuations destroy the order parameter before reaching the CEP. The symmetric phase has thus and extended stability regime and the multicritical point moves.}
    \label{fig:PhaseDiagram}
\end{figure*}

In this work we use symmetry based effective field theories -- a framework successfully applied beyond thermal equilibrium, predicting for instance the experimentally observed KPZ scaling in driven dissipative condensates \cite{Altman2015, Fontaine2021} or ordering below two dimensions in active matter theories of flocking \cite{Toner1995, Vicsek95} to name a few -- to address the interplay of exceptionality and critical fluctuations. We show that paradigmatic $O(N)$ models, which have acted as work horse theories for critical behavior at thermodynamic equilibrium, host limit-cycle phases as well as \ac{CEP}s once suitably driven out of equilibrium. They thus allow one to determine the universal phenomenology at such points in the spirit of effective field theory. One main result is, that even though the strongly enhanced Gaussian fluctuations in the vicinity of a \ac{CEP} suggest the absence of any transition, there is a fluctuation induced first order transition. The effect reveals itself only once going beyond Gaussian approximations, i.e. once one includes nonlinearities and stochastic fluctuations. Only then, a full picture of the transition emerges even on a qualitative level.

The symmetry based effective field theory approach also highlights the mechanisms and minimal requirements for the described phenomena. It allows us to show that limit-cycle phases and \ac{CEP} transitions as analysed here can emerge in very different physical systems, ranging from previously discussed active matter set ups over open quantum systems to driven solid state platforms. It thus constitutes a bridge connecting these different areas of nonequilibrium physics.

Furthermore, within our field theoretic model, we can show how Goldstone modes arise as collective excitations in the limit-cycle phase. Its dynamic nature lets them oscillate at finite frequency while still having divergent life times. An enlarged number of Goldstone modes underpins that the rotating order can be characterized by spontaneously broken symmetries.

In the following, we provide an overview of the setting, and describe our key results, albeit in a slightly different order than in the subsequent main part of the paper. 

\subsection{Key results and synopsis}

\paragraph*{Model  --} We study $N$-component order parameter models with an $O(N)$ symmetry in $d+1$ dimensions. We display here a variant of the model which is incomplete and will be extended below, but allows us to discuss all scales relevant to the universal aspects of the problem: 
\begin{align}\label{eq:model0}
    \Big(\partial_t^2+(2\gamma + u\rho  -Z\nabla^2) \partial_t + r+\lambda\rho-v^2\nabla^2 \Big)\boldsymbol{\phi}+\boldsymbol{\xi}=0,
\end{align}
where $\boldsymbol{\phi}$ is the $N$ component vector field, $\rho = \boldsymbol{\phi}^T\boldsymbol{\phi}$, and $\boldsymbol {\xi}$ is a Gaussian white noise with zero mean and variance $D$.
It has two important characteristics, the conspiracy of which is at the root of exceptional critical points discussed below: First, it stands in between equilibrium relativistic $O(N)$ models \cite{ZinnJustin} and Hohenberg-Halperin models for equilibrium dynamical criticality \cite{Hohenberg1977}: It shares with the first class an inertial, second order time derivative term, and with the second a damping, first order time derivative term. The inertial term is neglected near equilibrium critical points \cite{Hohenberg1977} since it is irrelevant in the sense of the renormalization group, but will prove of key importance at a \ac{CEP}.
As a second key ingredient, the model is driven out of equilibrium.  This is manifest by a coupling term $u$ which has no potential form and corresponds therefore to a nonconservative damping force. Physically these result from driving and/or coupling it to different baths which are not in global thermal equilibrium with each other, and technically by breaking the symmetry behind detailed balance explicitly \cite{Taeuber2014,Sieberer2016}. As a hallmark physical feature compared to equilibrium $O(N)$ models, we establish the emergence of a ``time crystalline'' limit-cycle phase, see Fig.~\ref{fig:PhaseDiagram} and the discussion below, where a long range many body order parameter traces out a closed orbit. 
It is described by a rotating field, $
\boldsymbol{\phi}_{s}=\sqrt{\rho_{0}}\cdot \left(\cos{E t},\, \sin{E t},0,...\right)^T$ and occurs only in the presence of an effective antidamping, $\delta \equiv 2\gamma + u\rho_0<0$ within mean-field theory (i.e., ignoring the noise terms in Eq.~\eqref{eq:model0}). In the same approximation one finds for the order parameter amplitude $\rho_0 = \boldsymbol{\phi}^T_s\boldsymbol{\phi}_s=-2 \gamma/u$ and $E= \sqrt{r-\frac{2\gamma\lambda}{u}}$.
Most importantly, the transition into this new phase proceeds via a \ac{CEP}, the nature of which we analyze in detail.

\paragraph*{\ac{CEP}s only occur out-of-equilibrium} -- We establish that,  while \ac{EP}s in general are not fundamentally tied to a system being driven out of equilibrium, CEPs are. One simple example of an equilibrium EP is provided by the damped harmonic oscillator at the point where it transits from under- to overdamped dynamics, including in the presence of noise fluctuations \cite{KrohaWeitz2021}. Such a transition does not realize a critical point in the sense of divergent length and time scales -- in fact, it is only detected in dynamical observables like dynamical susceptibilities by the absence of oscillations in the underdamped regime, but goes unseen in any static observable. The impact of fluctuations near such an equilibrium EP has recently been studied for a damped, noisy anharmonic oscillator \cite{KhedriZilberberg2022}. In contrast, below in  Sec.~\ref{ssec:nonthermal} we will demonstrate that a \ac{CEP} can be realized only if thermal equilibrium conditions are broken explicitly. In fact, the phenomenology revealed in this work can be traced back to a superthermal mode occupation near a CEP \cite{Hanai2020}, which is excluded at thermodynamic equilibrium, and so is intimately tied to the non-equilibrium nature of the problem.

\begin{figure}[t]
\center{}
   \includegraphics[width=0.95 \linewidth]{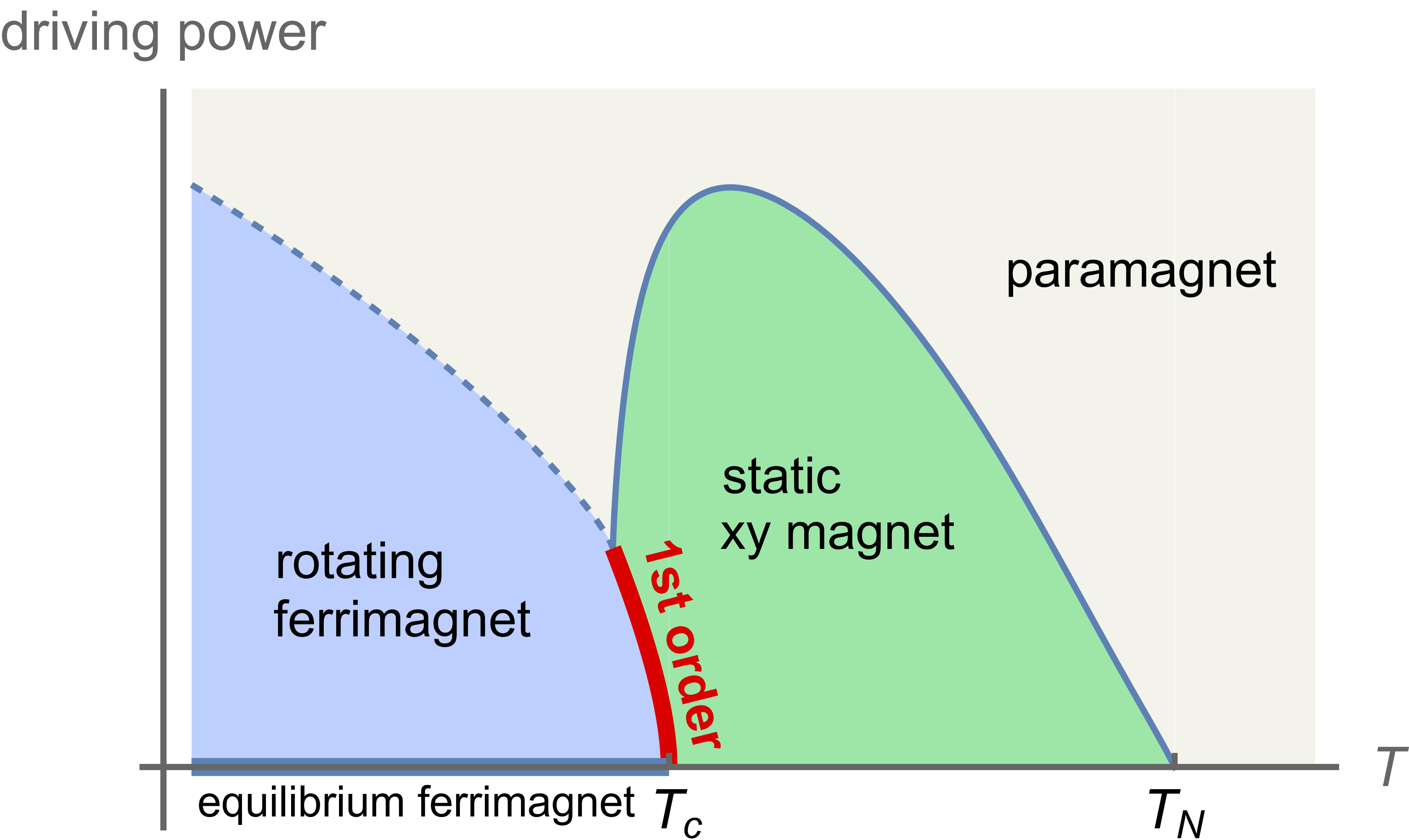}
   \caption{Schematic phase diagram of a driven ferrimagnet as function of temperature and the power of an external driving source, e.g., a laser or an oscillating magnetic field. We assume that in  equilibrium the system displays antiferromagnetic xy order for $T<T_N$ and becomes a ferrimagnet for $T<T_c<T_N$ by developing an extra out-of-plane ferromagnetic component.
Driving induces in the ferrimagnetic phase a rotation of the xy order parameter.
The transition is governed by a critical exceptional point (CEP) with its characteristic first-order phase transition (red line). The enhancement of fluctuations close to the CEP bends the transition line between paramagnet and $xy$ order down culminating in a multicritical point where all transition lines meet. For larger driving strength also a direct transition from the paramagnetic into the rotating ferrimagnetic phase will occur. Thus, all phases and phase transitions of the effective model, Fig.~\ref{fig:PhaseDiagram_b}, can be realized.
  } \label{fig:phasediagramFerri} 
\end{figure}

\paragraph*{Microscopic origin: Physical realizations} -- The effective field theory approach also allows us to identify possible realizations. We show that strikingly nonthermal macroscopic behavior associated to \ac{CEP}s and limit-cycle phases is not limited to active matter~\cite{Fruchart2021}, but can also emerge in solid state platforms only weakly driven out of equilibrium. In particular, we demonstrate that a system with in-plane ferro- or antiferromagnetic order near a ferrimagnetic instability at equilibrium maps to our model for $N=2$, even when only weakly driven out of equilibrium by, for example, terahertz radiation (see Sec.~\ref{sec:implementation}). Our results for this system are summarized on Fig.~\ref{fig:phasediagramFerri}.

The  model \eqref{eq:model0} is best viewed in the spirit of an effective field theory, see also Fig.~\ref{fig:concpept}: It results via coarse graining from a more microscopic model with the same symmetries, such as $O(N)$ or spatial rotation symmetry, and likewise broken detailed balance indicating non-equilibrium conditions. This encompasses a large class of physical setups. While we discuss the possible physical implementation in a driven ferrimagnet in more detail, we furthermore connect non-reciprocal phase transitions found in driven-dissipative condensates~\cite{Hanai2019,Hanai2020} and active matter scenarios \cite{Fruchart2021} to our mesoscopic model, and show that their universal phenomenology is described by our mechanism. We then discuss possible realizations in certain microscopic Lindblad quantum dynamics \cite{Ben_Arosh_2021}.

\textit{Extracting macrophysics: Evaluation strategy} --  Specifically close to a phase transition, one has to expect a strong impact of fluctuations: Both a deterministic approximation discarding noise, and the neglect of interaction effects, become invalid. To address these challenges, we first map the mesoscopic Langevin model Eq.~\eqref{eq:model0} to an equivalent Martin-Siggia-Rose-Janssen-DeDominicis functional integral \cite{DOMINICIS1976,Martin1973,Janssen1976}. We then extract the macroscopic physics of the interacting, fluctuating problem by computing the effective action. The latter might be thought of an action with the same symmetries as the mesoscopic model, but with the effects of fluctuations included via the renormalization of its parameters, i.e. the set of coupling constants. Beyond its practical value of systematically accounting for fluctuation effects, it is a very handy object, which leverages many properties usually discussed for the bare (mesoscopic) action to the full theory -- this includes, for example, the exact counting of Goldstone modes in the limit-cycle phase, or the implementation of exceptional points as a property of the fully renormalized single-particle retarded response. We will thus put it to work to distill the principles and universal mechanisms governing the macroscopic physics of non-equilibrium $O(N)$ models.

\paragraph*{Limit-cycle phase} --
As indicated earlier, our model hosts a limit-cycle phase where a long range order parameter dynamically traces out a closed orbit, a phenomenon clearly ruled out by the laws of thermodynamics and thus intrinsically nonequilibrium. We refer to it as the limit-cycle or rotating phase. Phases in thermal equilibrium can be classified by their symmetry -- in the present case, there  is a disordered phase with full $O(N)$ symmetry which is spontaneously broken in the statically ordered phase to $O(N-1)$. Goldstone theorem then predicts the emergence of $N-1$ soft, gapless modes, corresponding to the broken symmetry generators.

We show that the nonthermal limit-cycle phase can be classified by symmetry, as well. It constitutes a further symmetry breaking of $O(N)$ to $O(N-2)$, rather than for instance dynamically restoring a broken symmetry as argued previously \cite{Fruchart2021}. This is most clearly exemplified by the emergence of $2N-3$ Goldstone modes in the limit-cycle phase, due to the breaking of the same number of generators by the spontaneous choice of a plane within which the order parameter rotates. We show this explicitly and non-perturbatively within the effective action framework based on Ward identities. This result can thus be considered an exact property of the fully fluctuating and interacting problem.

\begin{figure}
    \centering
    \begin{tikzpicture}[scale=0.75]
        \draw[-Triangle,very thick,red] (-1.5,4)--(-1.5,-4) node[left,black]{length};
        \draw[black] (-3.25,3.5) node[align=center]{Microscopic};
        \draw[black] (-3.25,2.5) node[align=center]{$O(N)$ symmetry\\ Equilibrium \\broken};

        \draw[black] (-3.25,0) node[align=center]{mesoscopic};
        \draw[black] (-3.25,-3) node[align=center]{macroscopic};
        
        \draw[black] (0.25, 3) node[draw,align=center, minimum width=1.8cm, minimum height=25pt]{Driven \\ ferrimagnets};
        \draw[black] (3., 3) node[draw,align=center, minimum width=1.8cm, minimum height=25pt]{Driven \\condensates};
        
        \draw[black] (5.75, 3) node[draw,align=center, minimum width=1.8cm, minimum height=25pt]{Active \\ matter};
        \draw[line width=10pt,-{Triangle[length=15pt,width=20pt]},gray] (0.25,2.25)-- (0.75,0.5);
        \draw[line width=10pt,-{Triangle[length=15pt,width=20pt]},gray] (3.,2.25)-- (3,0.5);
        \draw[line width=10pt,-{Triangle[length=15pt,width=20pt]},gray] (5.75,2.25)-- (5.25,0.5);
        \draw (1.75,1.75) node[align=center]{coarse};
        \draw (4.25,1.75) node[align=center]{graining};
        
        \draw[black] (3,0) node[draw]{Non-equilibrium $O(N)$ MSRJD action};
        \draw[line width=10pt,-{Triangle[length=15pt,width=20pt]},gray] (3.,-0.5)-- (3,-2.5);
         \draw (3,-1.25) node[align=center]{Fluctuation \qquad corrections };
        \draw[black] (3, -3) node[draw]{Non-equilibrium $O(N)$ effective action};
    \end{tikzpicture}
    \caption{Concept of the approach. We consider quantum or classical problems which, on a microscopic level, feature an $O(N)$ symmetry and are driven out of equilibrium. After coarse graining, in the presence of noise these reduce generically to semiclassical noisy non-equilibrium $O(N)$ models described by a MSRJD action. Taking this mesoscopic action as a starting point, we compute the effective action which includes fluctuations, which is particularly important close to a critical point, such as a CEP.}
    \label{fig:concpept}
\end{figure}

For $N>2$, there are $2(N-2)$ Goldstone modes that differ from their overdamped equilibrium counterparts which disperse as $\omega\sim-i\vecq^2$. While they still correspond to excitations whose life time diverges for long wavelengths $\vecq\rightarrow 0$ and are generated by broken symmetry generators, they have a finite frequency corresponding exactly to the frequency $E$ of the limit cycle itself, i.e. $\omega=-iZ\vecq^2\pm E$.\\
The case $N=2$ is special, since the difference between $SO(2)$ and the required $O(2) = SO(2) \ltimes \mathbb{Z}_2$ symmetry  (semidirect product, the elements of $SO(2)$ and $\mathbb{Z}_2$ do not commute here), becomes crucially important~\footnote{The case of $SO(3)$ can in principle host a symmetry breaking limit cycle. It is however not captured by our analysis, since the Lorentz force like term is omitted. For larger $N$ the difference between $SO(N)$ and $O(N)$ is only manifest in higher order interactions which are expected to be irrelevant under RG.}. The symmetry that is broken when transiting from the statically to the dynamically ordered phase is the discrete $\mathbb{Z}_2$; physically, it corresponds to choice of direction in which the angular limit cycle is traversed, with angular velocity $\pm E$ respectively (for larger $N$, the sign of $E$ is unphysical, because limit cycles with opposite traversing directions can be smoothly connected by a rotation). The requirement of full $O(2)$ symmetry rules out a KPZ nonlinearity within the critical theory, we give more details in ~\ref{ssec:Sym_break}.

\paragraph*{Exceptional point phase transition --} What is the fate of universal fluctuations if an \ac{EP} is tuned to criticality? The mean field phase diagram \ref{fig:PhaseDiagram_a} displays three transition lines $A,B,C$. The transition $A$ between disordered and statically ordered phase belongs to the model A universality class of Hohenberg and Halperin \cite{Halperin1974} and shows no deviations from equilibrium models. The transition $B$ represents an instance of finite frequency criticality, a scenario developed in \cite{Scarlatella2019}, with a yet to be determined and potentially non-equilibrium universality class. The transition line $C$ separating static and rotating order is described by a critical exceptional point. Hence, we focus on this transition line $C$. The effective damping ($\delta=2\gamma-u\rho_0$ in mean field) of the phase fluctuations is tuned to zero at the transition, spoiling the canonical power counting of model A. More precisely, approaching the critical point, two poles of the Green's function coalesce and become gapless simultaneously  -- an exceptional point is made to coincide with a critical point. This critical point is reached via a single fine tuning of parameters as visible in the figure -- this is assisted by the gapless Goldstone modes as we elaborate in the main text, see also \cite{Fruchart2021}.

The fate of the \ac{CEP} transition is then dominated by two aspects:\\
(i) Non-analytic spectral properties. The \ac{CEP} is characterized by a complex dispersion 
\begin{eqnarray}
    \omega_{1,2}(\vecq)  = -\frac{i}{2}\Big( Z \vecq^2  + \delta\Big) \pm v |\vecq |,
\end{eqnarray}
upon approaching the phase transition from the statically ordered phase. The parameters of Eq.~\eqref{eq:model0} thus represent a  diffusion coefficient $Z$ and a propagation velocity $v$. The effective damping rate $\delta$ acts as a gap measuring the distance from the phase transition -- criticality emerges as $\delta \to 0$. Exceptionality is encoded in the coalescence of modes at $\vecq=0$. The scale $v$ of dimension momentum is associated to the characteristic non-analytic momentum dependence found near exceptional points \cite{BergholtzBudichKunstRMP,AshidaGongUeda}, and is of key importance for the effects found below.  It sets an obstruction to homogeneous scaling with a dynamical exponent $z=2$ with $\omega \sim -i \vecq^2$. Rather, here a mode with $z=1$ propagation behavior and $z=2$ dissipation emerges, which reflects the equal importance of the reversible $\partial_t^2$ and the irreversible $\partial_t$ terms in Eq.~\eqref{eq:model0} near the exceptional critical point.
\\
(ii) Superthermal mode occupation. A second characteristic of the \ac{CEP} is not visible in spectral properties and in a deterministic approximation (zero noise level $D=0$): The mode occupation at the \ac{CEP} is enhanced compared to thermal equilibrium~\cite{Hanai2020}. This is measured by the critical equal-time correlation function (or Keldysh Green's function), which is given by
\begin{align}
   \left\langle\boldsymbol{\phi}(-\vecq,t)\boldsymbol{\phi}(\vecq,t)\right\rangle= G^K(\vecq,t=0)\sim\frac{D}{\vecq^4},
\end{align}
to be compared to a scaling $\sim \vecq^{-2}$ at equilibrium, where it is fixed by the fluctuation-dissipation theorem. These giant fluctuations occur because the damping at $\vecq=0$, which suppresses fluctuations, is tuned to zero, while the noise level remains finite. This is not possible in equilibrium where fluctuation-dissipation  theorems guarantee that noise and dissipation are proportional to each other.

(i) and (ii) are both tied to exceptionality: (i) is a spectral property, associated to the propagation velocity $v$, and (ii) a statistical effect, associated to the noise level $D$. Both effects have a drastic impact on the CEP transition, which can go two ways in the fully interacting and fluctuating system:\\
(I) The \emph{statistical effect} of enhanced occupation (ii) due to Gaussian fluctuations melts down preexisting order in physical dimensions $d<4$ before the \ac{CEP} is reached \cite{Hanai2020}. This mechanism bends the model A transition line $A$ as depicted in Fig.~\ref{fig:PhaseDiagram_b}. We show in Sec~\ref{sec:ExceptionalRG} that after the crucial step of taking non-Gaussian fluctuations into account, there is a second case:\\
(II) \emph{Interactions} conspire with the non-analytic spectral properties (i) to render the \ac{CEP} phase transition into a fluctuation induced first order transition. This occurs through a mechanism which is -- albeit in a physically rather different system -- technically reminiscent of Brazovskii's seminal work \cite{Brazovskii75,Hohenberg1995}.

Technically, the above conspiracy results in a breakdown of canonical power counting and gradient expansions of the effective action functional (see also App.~\ref{app:scaling} for details on this breakdown). Interestingly, this leads to a suppression of two-loop effects at long wavelength. In turn, we can  utilize this insight to perform a full resummation of the entire perturbative series in the vicinity of the \ac{CEP} by means of Dyson Schwinger equations. We exhibit this mechanism in detail for the case $N=2$, and then generalized to arbitrary $N$. This furthermore unravels that the upper critical dimension, above which there is a second order CEP transition with correlation length exponent $\nu=\frac{1}{2}$ described by the Gaussian fixed point (see Sec~\ref{sssec:PhaseTrans}) is $d_c=4$.

Below the upper critical dimension $d<4$, there emerges a new scale $\rho_0g$, separating the symmetry restoring transition (I) and the interaction dominated fluctuation induced phase transition (II). It is composed of the amplitude of the order parameter $\rho_0$ and $g$, the effective nonconservative interaction of the gapless phase fluctuations in the broken phase, stemming from nonthermal nonlinearities like $u$ (cf. Eq.~\eqref{eq:model0}) in the original model. For small $\rho_0 g\ll1$ there is symmetry restoration (I) while for $\rho_0g\gg1$ the fluctuation induced first order transition (II) takes place. Combining these mechanisms, we are lead to the phase diagram displayed in Fig.~\ref{fig:PhaseDiagram_b}. Both transitions $A$ and $C$  meet at a multicritical point at a critical value $(\rho_0 g)_c\sim O(1)$.

The remainder of this article is structured as follows: In Sec.~\ref{sec:MeanField} we introduce the model and discuss its mean-field phase diagram. We introduce the field theoretic methodology, i.e. MSRJD path integrals, that we use for the systematic study of fluctuations in these phases in Sec.~\ref{sec:MF+Gauss}. We proceed to discuss exceptional and critical exceptional points and their properties including their incompatibility with thermal equilibrium conditions in Sec.~\ref{sec:EPs}. We then study the slow long wavelength dynamics of fluctuations in the statically ordered and rotating phase, Sec.~\ref{sec:fluctuations}. Here, we show how the Goldstone theorem and the respective modes play out in the dynamical rotating phase and study the Gaussian fluctuations in all three phases. This allows us to discuss the critical properties of the transitions on the Gaussian level valid above the upper critical dimensions. The impact of interactions beyond the Gaussian level at the \ac{CEP} transition is analysed in Sec.~\ref{sec:ExceptionalRG}, revealing the symmetry restoration as well as the fluctuation induced first order transition and the interaction scale $\rho_0g$ governing the competition of both mechanisms. We then give three examples for possible realization schemes including driven magnetic systems in Sec.\ref{sec:implementation}, and conclude afterwards.

\section{Warmup: Non-equilibrium \texorpdfstring{$O(N)$}{O(N)} models in \texorpdfstring{$(0+1)$}{(0+1)} dimensions}\label{sec:MeanField}

\subsection{The model}
We will study the phases and phase transitions of (semi)classical $O(N)$ order parameter fields $\boldsymbol{\phi}(\vecx,t)\in \mathbb{R}^N$ subject to a rotationally invariant stochastic Langevin time evolution in $d+1$ dimensions. To describe its basic physics, we first focus on the purely deterministic $(0+1)$ dimensional case. This may be considered as a mean-field theory for the order parameter of the full model, where both the spatial degrees of freedom and the noise are ignored. (Of course, both these ingredients neglected here will be needed to describe the full nonthermal phase diagram below.) More precisely, we consider an $O(N)$ symmetric $\boldsymbol{\phi}^4$-type model, i.e. with up to cubic interaction terms in the equation of motion,

\begin{align}\label{eq:0+1D_Model}
    (\partial_t^2+2\gamma\partial_t+r)\,\boldsymbol{\phi}+\lambda \rho\,\boldsymbol{\phi}+u\rho\,\partial_t\boldsymbol{\phi}+u'\boldsymbol{\phi}\,\partial_t\rho=0,
\end{align}
where $ \rho = \boldsymbol{\phi}^T\boldsymbol{\phi}$.
The first term amounts to an $N$-component damped harmonic oscillator with damping $\gamma$ and a 'mass' $r$. Overall stability is guaranteed if $u>0,\,\lambda>0$.

Let us then neglect the non-linearities $u$ and $u'$ for the moment. In this case, the equation of motion describes the motion of a particle with an inertial $\sim\partial_t^2$ and a damping $\sim\partial_t$ term in an anharmonic potential. For $r>0,\lambda>0$, the stable state is described by $\boldsymbol{\phi}_s=0$, but when the mass $r$ is tuned through zero, the potential takes a sombrero shape, and a finite expectation value with amplitude $\rho_0\equiv\boldsymbol{\phi}_s^T\boldsymbol{\phi}_s=\frac{-r}{\lambda}$ becomes the stable solution. The direction of the vector $\boldsymbol{\phi}_s$ is chosen spontaneously. The ordering phase transition of this model (once noise is included, and in higher spatial dimension), is described by the Model A universality class of the Hohenberg-Halperin classification \cite{Hohenberg1977}. In fact, dropping the inertial term from scratch (but restoring noise and spatial dimensions), the model matches Model A for an $N$-component order parameter. More generally, the inertial term is irrelevant in the sense of the \ac{RG} for this regime of parameters, and thus does not affect the (mean-field) universal critical behavior. 

Let us now restore the parameters $u$ and $u'$. In contrast to the interaction $\lambda$, these non-linearities cannot be generated by variation of a scalar potential (also referred to as nonconservative for that reason), and typically do not play a role in equilibrium systems~\cite{Taeuber2014}. They amount to nonlinear self-dampings of the field and constitute the simplemost, i.e. lowest order in field amplitude and time derivatives, nonconservative terms that are allowed by symmetry. 
Similar to the $\boldsymbol{\phi}^4$ potential allowing for negative values of $r$, the presence of these couplings provides a mechanism to shift the damping $2\gamma$ and tune it through zero, to trigger a transition into a new, stable phase, where the order parameter is rotating at a constant angular velocity. No matter its precise origin, a negative  damping $\gamma$, i.e. antidamping or pumping, clearly reflects non-equilibrium conditions and does not occur in equilibrium dynamics. This rationalizes that the phase induced by an antidamping defies thermal equilibrium, its nonequilibirum nature being most clearly reflected by the fact that the stable state is time-dependent.

\subsection{Phase Diagram} 

To establish the phase diagram, we search for stable state solutions of Eq.~\eqref{eq:0+1D_Model}. As indicated above, in the nonthermal antidamping regime, there is no stable steady (time-independent) state, but a time-dependent limit cycle where the order parameter rotation is stable. We will refer to this as the rotating phase. We thus make the stable state ansatz 
\begin{equation}
\boldsymbol{\phi}_{s}(\vecx,t)=\sqrt{\rho_{0}}\cdot \Big(\cos{E t},\, \sin{E t},0,...\Big)^T
\end{equation}
with $\rho_{0}>0,\,E\in\mathbb{R}$. For $\gamma<0$, Eq.~\eqref{eq:0+1D_Model} has indeed a stable solution with $E\neq0$ corresponding to the rotating phase. 
The three phases (disordered, statically ordered and dynamically ordered / rotating) that we anticipated are captured by this ansatz, with properties summarized in Table \ref{tab:phases}. 

\begin{table*}
\renewcommand{\arraystretch}{1.5}
\setlength{\tabcolsep}{10pt}
\begin{tabular}{ | p{2.5cm} | p{4cm} | p{4cm}| p{4cm}  |}
\hline
    &\textbf{Disordered phase} & \textbf{Ordered, static phase} &\textbf{Rotating phase}\\\hline
    \textbf{expectation values} & $\rho_0=0,\, E=0$ & $\rho_0\neq0,\, E=0$ & $\rho_0\neq0,\, E\neq0$ \\
    \hline
    \textbf{criterion} & $r>0$, $\gamma>0$& $r<0$, $\delta=2\gamma+u\frac{-r}{\lambda}>0$&$\delta<0$ and $r_{\text{eff}}=r+\lambda\frac{-2\gamma}{u}>0$\\
    \hline
    \textbf{Symmetry breaking pattern} & Disordered, all symmetries remain intact & $O(N)$ is broken down to $O(N-1)$&The field rotates at angular velocity $E$. $O(N)$ is broken down to $O(N-2)$.\\
    \hline
\end{tabular}
    \caption{The three stable phases of the $0+1$ dimensional $O(N)$ theory with nonconservative interactions. The amplitude in the statically ordered phase is $\rho_0=\frac{-r}{\lambda}$. In the rotating phase, the order parameter has an amplitude $\rho_0=-\frac{2\gamma}{u}$ and rotates at angular velocity $E =\sqrt{r_{\text{eff}}}$ with $r_{\text{eff}}=r-\frac{2\gamma\lambda}{u}$. The transition between the statically ordered phase ($r<0$) and the rotating phase occurs upon tuning the effective damping $\delta=2\gamma-\frac{r}{\lambda}$ through zero. 
    }
   \label{tab:phases}
\end{table*}

The phase transition from the disordered phase $\boldsymbol{\phi}_s=0$ can be explained by the equilibrium $\boldsymbol{\phi}^4$ theory. Disregarding the nonthermal nonlinearities $u$ and $u'$, the theory is described by damped motion in a potential 
\begin{align}
    V=\frac{r}{2}\boldsymbol{\phi}^2+\frac{\lambda}{4}\boldsymbol{\phi}^4,
\end{align}
which turns into the well known sombrero hat if $r<0$. In that case, the disordered solution $\boldsymbol{\phi}_s=0$ is unstable, the system undergoes spontaneous symmetry breaking and acquires a nonzero expectation value $\boldsymbol{\phi}_s^2=-\frac{r}{\lambda}$. The existence of this stable solution requires the nonlinearity $\sim \lambda\boldsymbol{\phi}^4$ with $\lambda>0$, otherwise the instability at the origin $\boldsymbol{\phi}=0$ cannot be cured.\\
In a similar manner, a finite $u>0$ allows for a new transition mechanism. A negative damping $\gamma<0$, which usually leads to unstable solutions and does not occur in equilibrium, can now be compensated for by a finite value $\boldsymbol{\phi}_s^2=\rho_0$ such that $2\gamma+u\rho_0\geq0$. Tuning through this transition from the disordered phase, $r>0$, the transition occurs at $\gamma=0$. Afterwards the field saturates to its stable state with $-2\gamma=u\rho_0$. In this case we have however $r+\lambda\rho_0>0$ even  if we neglect $\lambda$, which can be compensated by a rotation with angular velocity
\begin{align}
    E=\sqrt{r+\lambda\rho_0}=\sqrt{r-\frac{2\lambda\gamma}{u}}.\label{eq:MF_frequency}
\end{align}
Since $r$ remains finite at this transition, the angular velocity $E$ jumps at the transition while the field amplitude is continuous.

We can approach this phase also from the statically ordered phase, $r<0$. Since there already exists a finite amplitude $\rho_0=-\frac{r}{\lambda}$ in this phase, there is an effective damping for small fluctuations $\delta=2\gamma+u\rho_0$. As soon as this effective damping becomes negative, i.e. $2\gamma<u\frac{r}{\lambda}$, the order parameter grows again and saturates at $\rho_0=\frac{-2\gamma}{u}$. The order parameter rotates at the frequency \eqref{eq:MF_frequency} dictated by the effective mass $E^2=r_{\text{eff}}=r-\frac{2\lambda\gamma}{u}$. We see that the rotational frequency goes to zero as one approaches the phase transition from the rotating phase on the mean-field level. We thus consider the angular velocity $E$ as the order parameter of the transition between the statically ordered and the rotating phase.

\subsection{Relation to the van der Pol oscillator}

In the remainder of this work, we consider the case $N\geq2$, as for $N=1$ the transition in a limit-cycle phase does not occur through a \ac{CEP}. For $N=1$ the mean-field model reduces to the paradigmatic van der Pol oscillator, with $u$ and $u'$ taking the same form then. The van der Pol oscillator is well known to support stable oscillations of the amplitude $\phi$ for $N=1$ in the parameter region where the mean-field phase diagram supports a rotating phase for $N>2$. In fact the $N\geq2$ model can also support van der Pol oscillations of the amplitude, rather than the above described angular rotations in this region. However, this phase is destabilized by large enough values of $u'$. A more detailed discussion of this phase and its stability can be found in App.~\ref{app:vdp}. We concentrate on the parameter regime where the stable limit cycle is due to rotation of the angular (or phase) variables, to assess the phase transition passing through a \ac{CEP}.\\
In turn, this model also provides the setting to study universal behavior at the phase transition of a generalization of the van der Pol oscillator to spatially extended fields subject to stochastic noise.

\section{Field theoretic set up in \texorpdfstring{$d+1$}{d+1} dimensions}\label{sec:MF+Gauss}

\subsection{Langevin equation description}

We now extend the $0+1$ dimensional model to $d$ spatial dimensions and furthermore, we restore the stochastic element in the dynamics. This enables us to systematically analyse the impact of fluctuations in all three phases and especially at the transitions.
We thus consider a temporally and spatially varying field $\boldsymbol{\phi}(\vecx,t)\in\mathbb{R}^N$, where $\vecx$ is the $d$ dimensional spatial coordinate. Following the paradigm of effective field theory, we write the lowest order couplings allowed by symmetry as done above, as well as the lowest order spatial derivatives to capture the long wavelength dynamics. In addition to the $O(N)$ symmetry of the field, we assume rotational symmetry in space, and arrive at 
\begin{widetext}
\begin{align}\label{eq:Langevin}
    \Big(\partial_t^2+(2\gamma-Z\nabla^2+u\rho(\vecx,t))\partial_t+r+\lambda\rho(\vecx,t)-v^2\nabla^2 +u'\partial_t\rho(\vecx,t)\Big)\boldsymbol{\phi}(\vecx,t)+\xi(\vecx,t)=0.
\end{align}
\end{widetext}
$Z,\, v>0$ are phenomenological parameters determining diffusion and coherent propagation of fluctuations in space in the various phases. $\xi(\vecx,t)$ is a Gaussian white noise with zero mean and variance
\begin{align}\label{eq:gaussianwhite}
\langle\xi_i(\vecx,t)\xi_j(\vecx',t')\rangle=2D\delta(t-t')\delta(\vecx-\vecx')\delta_{ij}.
\end{align}
Let us elaborate here on the importance of noise, which is ubiquitous and non-negligible in systems with many microscopic (or mesoscopic) degrees of freedom. This can be gleaned from an equilibrium situation, where we have a fluctuation-dissipation relation relating the noise level $D$ to the damping rate $\gamma$ according to $D= 4\gamma k_\text{B} T$, $k_\text{B}$ the Boltzmann constant. Noise is non-negligible when the typical frequency of the system, say $r$, is on the same order as $k_\text{B}T$ \cite{Kamenev2011}; due to the smallness of the Boltzmann constant, this only applies to situations with microscopic degrees of freedom as anticipated above. In particular, for a vanishing mass scale (e.g. $r\to 0$) as in the presence of soft Goldstone modes or near a critical point, noise always becomes important. Non-equilibrium conditions are achieved by adding driving mechanisms, but do not alter this picture qualitatively. Quantum fluctuations might also be present, but are generically overwritten by statistical (equilibrium or non-equilibrium) fluctuations of the type described above in the low frequency, long wavelength limit, which justifies the semiclassical limit within which we study our $O(N)$ models \cite{Sieberer2016}. 

Within the effective field theory paradigm, we thus expect  this model to describe the low frequency, long wavelength fluctuations of a system with the assumed symmetries. As mentioned above, if one restricts to the dynamics of spatially homogeneous field configurations and neglects the noise, one recovers the $0+1$ dimensional model discussed prior. Thus, the phase diagram in Fig. \ref{fig:PhaseDiagram_a} is the mean-field phase diagram of the $d+1$ model.



\subsection{MSRJD representation and effective action}

To study the impact of noise induced fluctuations systematically, we turn to the equivalent description of the Langevin equation \eqref{eq:Langevin} in terms of a path integral following the \ac{MSRJD} construction \cite{DOMINICIS1976,Martin1973,Janssen1976}. A Langevin equation 
\begin{align}
    \mathcal{L}[\boldsymbol{\phi}]\boldsymbol{\phi}(\vecx,t)+\boldsymbol{\xi}(\vecx,t)=0
\end{align}
with Gaussian white noise defined in Eq.~\eqref{eq:gaussianwhite} corresponds to a path integral
\begin{align}\label{eq:zpathint}
    Z[\boldsymbol{j},\tilde{\boldsymbol{j}}]=\int\mathcal{D}\boldsymbol{\phi}\mathcal{D}i\tilde{\boldsymbol{\phi}} e^{-S[\boldsymbol{\phi},\tilde{\boldsymbol{\phi}}]+\int_{X}\tilde{\boldsymbol{j}}^T\boldsymbol{\phi}+\boldsymbol{j}^T\tilde{\boldsymbol{\phi}}}
\end{align}
with the action 
\begin{align}
    S[\tilde{\boldsymbol{\phi}},\boldsymbol{\phi}]=\int_{X}\tilde{\boldsymbol{\phi}}^T(X)\mathcal{L}[\boldsymbol{\phi}]\boldsymbol{\phi}(X)-D\tilde{\boldsymbol{\phi}}^T(X)\tilde{\boldsymbol{\phi}}(X).
\end{align}
$\boldsymbol{\phi}$ is the $N$-component order parameter field also entering the Langevin equation, and we introduced $X=(\vecx,t)$ to streamline notation. $\tilde{\boldsymbol{\phi}}$ is an $N$-component auxiliary variable, associated to the noise, often referred to as response or quantum field. The path integral $Z[\boldsymbol{j},\tilde{\boldsymbol{j}}]$ generates the noise averaged correlation and response functions of the Langevin dynamics by taking derivatives with respect to the source fields $\boldsymbol{j},\,\tilde{\boldsymbol{j}}$, and evaluating at vanishing sources. In particular, the (retarded) two-point response function and correlation function are, again using a shorthand notation $Q=(\vecq,\omega)$
\begin{align}
    \chi_{ij}^R(Q,Q')&=\frac{\delta^2 \ln Z}{\delta \tilde j_i(Q)\delta j_j(Q')}\Big|_{\boldsymbol{j}=\tilde{\boldsymbol{j}} = 0}\equiv G_{ij}^R(Q)\delta(Q+Q'),\\
    \mathcal{C}_{ij}(Q, Q')&=\frac{\delta^2 \ln Z}{\delta \tilde j_i(Q)\delta \tilde j_j(Q')}\Big|_{\boldsymbol{j}=\tilde{\boldsymbol{j}} = 0}\equiv G_{ij}^K(Q)\delta(Q+Q'),
\end{align}
where we used time and space translation invariances. The rotating phase has a time-dependent stable state which generically breaks this structure, but we will see that in the proper comoving frame it is recovered. 

These objects represent the \emph{full} two-point Green functions of the theory, including all corrections due to nonlinearities and noise. Absent spontaneous symmetry breaking, they are $\propto \delta_{ij}$ by $O(N)$ symmetry.  The full Green function in Fourier space $\mathcal G(Q)$ is a $2\times2$ matrix in the Nambu space $\boldsymbol{\Phi}=(\boldsymbol{\phi},\tilde{\boldsymbol{\phi}})^T$ and has the form
\begin{align}
    \mathcal{G}(Q)=\begin{pmatrix}
        G^K(Q) & G^R(Q)\\
        G^A(Q) & 0
    \end{pmatrix}.
\end{align}
We introduce here a notation borrowed from Keldysh field theory, with retarded ($G^R$), advanced ($G^A = (G^R)^\dag$) and Keldysh ($G^K$) component for the Green function. It highlights the connection to the Keldysh formalism for quantum systems out of equilibrium, from which the \ac{MSRJD} path integral emerges as a semiclassical limit, see e.g. \cite{Sieberer2016} for a review. 

While the path integral for the dynamical partition function, Eq.~\eqref{eq:zpathint}, encodes all information of the problem, we transit here to another object -- the effective action (see \cite{Amit2005} for an in-depth discussion of this object, and~\cite{Taeuber2014,Sieberer2016} for the nonequilibrium effective action). It encodes the same information but organizes it in a way that is beneficial for the analysis of the present problem, both conceptually and in terms of practical calculations. For example, it allows for a simple proof of Goldstone's theorem, and the construction of the associated soft modes including in the rotating phase. It will also enable us to develop a quantitative potential picture for the fluctuation induced first order transition. 

The effective action functional is defined as the Legendre transform of the generating functional for connected correlation functions, $W[\boldsymbol{j},\tilde{\boldsymbol{j}}]=\ln{Z[\boldsymbol{j},\tilde{\boldsymbol{j}}]}$:  $\Gamma[\boldsymbol{\varphi},\tilde{\boldsymbol{\varphi}}] =\sup_{j,\tilde j}[ - W[\boldsymbol{j},\tilde{\boldsymbol{j}}] + \int_X \boldsymbol{j} \tilde{\boldsymbol{\varphi}} + \tilde{\boldsymbol{j}} \boldsymbol{\varphi}]$.
Similarly to a classical action, the effective action induces an equation of motion. Its solution $\boldsymbol{\varphi}_s$ yields the physical field expectation value, with $\boldsymbol{\varphi}_s\neq0$ signalling macroscopic occupation/condensation, while $\tilde{\boldsymbol{\varphi}} =0$ when evaluated at the physical point due to
probability conservation \cite{Kamenev2011}. The full equation of motion is given by 
\begin{align}
\label{eq:effEoM}
    \frac{\delta\Gamma}{\delta\boldsymbol{\varphi}}\Big|_{\boldsymbol{\varphi}=\langle\boldsymbol{\phi}\rangle,\tilde{\boldsymbol{\varphi}}=\langle\tilde{\boldsymbol{\phi}}\rangle}=\frac{\delta\Gamma}{\delta\tilde{\boldsymbol{\varphi}}}\Big|_{\boldsymbol{\varphi}=\langle\boldsymbol{\phi}\rangle,\tilde{\boldsymbol{\varphi}}=\langle\tilde{\boldsymbol{\phi}}\rangle}=0.
\end{align}
The effective action has an intuitive path integral representation as 
\begin{align}
\label{eq:effective_action}
    \Gamma[\boldsymbol{\varphi},\tilde{\boldsymbol{\varphi}}]=-\ln \int\mathcal{D}\boldsymbol{\phi}\mathcal{D}i\tilde{\boldsymbol{\phi}} e^{-S[\boldsymbol{\phi}+\boldsymbol{\varphi},\tilde{\boldsymbol{\phi}}+\tilde{\boldsymbol{\varphi}}]+\frac{\delta\Gamma}{\delta\tilde{\boldsymbol{\varphi}}}\tilde{\boldsymbol{\phi}}+\frac{\delta\Gamma}{\delta\boldsymbol{\varphi}}\boldsymbol{\phi}},
\end{align}
with $\tilde{\boldsymbol{j}}=\frac{\delta\Gamma}{\delta\boldsymbol{\varphi}},\, {\boldsymbol{j}}=\frac{\delta\Gamma}{\delta\tilde{\boldsymbol{\varphi}}}$. Eq. \eqref{eq:effective_action} states that the effective action obtains from the bare action by summing over all possible configurations of the Nambu field $\boldsymbol{\Phi} = (\boldsymbol{\phi}, \tilde{\boldsymbol{\phi}})^T$. Conversely, omitting fluctuations in a mean-field approximation reproduces the bare action, $\Gamma[\boldsymbol{\varphi},\tilde{\boldsymbol{\varphi}}]= S[\boldsymbol{\varphi},\tilde{\boldsymbol{\varphi}}]$. The representation makes it transparent that the effective action shares the symmetries of the bare one absent sources.

The second derivative with respect to the  Nambu field $(\boldsymbol{\varphi}(Q),\tilde{\boldsymbol{\varphi}}(Q))^T$ around a time and space translation invariant solution of the equations of motion satisfies
\begin{align}
\nonumber
    \Big(\Gamma^{(2)}(Q,Q')\Big)^{-1}&=\begin{pmatrix}
        0 & \Gamma^A(Q)\\
        \Gamma^R(Q) & \Gamma^K(Q)
    \end{pmatrix}^{-1}\delta(Q+Q')\\
    &=\begin{pmatrix}
        G^K(Q) & G^R(Q)\\
        G^A(Q) & 0
    \end{pmatrix}\delta(Q+Q'),
\end{align}
and thus gives the full Green function of the theory in $\vecq,\omega$-space including the retarded and advanced responses $G^{R/A}(Q)=\Big(\Gamma^{R/A}(Q)\Big)^{-1}$ and the correlation function $G^K(Q)=-G^R(Q)\Gamma^K(Q)G^A(Q)$~\footnote{Note that a breaking of time translation invariance leads to Green functions that are not diagonal in frequency space.}.

Higher order field derivatives of $\Gamma$ give the full one-particle irreducible (1PI), or amputated, correlators.
To streamline equations in the remainder of the text, we introduce the following notation for field derivatives of the effective action evaluated on $\boldsymbol{\varphi}=\langle \boldsymbol{\phi} \rangle$:
\begin{align}
\nonumber
   & \Gamma^{(m,n)}_{i_1 ... i_{n+m}}(X_1, ... , X_{n+m}) \equiv\\
    &\frac{\delta^{m+n}\Gamma}{\delta\tilde{\varphi}_{i_1}(X_1)...\delta\tilde{\varphi}_{i_m}(X_m)\delta\varphi_{i_{m+1}}(X_{m+1})...\delta\varphi_{i_{n+m}}(X_{m+n})}.
\end{align}

Following this construction, the bare MSRJD action $S[\boldsymbol{\phi},\tilde{\boldsymbol{\phi}}]$ corresponding to our model \eqref{eq:Langevin} is given by
\begin{widetext}
\begin{align}\label{eq:action_mic}
  S[\boldsymbol{\phi},\tilde{\boldsymbol{\phi}}]&=S_0[\boldsymbol{\phi},\tilde{\boldsymbol{\phi}}]+S_{\text{int}}[\boldsymbol{\phi},\tilde{\boldsymbol{\phi}}],\\
  S_0[\boldsymbol{\phi},\tilde{\boldsymbol{\phi}}]&=\int_{X} \tilde {\boldsymbol{\phi}}(X)^T\Big(\partial_t^2+(2\gamma-Z\nabla^2)\partial_t+r-v^2\nabla^2 \Big)\boldsymbol{\phi}(X)-D\tilde {\boldsymbol{\phi}}(X)^T\tilde {\boldsymbol{\phi}}(X) \\
    S_0[\boldsymbol{\phi},\tilde{\boldsymbol{\phi}}]&=\frac{1}{2}\int_{Q}(\boldsymbol{\phi}(-Q),\tilde{\boldsymbol{\phi}}(-Q))\mathcal{G}^{-1}_0
    \begin{pmatrix}
        \boldsymbol{\phi}(Q)\\
        \tilde{\boldsymbol{\phi}}(Q)
    \end{pmatrix},\\
    \label{eq:G0gauss}
    \mathcal{G}^{-1}_0&=\begin{pmatrix}
        0 & -\omega^2+i\omega(2\gamma+Z\vecq^2)+r+v^2\vecq^2\\
        -\omega^2-i\omega(2\gamma+Z\vecq^2)+r+v^2\vecq^2 & -2D
    \end{pmatrix},\\
    S_{\text{int}}&=\int_{X}\lambda\tilde{\boldsymbol{\phi}}(X)^T \boldsymbol{\phi}(X)\rho(X)
    +u\tilde{\boldsymbol{\phi}}(X)^T \partial_t\boldsymbol{\phi}(X)\rho(X)
    +u'\tilde{\boldsymbol{\phi}}(X)^T \boldsymbol{\phi}(X)\partial_t\rho(X),
\end{align}
\end{widetext}
where $ \mathcal{G}^{-1}_0$ is the bare inverse Green function. 
As mentioned above, the effective action will share the symmetries of the bare one, but encode the effects of fluctuations in terms of renormalized parameters. In a gradient approximation, the effective action maintains the functional form of the microscopic action but with renormalized effective couplings. 
We denote the renormalized versions of action parameters with bars in the following, e.g. $\bar\gamma$ denotes the renormalized damping. Since the path integral cannot be performed exactly in general, one has to resort to e.g. perturbation theory, resummations, or renormalization group techniques to derive corrections to the bare couplings from the microscopic action $S$. In this way one could obtain an improved phase diagram in terms of the renormalized and a priori unknown parameters of the effective action, which includes noise and interaction effects. Here we do not aim for precision estimates of the nonuniversal fluctuation corrections to transition lines. We will rather focus on universal aspects of the fluctuations corrections below.

\section{Exceptional points and critical exceptional points}\label{sec:EPs}

\subsection{Modes, dispersions and critical points}\label{ssec:modes_and_EPs}

\begin{figure}
\centering
\subfloat[][]{%
  \includegraphics[width=0.23\textwidth]{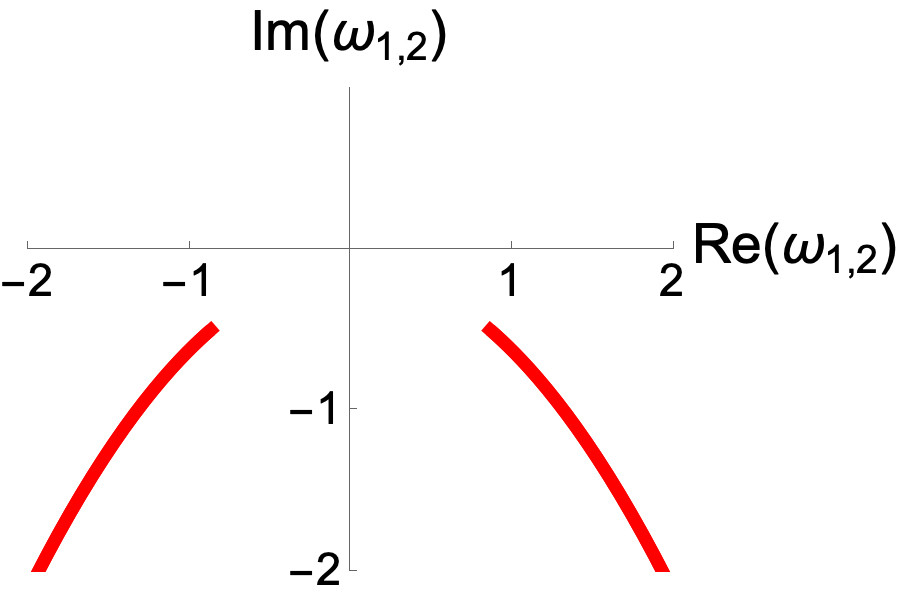}%
}\hfill
\subfloat[][]{%
  \includegraphics[width=0.23\textwidth]{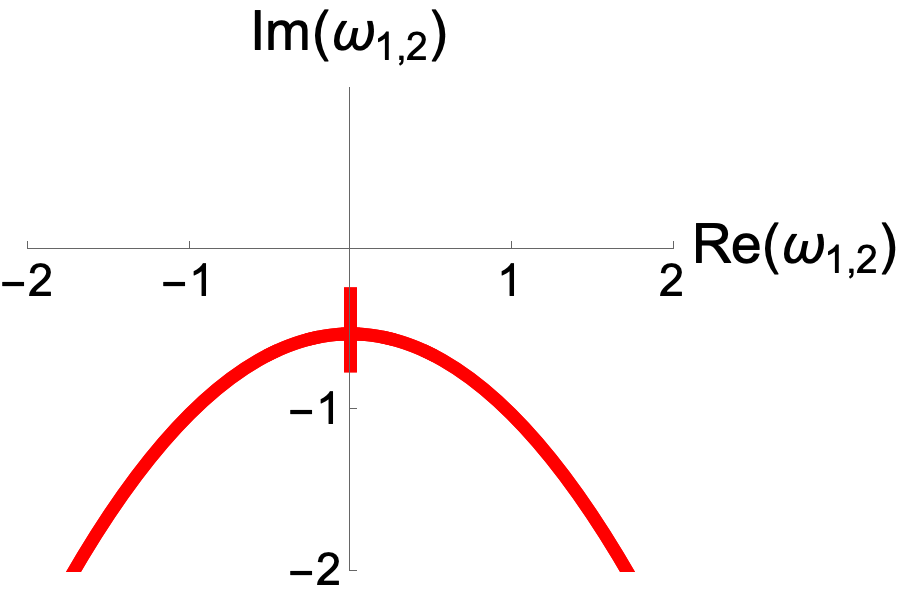}%
}\hfill
\subfloat[][]{%
  \includegraphics[width=0.23\textwidth]{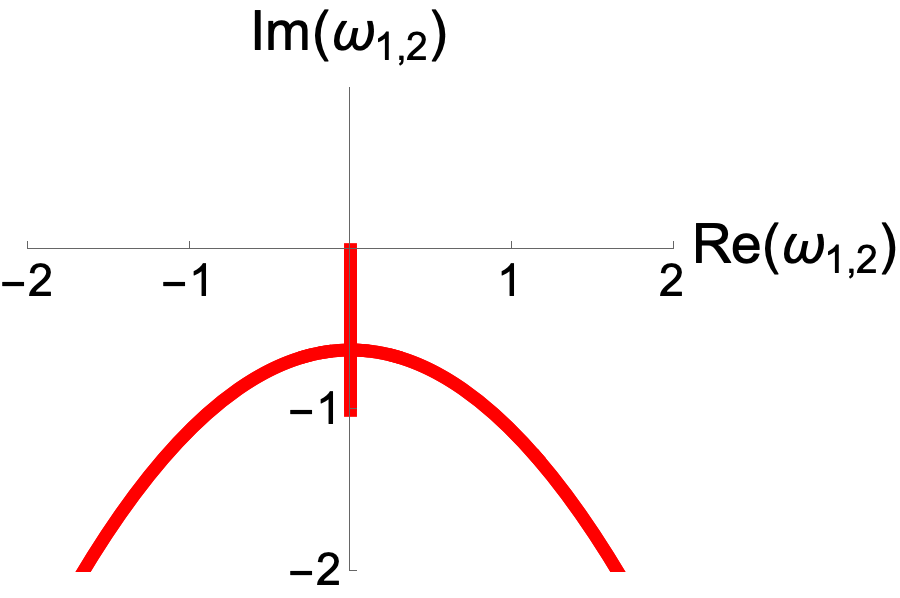}%
}\hfill
\subfloat[][]{%
  \includegraphics[width=0.23\textwidth]{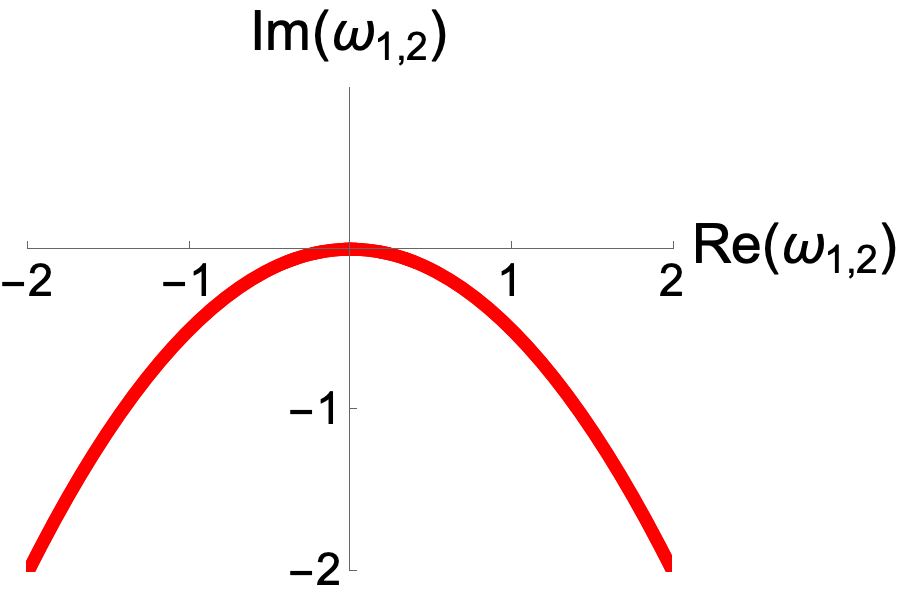}%
}
\caption{Position of the poles of the retarded response of a scalar field in the complex plane. The dispersions are parametrized as $\omega_{1,2}=-\frac{i}{2}(2\gamma+Z\vecq^2)\pm\sqrt{r+v^2\vecq^2-(2\gamma+Z\vecq^2)^2/2}$, with $v=Z=1$.
a) Purely underdamped motion $r=2\gamma$, all poles have a finite distance from the real and imaginary axis. b) Underdamped excitations exist, constituting a line of poles on the imaginary axis. The gap between real axis and the pole spectrum remains finite $r=0.4\gamma$. At large enough wavevectors, there is an EP separating the underdamped from the overdamped regime. It can be clearly detected by the pole lines with finite real part terminating nonanalytically in the line of overdamped excitations. c) Gapless (critical) excitation spectrum, $2\gamma>0, r=0$. The line of underdamped poles touches the zero in the complex plane. At finite damping, a gapless spectrum always has an underdamped regime at low momenta. d) \ac{CEP} spectrum, $2\gamma=r=0$. The EP, where the underdamped motion terminates sits at the zero in the complex plane. At finite momenta, all excitations are underdamped. Real and imaginary parts of the dispersions scale differently with momentum.}
\label{fig:poles}
\end{figure}

We now define the notion of an EP and a \ac{CEP} within the above formulation. Before doing so, we briefly fix some further basic conventions and nomenclature for the remainder of this work. We can access the \emph{mode spectrum} around a given stable state $\boldsymbol{\varphi}_s$ by linearizing the coarse grained equation of motion around its solution
\begin{align}
\label{eq:LinEq}
    \sum_j\Gamma^{R}_{ij}(\vecq,t)\Big|_{\boldsymbol{\varphi}=\boldsymbol{\varphi}_s}\delta\varphi_j(\vecq,t)=0,
\end{align}
where we have assumed that the equation of motion is Markovian, i.e. depends only on one time variable, as it is the case for this work. The set of linearly independent solutions $\Big(\boldsymbol{\delta\varphi}_\alpha(\vecq,t)\Big)_{\alpha=1,..}$ are the excitation modes. Put differently, the modes span the kernel of the inverse Green function in time and momentum space $\Gamma^{R}(\vecq,t)$. If $\Gamma^{R}(\vecq,t)$ is not explicitly time dependent but only contains time derivative operators, the modes usually take the $\boldsymbol{\delta\varphi}_\alpha(\vecq,t)=e^{-i\omega_\alpha(\vecq)t}\boldsymbol{\delta\varphi}_\alpha(\vecq,0)$, where $\omega_\alpha(\vecq)$ are the mode dispersions. The dispersions are also the roots of $\det\Gamma^R(\omega,\vecq)=0$ and equivalently the poles of the retarded Green function in frequency space. The real part of a dispersion gives the \emph{frequency} or inverse \emph{period} at which the corresponding mode oscillates, while the imaginary part yields how fast the mode dissipates, i.e. its inverse \emph{life time}. See also Fig.~\ref{fig:poles} for an illustration.

For the solution $\boldsymbol{\varphi}_s$ to be stable, no dispersion can have a positive imaginary part since this corresponds to an exponentially growing fluctuation. Therefore, an instability towards a new phase occurs if one tunes some parameter such that a dispersion is at the verge of moving into the upper complex half plane, i.e.\ when the imaginary part of the dispersion goes to zero. The system reaches a \emph{critical point} and a continuous phase transition takes place, indicated by a divergence of e.g.\ the two-point correlation function at equal-time $G^K(q,t=0)$. Typically, continuous transitions occur for a vanishing dispersion $\omega(\vecq)=0$ but an instability at finite frequency can however occur, too. This corresponds, for example, to the cases $\textbf{I}_0$ ($\vecq_c=0$) and $\textbf{III}_0$ ($\vecq_c\neq0$) in the classification of instabilities in noiseless systems by Cross and Hohenberg \cite{Cross1993}. 

In the simplest case of a single scalar field variable, the linearized renormalized equation of motion at low frequencies reduces to the damped harmonic oscillator
\begin{align}\label{eq:general_EOM}
    \Big(\partial_t^2+2\bar\gamma(\vecq)\partial_t+\bar r(\vecq)\Big)\boldsymbol{\delta\varphi}(\vecq,t)=0
\end{align} 
or, equivalently
\begin{align}\label{eq:general_quadratic_action}
    \Gamma^{R}(\vecq,\omega)=(-\omega^2-2\bar\gamma(\vecq)\omega+ \bar r(\vecq)).
\end{align}
The modes are 
\begin{align}
    \delta\varphi_{1,2}(\vecq,t)=e^{-i\omega_{1,2}(\vecq)t}
\end{align}
with dispersions
\begin{align}\label{eq:general_dispersions}
    \omega_{1,2}(\vecq)=-i\bar\gamma(\vecq)\pm\sqrt{\bar r(\vecq)-\bar\gamma(\vecq)^2},
\end{align}
As an example at the mean-field level, we have $r(\vecq)=v^2 \vecq^2+r$ and $\gamma(\vecq)=\gamma+\frac{Z}{2} \vecq^2$ using~\eqref{eq:G0gauss}. 

Stability, i.e. a finite lifetime for both modes, demands that $r>0,\,\gamma>0$. If one tunes the \emph{mass term} $r$ to zero, one dispersion becomes gapless $\omega_1=0$ while the other remains decaying $\omega_2=-2i\bar\gamma(\vecq)$. The first becomes unstable upon tuning the mass $r$ negative. 
In our case, we reach the critical point describing the phase boundary A of the phase diagram Fig.~\ref{fig:PhaseDiagram}. Tuning the damping $\gamma$ negative also induces an instability. However, it does not proceed through a point where the dispersions vanish in the complex plane, but both dispersions maintain a finite real part $\omega_{1,2}=\pm \sqrt{r}$ at $\gamma=0$. It corresponds to the phase transition B in Fig.~\ref{fig:PhaseDiagram}.

\subsection{(Critical) exceptional point}\label{ssec:EPFluct}

We first consider the case of a single damped oscillator, $N=1$. A special point occurs when there is a wavevector $\vecq^*$ at which $\bar\gamma^2(\vecq^*)= \bar r(\vecq^*)$ and both formerly independent modes coalesce. At this point a new linearly independent solution emerges: $\boldsymbol{\delta\varphi}_{EP}(\mathbf{q}^*,t)=t e^{-i\omega(\vecq^*)t} \boldsymbol{\delta\varphi}_{EP}(\mathbf{q}^*,0)$. This marks an \emph{exceptional point}. The damped harmonic oscillator's EP separates a purely dissipative, overdamped regime, where both dispersions are imaginary without a real part, and an underdamped regime where excitations oscillate due to a finite real part of their dispersions. Clearly, at an EP the square root appearing in~\eqref{eq:general_dispersions} vanishes and therefore the EP occurs at a nonanalyticity of the dispersion relations.

Equivalently, it is also possible to rewrite Eq.~\eqref{eq:general_EOM} as a first-order linear differential equation of the form $\partial_t \delta\boldsymbol{\varphi}=M\delta\boldsymbol{\varphi}$. An exceptional point, i.e.\ a coalescence of modes, is then defined as a point in parameter space where the $2\times 2$ matrix $M$ is not diagonalizable in internal indices, making 
contact with the more usual definition of EP~\cite{Heiss2004,AshidaGongUeda,BergholtzBudichKunstRMP} (see also App.~\ref{app:Definition_CEP}).

We say that there is a \emph{critical exceptional point}, if the dispersion at which the EP occurs is gapless, i.e. when $\bar\gamma(\vecq^*)=\bar r(\vecq^*)=0$.
For $\vecq^*=0$, we then have, at the \ac{CEP},
\begin{align}
    \label{eq:CEPdef}
    \Gamma^R(\vecq^*=0,\omega)=-\omega^2,
\end{align}
underlying the necessity to keep the second order time derivative.
We emphasize again that a \ac{CEP} is hence a property of the full renormalized inverse retarded Green function.

We now generalize the notion of a \ac{CEP} to the dynamics of $N$ component fields. A full retarded Green function that can be diagonalized in field space, 
\begin{align}
\label{eq:GamDiag}
\Gamma^{R}_{ij}(Q)=\Gamma^{R}_i(Q)\delta_{ij},
\end{align}
where $\Gamma^{R}_i$ are of the form~\eqref{eq:general_quadratic_action} therefore displays a \ac{CEP} if and only if the full diagonalized inverse Green function has at least one element $\Gamma^{R}_i$ which verifies~\eqref{eq:CEPdef}. We show in App.~\ref{app:Definition_CEP} that the case of a CEP occurring through a nondiagonalizable Green function can always be mapped to this case in the vicinity of the CEP. Since the dynamics is diagonal and thus decoupled, we now drop the index $i$ and concentrate on the pair of modes becoming critical and  exceptional simultaneously.
In our case at mean-field, the inverse Green function is diagonal and all its elements take the form 
\begin{align}
    \label{eq:GamRCEP}
    \Gamma^R(\vecq,\omega)=-\omega^2-Z i \omega \vecq^2+ v |\vecq|,
\end{align}
 and the dispersions at the \ac{CEP} are
\begin{align}
\label{eq:dispCEP}
    \omega_{1,2}(\vecq)=-i\frac{Z}{2}\vecq^2\pm v|\vecq|.
\end{align}

Reaching a \ac{CEP} generically requires two fine tunings, both $\bar \gamma(\vecq=0)$ and $\bar r(\vecq=0)$ have to be tuned to zero. We will show however in Sec.~\ref{sec:MF+Gauss} that the transition between the static and the rotating phase constitutes a \ac{CEP}. There is only one fine tuning necessary as the vanishing of $\bar r(\vecq=0)$ for phase fluctuations in the static ordered phase is guaranteed by Goldstone's theorem. The idea to generate \ac{CEP}s with only one fine-tuning by considering systems with a Goldstone mode was first put forward in~\cite{Hanai2020,Fruchart2021}.

\subsection{Superthermal mode occupation}\label{subsec:superth}

The discussion of (critical) exceptional points above makes it clear that these are spectral properties, related to the retarded Green function. Now we study the consequences of such points for the statistical properties, i.e. mode occupation numbers. These are encoded in the full equal-time correlation function or Keldysh Green function. The \ac{CEP} is signalled by a vanishing of two coalescing modes $\omega_{1,2}(\vecq)$ as $\vecq\rightarrow 0$. Near the \ac{CEP}, the Keldysh Green function associated to the coalescing critical modes takes the form
\begin{align}\label{eq:gkex}
    G^K(Q)=\frac{2\bar D(Q)}{|\omega-\omega_1(\vecq)|^2|\omega-\omega_2(\vecq)|^2},
\end{align}
where $\bar D(Q)\equiv \Gamma^K(Q)$ is a generic frequency and momentum dependent noise kernel of the respective field direction.

To determine the physics at low frequencies and momenta, we can restrict the discussion to $\bar D(Q \to 0)\equiv \bar D$, which absent fine tuning is larger than zero, corresponding to a generic Markovian noise level
~\footnote{The constant noise level also distinguishes the \ac{CEP} from the Goldstone fixed point of models with conserved currents such as in the Hohenberg-Halperin E, F and G~\cite{Hohenberg1977,Taeuber2014}, where the spectrum of the Goldstone excitations in the ordered phase coincide with that of a \ac{CEP}. However, their noise kernel has to vanish as $D(\vecq)\sim\vecq^2$ due to conservation laws and thus there is no enhanced fluctuations as at a \ac{CEP} transition. These scaling regimes do not describe transitions but the fixed points of symmetry broken phases itself.}.

This general property of a \ac{CEP} reproduces the structure pointed out in \cite{Hanai2020}.
There are two poles at $\omega=0$ that multiply, causing a significantly enhanced infrared divergence of the correlation function, irrespective of the precise forms of the dispersions. This can be easily seen by inspecting the equal-time Keldysh Green-function obtained from~\eqref{eq:gkex},
\begin{align}
    G^K(\vecq,t=0)\sim\frac{\bar D}{\bar \gamma(\vecq)\bar r(\vecq)},
\end{align}
since both $\gamma$ and $r$ go to zero precisely at the \ac{CEP}.

With the mean-field dispersions~\eqref{eq:dispCEP}, the equal-time correlation function is given by
\begin{align}
    G^K(\vecq,t=0)\sim\frac{D}{\vecq^4},
\end{align}
which has a significantly stronger infrared divergence as in the vicinity of a usual (Gaussian) critical point where $G^K(\vecq,t=0)\sim\vecq^{-2}$ e.g.\ at the phase boundary A and B of the phase diagram Fig.~\ref{fig:PhaseDiagram}, where respectively $r$ and $\gamma$ are fine-tuned to 0. In particular, it is superthermal: the fluctuation-dissipation relation (see next subsection) implies generally that $G^K(\vecq,t=0)\sim\vecq^{-2}$. This is a hint that a \ac{CEP} is a genuine non-equilibrium feature.

\subsection{CEP exists only out-of-equilibrium}\label{ssec:nonthermal}

Here we show that indeed, a \ac{CEP} cannot occur at thermal equilibrium. In that circumstance, the full correlation and response functions obey a \ac{fdr}, which reads for the two-point functions ($k_\text{B}=1$)
\begin{align}
    G^K(Q)=\frac{2T}{i\omega}\Big(G^R(Q)-G^A(Q)\Big).
\end{align}\\
In  thermal equilibrium with global detailed balance, \ac{fdr}s have to hold not only for the full, renormalized two-point Green functions, but  also for all higher $n$-point correlations and responses as well. This leads to an infinite tower of relations to be checked. This can however be elegantly avoided, as the \ac{fdr} can be understood as a consequence of a symmetry of the MSRJD (or Schwinger-Keldysh) action and effective action~\cite{Janssen1976,Bausch1976a,Andreanov2006,Aron2010,Sieberer2016,Gyoergi1992, Aron2016, Enz79}. Rather than calculating all full $n$-point functions, it is sufficient to check if the MSRJD action has that symmetry to establish if the system is in thermal equilibrium or not. For $O(N)$ vector fields, this  thermal symmetry is given by
\begin{equation}
\label{eq:ThermalSymm}
\begin{split}
    &\boldsymbol\varphi(\vecx,t)\rightarrow \boldsymbol\varphi(\vecx,-t),\\
    &\tilde{\boldsymbol\varphi}(\vecx,t)\rightarrow \tilde{\boldsymbol\varphi}(\vecx,-t)+\beta\partial_t\boldsymbol\varphi(\vecx,-t).
\end{split}    
\end{equation}
There is one parameter in the transformation, which is associated to the temperature $\beta=\frac{1}{T}$, shared by all subsystems (all subsystems are in equilibrium with each other, sometimes referred to as detailed balance).
Force terms $\sim \tilde{\boldsymbol{\phi}}(\vecx,t)F[\boldsymbol{\phi}]$ in the Lagrangian generate the following additional contribution under the symmetry operation \eqref{eq:ThermalSymm}
\begin{align}
    \int_t\tilde{\boldsymbol\phi}(\vecx,t)F[\boldsymbol\phi]&\rightarrow\int_t\tilde{\boldsymbol\phi}(\vecx,t)F[\boldsymbol\phi]+\delta S,\\
    \delta S&=\int_t\beta\partial_t\boldsymbol{\phi}(\vecx,t)F[\boldsymbol\phi].
\end{align}
If now the force $F[\boldsymbol\phi]$ is conservative, i.e. $F[\boldsymbol\phi]=-\frac{\delta V}{\delta\boldsymbol\phi}$ we have
\begin{align}
    \delta S=\int_t \frac{dV[\boldsymbol\phi]}{dt}=0.
\end{align}
Thus, any conservative term is invariant under~\eqref{eq:ThermalSymm}. Non-conservative damping terms are allowed in equilibrium, however only if they come with associated noise terms with a strict relation for the coefficients, e.g. for the full momentum dependence of the damping
\begin{align}
\label{eq:thermalNoise}
    &\sim \int_{\mathbf{q},t}\bar \gamma(\vecq)\tilde{\varphi}_i(\partial_t\varphi_i- T\tilde{\varphi}_i), \\
    &\sim \int_{\mathbf{q},t}(2\bar\gamma+ \bar Z \vecq^2+ \dots)\tilde{\varphi}_i(\partial_t{\varphi}_i- T\tilde{\varphi}_i), 
\end{align}
so that the thermal symmetry is realized. The presence of the thermal symmetry is then equivalent to the existence of a fixed ratio between dissipative and fluctuating terms.

In other words, the quadratic part of the action~\eqref{eq:action_mic} is invariant under this transformation if the full renormalized damping $\bar\gamma(\vecq)$ and the full renormalized noise level $\bar D(\vecq)$ are proportional to each other with
\begin{align}\label{eq:betaD}
    T = \frac{\bar D(\vecq)}{ 4\bar\gamma(\vecq)},
\end{align}
where in a state of true thermal equilibrium the temperature is independent of the momentum $\vecq$.

If the system is driven out of equilibrium on a more microscopic level, such a fine tuning of parameters is unnatural.
However, thermal symmetry (i.e. equilibrium) can emerge under coarse graining at long wavelength, e.g.\ in the vicinity of phase transitions~\cite{Mitra2006,Foss-Feig2017,Overbeck2017,Taeuber2014}.
In particular, the effective long wavelength description of the symmetric phase, the static ordered phase and the phase transition between them are characterized by such an emergent thermal behavior, as we will show in the next section. 

This reasoning however breaks  down as a matter of principle as one tunes $\gamma$ through zero entering the rotating phase. Intuitively, this phase is clearly nonthermal, as it has a time dependent stable state and such a {\em perpetuum mobile} cannot occur in equilibrium. This behavior should extend to
phase boundaries of the rotating phase, and therefore in particular at a CEP.

More formally, for the damped harmonic oscillator~\eqref{eq:general_quadratic_action} with $\bar\gamma(\vecq=0) \neq 0$, it is always possible to realize the thermal symmetry~\eqref{eq:ThermalSymm}  with a temperature  $T=\frac{D(\vecq)}{4\gamma(\vecq)}$. But, by definition of a \ac{CEP} where the full renormalized damping at zero momentum is tuned to zero, $\bar\gamma(\vecq \to 0) \sim |\vecq|^{\alpha}$ with $\alpha>0$, in the presence of a finite noise level $\bar D(\vecq \to 0)\sim \bar D$, the dynamics has to break thermal equilibrium conditions. Indeed,~\eqref{eq:betaD} does not hold and the form~\eqref{eq:thermalNoise} is not realized.

\section{Statically and rotating ordered phases: Goldstone modes and low lying excitations}\label{sec:fluctuations}

After having introduced the methodology, we now include fluctuations around the mean-field phases discussed in Sec \ref{sec:MeanField}. We first analyse the symmetry breaking patterns characterising the three phases and show formally how Goldstone modes follow from symmetry breaking in the rotating phase. After assuming a certain form of the stable state field expectation value -- which differs in the static and rotating phases -- this discussion is exact. We afterwards discuss the spectrum of linearized fluctuations in all three phases, and thereby access the Gaussian fixed points describing the phase transitions. This discussion is exact in the long wavelength limit above the upper critical dimension, which we determine to be $d_c=4$ in Sec.~\ref{sec:ExceptionalRG}, and forms the basis for the loop fluctuation analysis, see also Sec. \ref{sec:ExceptionalRG}.

\subsection{Symmetry breaking patterns and Goldstone modes}\label{subsec:symmetry_breaking}

We start our discussion of the phase diagram by analysing the symmetry properties of the three phases, i.e. studying which part of the $O(N)$ symmetry is broken by the respective stable states. This leads to the emergence of $N-1$ Goldstone modes in the statically ordered phase as usual, and $2N-3$ Goldstone modes in the rotating phase. These statements are not confined to approximations but rely on general exact properties of the effective action for given symmetry breaking patterns. 

\subsubsection{Statically ordered phase} \label{sssec:SSB_Rotating}
The equation of motion \eqref{eq:Langevin} and the effective action~\eqref{eq:effective_action} are invariant under global $O(N)$ transformations of the field. Here we stress that it is actually invariant under rotations and reflections, i.e. $O(N)\cong SO(N)\ltimes\mathbb{Z}_2$ - the product is semi-direct since rotations and reflections generally do not commute. For $N>2$ this difference is not relevant for our purposes, it will however turn out to be of crucial importance in the case $N=2$. In the disordered phase the stable state order parameter $\boldsymbol{\varphi}_s=0$ does not transform under $O(N)$ and therefore the full symmetry group remains intact.\\
We now turn to the statically ordered sector, where we can parametrize the noise averaged stable state order parameter as
\begin{align}\label{eq:ssbstat}
   \boldsymbol{ \varphi}_s=\sqrt{\rho_0}(1,0,...,0)^T\equiv\sqrt{\rho_0} \hat e_1.
\end{align}
The direction is picked spontaneously, and without loss of generality we choose it to be aligned with the $1$-axis.
The symmetry group $SO(N)$ is generated by the skew symmetric real $N\times N$ matrices $T_{ij} = - T_{ji}$, which are parameterized as $(T_{ij})_{nm} = \delta_{in}\delta_{jm} - \delta_{im}\delta_{jn}$. Each of these generates a rotation of the two components $i,j=1,...,N$  into each other. In other words, it generates rotations in the plane spanned by $\hat e_{i},\hat e_j$. There are ${\binom{N}{2}}=\frac{N(N-1)}{2}$ such rotations/generators. The stable state breaks the $N-1$ generators that mix the first component with any other, thus there are $N-1$ Goldstone modes and the remaining $\frac{(N-1)(N-2)}{2}$ generators generate $SO(N-1)$. Note that the reflection symmetry of $O(N)$ remains intact as the stable state does not break e.g. $S=\text{diag}(1,...,1,-1)$, and thus the full unbroken symmetry group is $O(N-1)$. This is the usual symmetry breaking pattern also encountered in model A of the Hohenberg Halperin classification. We briefly discuss two special cases: \\
$N=2$: The stable state is a point on a circle. The unbroken symmetry is $O(1)=\mathbb{Z}_2$, a reflection along the axis defined by the order parameter. Therefore, the statically ordered state would not leave any symmetry intact, if the original symmetry were $SO(2)$ rather than $O(2)$.\\
$N=3$: The stable state is a point on a sphere with fixed radius. The unbroken subgroup $O(2)$ are the rotations around the axis defined by this point, while the two Goldstone modes correspond to the two directions in which one can move a point on a sphere. 

\subsubsection{Dynamically ordered rotating phase} 
We now turn to the rotating phase, where we parametrize the stable state as 
\begin{align}\label{eq:ssbdyn}
   \boldsymbol\varphi_s&=\sqrt{\rho_0}(\cos{Et},\sin{Et},...,0)^T\nonumber\\
    &\equiv\sqrt{\rho_0} (\cos{Et}\hat e_1+\sin{Et}\hat e_2).
\end{align}
The order parameter now traces out a two-dimensional plane, which again is picked spontaneously; we choose it to be the $1-2$ plane.\\
This stable state remains invariant under the rotations of the $3^{\text{rd}}$ to $N^{\text{th}}$ component into each other. Therefore, there are ${\binom{N-2}{2}}=\frac{(N-2)(N-3)}{2}$ unbroken generators constituting an unbroken $O(N-2)$ subgroup. For $N>2$, the $\mathbb{Z}_2$ parity part remains unbroken analogously to the case discussed above~\footnote{We stress that a counter rotating orbit with angular velocity $-E$ is smoothly connected by a $\pi$ rotation in e.g. the $2-3$ plane that is part of the unbroken subgroup $O(N-2)$ and does not need a chiral symmetry.}.

\begin{figure}
\centering
\subfloat[][]{%
  \includegraphics[width=0.2\textwidth]{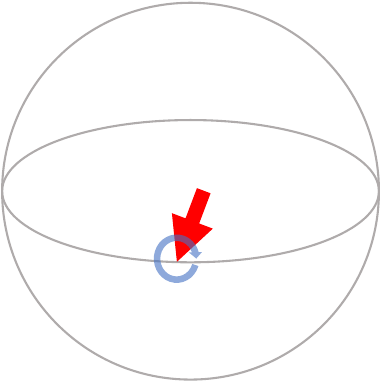}%
}\hfill
\subfloat[][]{%
  \includegraphics[width=0.2\textwidth]{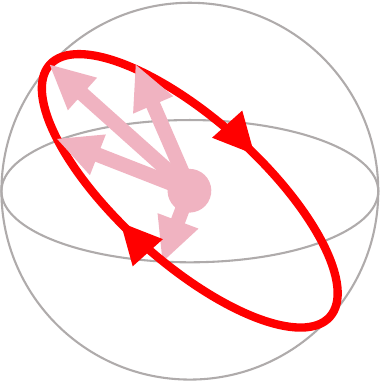}%
}
\caption{Visualization of the symmetry pattern for the $O(3)$ case. The spontaneously formed static order corresponds to a point on a sphere, see (a). The two Goldstone modes correspond to the two directions into which one can move the point on the sphere. The unbroken $O(2)$ group corresponds to rotations along the axis defined by the order parameter. In the rotating phase (b), the order parameter rotates along a spontaneously picked grand circle of the sphere. This breaks the previously unbroken $O(2)$ symmetry down to a $\mathbb{Z}_2$ reflection against the plane defined by the limit cycle. The three Goldstone modes correspond to the two directions in which the grand circle can be rotated on the sphere, plus a shift of the along the circle.}
\label{fig:SSB_Pattern}
\end{figure}

The rotating stable state breaks the $N-2$ generators that rotate the second component into any higher component. These lead to $N-2$ new Goldstone modes. The previously $N-1$ broken generators, rotating the first component into any other remain broken by the rotating stable state and also lead to gapless Goldstone modes. We therefore have $2N-3$ Goldstone modes in total in the rotating phase. \\
For $N=2$, the unbroken symmetry of the static phase is the remnant $\mathbb{Z}_2$, which is broken in the rotating phase. Thus the rotating phase is characterized  by a symmetry breaking  if the microscopic dynamics is fully $O(2)$ symmetric; an $SO(2)$ invariance is not sufficient for the purpose. For $N=3$, see Fig.~\ref{fig:SSB_Pattern}.

The rotating order can also be viewed as resulting from a combination of the spontaneous breaking of the internal $O(N)$ symmetry and the external symmetry of time translations: In the example above, the first produces Eq. \eqref{eq:ssbstat}, and the second allows for a time-dependent generator $\exp (Et T_{12})$, producing Eq.~\eqref{eq:ssbdyn}. We see here the reason for the simplicity of the limit cycle solutions: time translation symmetry breaking enables dynamics on the degenerate manifold available due to the breaking of a continuous internal symmetry. This activation mechanism for soft modes should be very general for systems driven out of equilibrium, where time translation symmetry breaking can occur. At the same time, it rationalizes why the van der Pol limit cycle is more complicated: time translation symmetry is broken, but there is no continuous internal symmetry which could be broken, and thus no Goldstone mode to be activated by it.

\subsubsection{Goldstone theorem for the rotating phase}\label{sssec:goldstone_theorem}

Above we have counted the Goldstone modes via the number of broken symmetry generators. Here we will show more formally how these broken  generators lead to gapless Goldstone modes, specifically in the rotating phase, and give them a geometric interpretation. \\
In spatially and temporally homogeneous states of matter, Goldstone modes are signalled by poles of the retarded Green function at zero momentum and frequency, at the origin of the complex frequency plane, describing spatially and temporally homogenous, non-decaying modes. Equivalently, they manifest in gapless zero modes of $\Gamma^{(1,1)}$. In our rotating stable state, the Green function is not diagonal in frequency space. We need to generalize this criterion to finding the linearly independent elements of the kernel of  the derivative operator $\Gamma^{(1,1)}(X',X)$ that do not decay over time, without fixing them to be fully time independent. Indeed, we will find  \emph{finite frequency} Goldstone modes that oscillate exactly at the frequency of the limit cycle $\omega = \pm E$. To this end, we now assume that the field expectation value takes the form of a rotating  configuration, 
 \begin{align}\label{eq:limcyc}
 \tilde{\boldsymbol{\varphi}}_s=0,\quad \boldsymbol{\varphi}_s=\sqrt{\rho_0} (\cos{Et}\hat e_1+\sin{Et}\hat e_2),
 \end{align}
 where $\hat{e}_{1,...,N}$ denote the basis vectors in field space.
First, we consider how a general field configuration transforms under an infinitesimal $O(N)$  rotation generated by $T_{1,i}$ by an angle $\theta_i^{(1)}$, for all $i=3,...,N$, which rotate out of the $\hat e_1 - \hat e_2$ plane while leaving $\hat e_2$ invariant. Their action on the field expectation value is given by
\begin{figure}[htbp]
\centering
\subfloat[\label{fig:GoldstoneShift}]{%
  \includegraphics[width=0.2\textwidth]{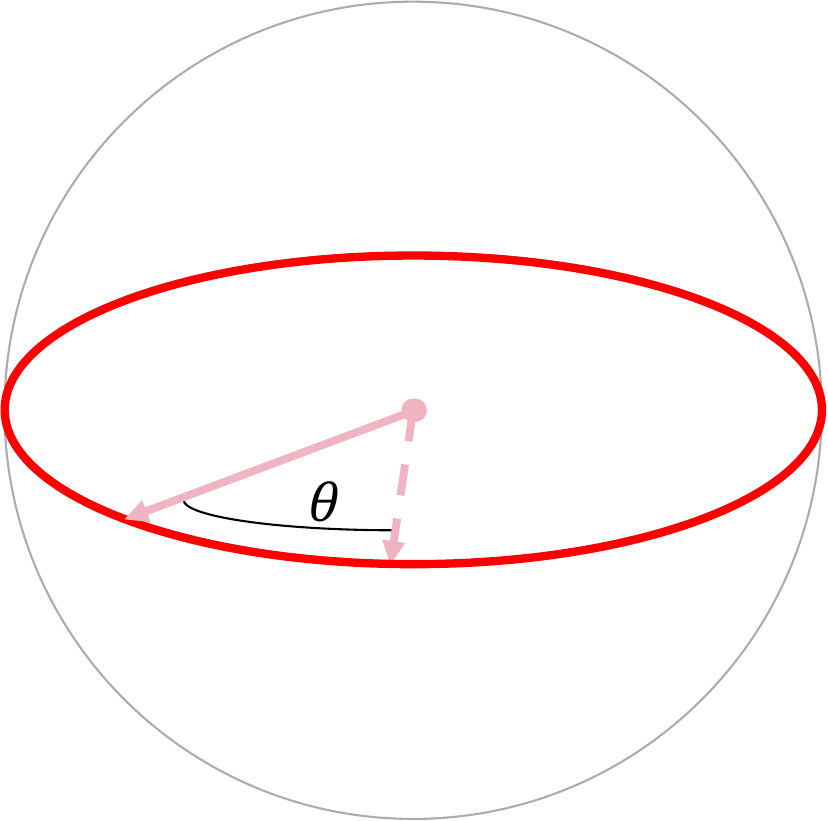}%
}\hfill
\subfloat[\label{fig:OldGoldstone}]{%
  \includegraphics[width=0.2\textwidth]{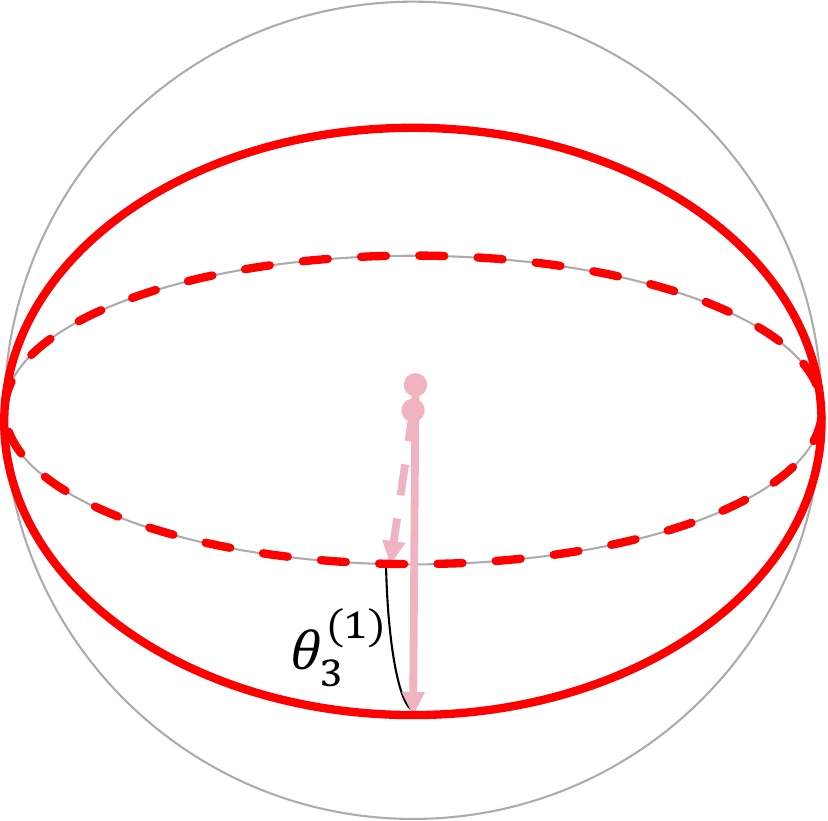}%
}\hfill
\subfloat[\label{fig:NewGoldstone}]{%
  \includegraphics[width=0.2\textwidth]{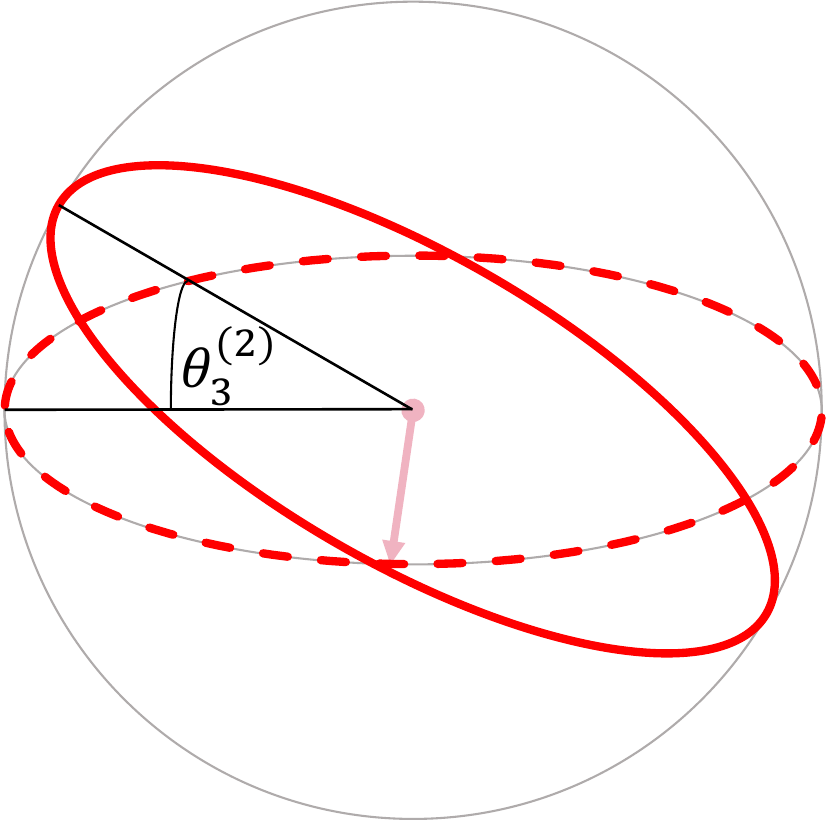}%
}
\caption{Visualization of the action of the different broken generators and the respective Goldstone modes for the case $O(N=3)$. (a) The generator $T_{1,2}$ shift the field along the limit cycle. (b) and (c) show the angles in which the plane of the limit cycle itself can be tilted. (b) corresponds to a $\cos{Et}$ wave around the original solution, while (c) corresponds to a $\sin{Et}$ wave.}
\label{fig:goldstones}
\end{figure}
\begin{align}
   &\boldsymbol\varphi\rightarrow\boldsymbol\varphi+\theta_i^{(1)}\,\varphi_1\,\hat{e}_i -\theta_i^{(1)}\varphi_i\hat{e}_1,\\
   &\tilde{\boldsymbol\varphi}\rightarrow\tilde{\boldsymbol\varphi}+\theta_i^{(1)}\,\tilde{\varphi}_1\,\hat{e}_i -\theta_i^{(1)}\tilde{\varphi}_i\hat{e}_1,
\end{align}
 and shown in Fig.~\ref{fig:goldstones} for $N=3$. 
Since the effective action is invariant under $O(N)$ transformations, it follows for an infinitesimal rotation that
\begin{align}
\nonumber
    \int_{\vecx,t}&\Big(\frac{\delta\Gamma}{\delta\varphi_i(\vecx,t)}\varphi_1(\vecx,t)+ \frac{\delta\Gamma}{\delta\tilde{\varphi}_i(\vecx,t)}\tilde{\varphi}_1(\vecx,t)\Big)\hat{e}_i\\
    \nonumber
    -&\Big(\frac{\delta\Gamma}{\delta\varphi_1(\vecx,t)}\varphi_i(\vecx,t)+\frac{\delta\Gamma}{\delta\tilde{\varphi}_1(\vecx,t)}\tilde{\varphi}_i(\vecx,t)\Big)\hat{e}_1=0 \\
    &\forall\, i=3,...,N.
\end{align}
If evaluated on the field expectation value, this equation is trivially true since $\Gamma^{(0,1)}=\Gamma^{(1,0)}=0$ precisely are the equations of motion. Information about the spectral component of the full Green function can however be gained by taking another derivative with respect to $\tilde{\varphi}_j(\vecx',t')$, $(j=1,..,N)$ and evaluating on the equation of motion, i.e. $\Gamma^{(0,1)}=0,\, \varphi_{i,0}=0,\,\varphi_{1,0}=\sqrt{\rho_0}\cos{Et}$  afterwards,
\begin{align}\label{eq:Ward_Takashi}
    \int_{\vecx,t}\Gamma^{(1,1)}_{ji}(\vecx',\vecx;t',t)\,\cos{Et}\,\hat{e}_i=0.
\end{align}

We therefore have identified $N-2$ spatially homogeneous linearly independent modes, one for every $i=3,...,N$, that do not decay and are elements of the kernel of the inverse Green function. The Goldstone modes associated to the breaking of the generators $T_{1,i}$ with $i=3,...,N$ in the rotating phase are identified as cosine waves. This corresponds to shifting the limit cycle as depicted in Fig.~\ref{fig:OldGoldstone}.
Taking a derivative with respect to $\boldsymbol{\varphi}$ of \eqref{eq:Ward_Takashi} does not lead to another constraint, since $\Gamma^{(0,2)}=0$ due to conservation of probability \cite{Sieberer2016}.

We now perform the analogous analysis for the broken generators $T_{2,i},\,i={3,...,N}$, which generate rotations out of the $\hat e_1 - \hat e_2$ plane while leaving $\hat e_1$ invariant, and act on the physical field expectation value as depicted in Fig.~\ref{fig:NewGoldstone}. Such rotations by an angle $\theta_i^{(2)}$ transform the fields as 
\begin{align}
   &\boldsymbol\varphi\rightarrow\boldsymbol\varphi+\theta_i^{(2)}\,\varphi_2\,\hat{e}_i -\theta_i^{(2)}\varphi_i\hat{e}_2,\\
   &\tilde{\boldsymbol\varphi}\rightarrow\tilde{\boldsymbol\varphi}+\theta_i^{(2)}\,\tilde{\varphi}_2\,\hat{e}_i -\theta_i^{(2)}\tilde{\varphi}_2\hat{e}_1,
\end{align}
which leads, in the same manner as before, to
\begin{align}
    \int_{t,\vecx}\Gamma^{(1,1)}_{ji}(\vecx',\vecx,t',t)\,\sin{Et}\,\hat{e_i}=0.
\end{align}
Therefore, the breaking of generators $T_{2,i},\, i=3,...,N$ leads to $N-2$ linearly independent sine waves as Goldstone modes. They correspond to the orbit of the rotating field after the plane of the original limit cycle has been rotated by the broken generators. This amounts to the respective sine and cosine fluctuations around the original limit cycle as visualized in Fig. \ref{fig:goldstones}. Due to the linear independence of sine and cosine functions, we arrive at a total of $2N-4$ Goldstone modes so far. 

We are left with the last broken generator $T_{1,2}$, which generates rotations in the $\hat e_1 - \hat e_2$ plane, i.e. shifts along the limit cycle, see Fig.~\ref{fig:GoldstoneShift}. More precisely, such a rotation by an angle $\theta_i^{(2)}$ transforms the fields as 
\begin{align}
   &\boldsymbol\varphi\rightarrow\boldsymbol\varphi+\theta\,\varphi_1\,\hat{e}_2 -\theta\varphi_2\hat{e}_1,\\
   &\tilde{\boldsymbol\varphi}\rightarrow\tilde{\boldsymbol\varphi}+\theta\,\tilde{\varphi}_1\,\hat{e}_2 -\theta\tilde{\varphi}_2\hat{e}_1,
\end{align}
and thus shifts the physical field expectation value along its orbit:
\begin{align}
    \boldsymbol\varphi_s\rightarrow\boldsymbol\varphi_s+\theta\,(-\sin{Et}\,\hat{e}_1+\cos{Et}\,\hat{e}_2).
\end{align}
This leads to 
\begin{align}
\label{eq:GoldthmPar}
    \int_{\vecx,t}\Gamma^{(1,1)}(\vecx',\vecx,t',t)\cdot(-\sin{Et}\,\hat{e}_1+\cos{Et}\,\hat{e}_2)=0,
\end{align}
where $\Gamma^{(1,1)}=\Big(\Gamma^{(1,1)}_{ij}\Big)$ is again an $N\times N$ matrix. This adds another linearly independent, spatially homogenous and non-decaying mode to the kernel of $\Gamma^{(1,1)}$. This Goldstone mode corresponds to a shift along a given limit cycle. 

We thereby arrive at a total of $2N-3$ Goldstone modes in the sense of excitations that do not decay. We remark that this counting is  nonperturbative and applies to the  renormalized Green functions. It only depends on the fact that $O(N)$ is an actual symmetry, and that it is broken in the form of Eq. \eqref{eq:limcyc}. It applies in the entire rotating phase. We will show explicitly how the Goldstone modes emerge in the dynamics of linearized fluctuations in the various phases, including the shift along the limit cycle in the rotating phase in Sec. \ref{subsec:gaussian_fluctuations}. 

The case of $N=2$ is  special in the sense that there is only one Goldstone mode in both the rotating and the ordered phase. Since the symmetry that is broken between the ordered and rotating phase for $N=2$ is $\mathbb{Z}_2$ ($O(2)$ is broken to $\mathbb{Z}_2$ in the ordered phase already), no additional Goldstone modes occur. This is confirmed by the counting laid out above.

The discussion shows explicitly what we stated above: Due to the emergence of the limit cycle, time translation invariance is spontaneously broken. However a time translation $t \to t + \Delta t$ and a rotation generated by $T_{1,2}$ by the angle $ \Delta t\cdot E $ are the same. The Goldstone mode generated by $T_{1,2}$ can equivalently be viewed as arising from the breaking of time translation symmetry. Below we will use this relation between time translations and internal rotations to write the action in a co-moving frame, where it becomes time-independent. 
The Goldstone modes then can also identified with poles of the retarded response in frequency space, which lie at real frequencies $\omega=\pm E$ for the fluctuations orthogonal to the limit cycle and at vanishing frequency for the fluctuations along the limit cycle.

\subsection{Linear fluctuations}\label{subsec:gaussian_fluctuations}
\subsubsection{Spectra}
After these exact considerations for the $\vecq=0$ excitations in the rotating phase, we now write the action of fluctuations around their respective mean-field solutions $\boldsymbol\varphi_s$ which also serve as a low frequency, long wavelength description of the phases. That is, we expand the action to quadratic order
\begin{align}
    S[&\boldsymbol\phi_s+\Delta\boldsymbol\phi,\tilde{\boldsymbol\phi}]\approx \int_{\vecx,t}(\Delta\boldsymbol\phi(\vecx,t),\tilde{\boldsymbol\phi}(\vecx,t))\mathcal{G}_0^{-1}\begin{pmatrix}
        \Delta\boldsymbol\phi(\vecx,t)\\
        \tilde{\boldsymbol\phi}(\vecx,t)
    \end{pmatrix}.
\end{align}
We note that, by reversing the MSRJD construction, this corresponds to expanding the Langevin equation to linear order around a respective mean-field solution. This allows us to access the spectrum of dispersions $\omega_i(\vecq)$, to derive the inverse bare Green function $\mathcal{G}_0^{-1}$ of fluctuations in the various phases, and to identify the \ac{CEP}s and their properties. 
In the static phase, we pass to a phase-amplitude representation
\begin{align}
    \label{eq:PhaseEqRep}
    \nonumber
    &\boldsymbol{\phi}=\sqrt{\rho_0+\delta\rho}\exp(\sum_{i=2}^{N}\theta_i T_{1,i})\hat{e}_1,\\
    &\tilde{\boldsymbol{\phi}}=\sqrt{\rho_0} \exp(\sum_{i=2}^{N}\theta_i T_{1,i})\tilde\chi,
\end{align}
where $\tilde\chi \in \mathbb{R}^{N}$ is parametrized as $\tilde\chi=(\tilde{\delta\rho},\tilde \theta_2,\dots,\tilde\theta_N)$, and expand to quadratic order. The amplitude sector is
\begin{equation}
\label{eq:ampStat}
\begin{split}
    S^0_\rho=\rho_0\int_{\vecx,t}&\tilde\rho(\partial_t^2+(\delta+2u'\rho_0-Z\nabla^2)\partial_t\\
    &+2\lambda\rho_0-v^2\nabla^2)\rho-D\tilde\rho^2,
    \end{split}
\end{equation}
with the relative amplitude fluctuation $\rho=\frac{\delta\rho}{2\rho_0}$ while the Gaussian fluctuations of the phases $\theta_2,...,\theta_{N}$ are described by
\begin{equation}
\label{eq:S0GoldstoneStat}
\begin{split}
    S^0_\theta=\rho_0\int_{\vecx,t}&\tilde\theta_i(\partial_t^2+(\delta-Z\nabla^2)\partial_t-v^2\nabla^2)\theta_i-D\tilde\theta_i^2.
    \end{split}
\end{equation}
The equal time correlation function and the dispersion gaps (i.e. $\Delta=\omega_{1,2}(\vecq\rightarrow0)\in\mathbb{C}$) corresponding to this quadratic action are displayed in the third row of table \ref{tab:linear_fluctuations}. This action also serves as a starting point for an effective long wavelength theory describing the transitions out of the statically ordered phase. 
The same procedure can be carried out in the rotating phase, where we parametrize the fluctuating field in a comoving frame as
\begin{align}
\varphi=\exp(EtT_{1,2})\sqrt{\rho_0+\rho}\exp(\sum_{i=2}^{N}\theta_iT_{1,i})\;\hat{e}_1,
\end{align}
where $E$ is the angular velocity of the limit cycle which we choose, without loss of generality, to lay in the $1-2$ plane as before. Working in the comoving frame allows us to retrieve an action which does not explicitly depend on time, and thus to use frequency space conservation and define modes as in~Sec.~\ref{ssec:modes_and_EPs}.   The resulting quadratic action is block diagonal with a diagonal part for the phase fluctuations orthogonal to the limit cycle $\theta_{3,...,N}$
\begin{equation}
\begin{split}
\label{eq:S0perp}
    S^0_\perp=\rho_0\int_{\vecx,t}&\tilde\theta_i(\partial_t^2-Z\nabla^2\partial_t-v^2\nabla^2+E^2)\theta_i-D\tilde\theta_i^2.
    \end{split}
\end{equation}
Its form is in agreement with the prediction from Goldstone theorem from \ref{sssec:goldstone_theorem}.

The quadratic action for the phase flucutations along the limit cycle and the amplitude fluctuations is however not diagonal. Its full form is given in App. \ref{app:PhaseAmp}. An effective theory for the phase fluctuations along the limit cycle $\theta_\parallel$ can however be obtained by performing the Gaussian integration over the gapped amplitude fluctuations. It yields, for small limit cycle frequencies $E$
\begin{equation}
\begin{split}
    S^0_\parallel=\rho_0\int_{t,\vecx}&\tilde\theta_\parallel(\partial_t^2+(|\delta|-Z\nabla^2)\partial_t-v^2\nabla^2)\theta_\parallel-D\tilde\theta_\parallel^2.
    \end{split}
\label{eq:S0par}
\end{equation}
The corresponding gaps and equal time correlators for all Goldstone modes are shown in the fourth row of table \ref{tab:linear_fluctuations}.\\
Once in the rotating frame, the Gaussian Green function displays a pole at vanising momenta and frequency for the $\theta_\parallel$ fluctuations like for a usual equilibrium Goldstone mode.

\begin{table*}
\renewcommand{\arraystretch}{3}
\setlength{\tabcolsep}{5pt}
\setlength{\extrarowheight}{0pt}
\begin{tabular}{ | c | c| c | c | }
\hline
    \makecell{\textbf{Phase}} & \makecell{\textbf{Fluctuation}} &\makecell{\textbf{Dispersion Gap}\\ $\Delta_{1,2}=\omega_{1,2}(\vecq\rightarrow0)$} & \makecell{\textbf{Equal time} \\\textbf{correlation function}}\\[5pt]\hline
    \makecell{Symmetric $\boldsymbol{\varphi}_s=0$} & \makecell{$\varphi_{1,...,N}$} & \makecell[l]{$\Delta_{1,2}=-i\gamma\pm({r-\gamma^2})^{1/2}\rightarrow\;$ gapped}& \makecell[l]{$G^K_{0,ij}\sim\frac{D\delta_{ij}}{(Z\vecq^2+2\gamma)(v^2\vecq^2+r)}\sim q^{0}$}\\[5pt]\hline
    \makecell{Static Order} & \makecell{$N-1$ phase \\fluctuations} &\makecell[l]{$\Delta_1=-i\frac{v^2}{\delta}\vecq^2\rightarrow\;$ gapless\\[2pt] $\Delta_2=-2i\delta\equiv -2i(2\gamma-\frac{r}{\lambda})\rightarrow\;$ gapped}& \makecell[l]{$G^K_{0,ij}\sim\frac{D\delta_{ij}}{v^2\vecq^2(Z\vecq^2+\delta)}\sim q^{-2}$}\\[5pt] \cline{2-4}
    \makecell{$\rho_0=-\frac{r}{\lambda}>0,\, E=0$} & \makecell{Amplitude\\ fluctuations} & \makecell[l]{$\Delta_{1,2}=-i\frac{\delta+u'\rho_0}{2}\pm\frac{(8\lambda\rho_0-(\delta+u'\rho_0)^2)^{1/2}}{2}$\\$\rightarrow\;$gapped} & \makecell[l]{$G^K_{0,\rho}\sim\frac{D}{(Z\vecq^2+\delta+u'\rho_0)(v^2\vecq^2+2\lambda\rho_0)}\sim q^{0}$}\\[5pt] \hline  
    \makecell{Rotating Order} & \makecell{Phase fluctuations \\along limit cycle} &\makecell[l]{$\Delta_1=-i\frac{v^2}{|\delta|}\vecq^2\rightarrow\;$gapless\\ $\Delta_2=-2i|\delta|\rightarrow\;$gapped}& \makecell[l]{$G^K_{0}\sim\frac{D}{v^2\vecq^2(Z\vecq^2+|\delta|)}\sim q^{-2}$}\\[5pt] \cline{2-4}
     \makecell{$\rho_0=-\frac{\delta}{u}>0,$\\ $ E^2=r+\lambda\rho_0>0$} & \makecell{$N-2$ phases\\ perpendicular to \\limit cycle} & \makecell[l]{$\Delta_{1,2}=-i\frac{Z}{2}\vecq^2\pm E$ $\rightarrow\;$  oscillating,\\ no decay}& \makecell[l]{$G^K_{ij}\sim\frac{D\delta_{ij}}{Z\vecq^2(v^2\vecq^2+E^2)}\sim q^{-2}$}\\[5pt]
     \hline
\end{tabular}
    \caption{Dispersion gaps and equal time correlators from linearized fluctuations in the three phases. The gaps reveal the counting of gapless modes. There are $N-1$ gapless phase fluctuation modes in the statically ordered phase while the amplitude fluctuations remain gapped in the entire phase, also at the transition into the limit cycle at $\delta\rightarrow0$. In the rotating phase, after adiabatically eliminating the amplitude fluctuations (see App. \ref{app:PhaseAmp}), there is one gapless mode for phase fluctuations along the rotating limit cycle and $2(N-2)$ modes that do not dissipate but oscillate at the frequency of the limit cycle for phase fluctuations perpendicular to the limit cycle, all in agreement with the Goldstone theorem. The scaling behavior of the equal time correlators is discussed in the main text. 
    }
   \label{tab:linear_fluctuations}
\end{table*}
The fact that the correlator diverges as $\vecq^{-2}$ in the entire phase does not indicate an instability but is a hallmark of the Goldstone nature of the phase fluctuations. Below the lower critical dimension $d_l=2$, this divergence leads to infrared divergences of the loop corrections due to phase fluctuations, destroying long range order as a consequence of the Mermin-Wagner theorem.

In the statically ordered as well as in the symmetric phase the Gaussian action satisfies the thermal symmetry or equivalently the \ac{fdr} with respective effective temperature, $T=\frac{D}{4\gamma}$ and $T=\frac{D}{2\delta}$. The quadratic sector is thermal in both phases, whereas nonconservative interactions can induce nonthermal behavior only for large fluctuations at finite frequencies or momenta. We thus face a case of an approximate, emergent equilibrium behavior despite the microscopic violation of equilibrium conditions.  
In the rotating phase instead, the part of the action describing the amplitude fluctuations and fluctuations of the phase along the limit cycle is explicitly time dependent if one does not go into the rotating frame, and not invariant under the thermal symmetry~\eqref{eq:ThermalSymm}. When $N>2$, for the fluctuations tilting the limit cycle, see Fig.~\ref{fig:goldstones}, the damping vanishes at zero momenta and the effective action cannot be of the form~\eqref{eq:thermalNoise}. The thermal symmetry is broken. The \ac{fdr} are not satisfied either by the Gaussian Green functions. Thermal symmetry is violated even in the quadratic sector and there is no effective thermal equilibrium emerging at long wavelengths. For $N=2$, the quadratic sector of the phase fluctuations appears thermal in the rotating frame. There is however a KPZ nonlinearity breaking the thermal symmetry beyond the Gaussian level, see ~\ref{ssec:Sym_break}.

\subsubsection{Phase transitions}\label{sssec:PhaseTrans}
In addition to the spectrum, we can discuss the universal behavior at the phase transitions above their respective upper critical dimensions $d_c$, where the Gaussian approximation is exact.

The entire spectrum is gapped in the symmetric phase, i.e. the poles of the dispersions are located  in the lower complex half plane with finite distance from the real axis. Thus fluctuations decay exponentially with time, and the correlator remains analytic for vanishing momenta. There are two limiting cases where the bare correlator diverges algebraically, marking critical points. Upon tuning the mass term $r$ to zero, one dispersion becomes gapless, $\omega_1(\vecq)\approx-\frac{i}{\gamma^2}v^2\vecq^2$  and the equal time correlator diverges as $\vecq^{-2}$. Furthermore, the Gaussian exponent for the divergence of the correlation length is $\nu=\frac{1}{2}$. At this point the phase transition into the ordered phase occurs. This transition is in fact the equilibrium model A transition of the Halperin-Hohenberg classification~\cite{Halperin1978} because  the additional microscopic breaking of equilibrium conditions we add are all irrelevant. Indeed, the most relevant non-linearity is the usual $(\boldsymbol{\phi}^4)$ $\lambda$ interaction which has dimension mass $[\lambda]=4-d$ (therefore $d_c=4$). Power counting reveals that the inertial term $\partial_t^2$ has dimension $-2$ and the interactions $u,u'$ have $[u]=[ u']=2-d$ at the Gaussian fixed point. Being irrelevant for $d>2$, their bare values play no role both at the critical Gaussian fixed point above $d_c$ and at the Wilson-Fisher fixed point below it.  Model A transition and exponents are thus recovered. 

The bare correlator also displays an algebraic singularity $\sim\vecq^{-2}$ at the transition into the rotating phase at $2\gamma=0$. At this point, the imaginary part of the dispersions vanishes, indicating an instability and a second-order phase transition. This occurs at a finite frequency $\pm\sqrt{r}$. This is an example of a finite frequency critical point that can occur outside of thermal equilibrium and is the generalization of the $\textbf{III}_s$ scenario from Cross and Hohenberg \cite{Cross1993} to noisy dynamics. A finite frequency transition has for instance been studied in \cite{Scarlatella2019}. This transition however does not proceed via a \ac{CEP} due to the finite mass $\sim r$ at the transition, and is not in the focus of this work. Its analysis below the upper critical dimension (which is suggested to be four by a simple analysis of the perturbative corrections) remains open for future work.

The multicritical point $r=2\gamma=0$, where both transition lines coincide, is a CEP as shown in \ref{ssec:EPFluct}. 
This multicritical point can only be reached upon double fine-tuning and is not the focus of this work. 
A simple analysis of the Gaussian theory and one-loop divergences suggests an upper critical dimension $d_c=6$ above which the Gaussian fixed point is stable, but a detailed \ac{RG} analysis of its universal fluctuations, is reserved for future work. 

Using the phase-amplitude description of the broken phases, we can approach the transition from the static into the rotating phase. It occurs upon tuning the effective damping
\begin{align}\label{eq:delta}
    \delta=2\gamma-\frac{ur}{\lambda}
\end{align}
through zero. 
This marks it as a \ac{CEP} as defined in Sec.~\ref{ssec:modes_and_EPs}, since the modes becoming critical have no mass-like contribution  to begin with due to their Goldstone nature.

Furthermore, the amplitude fluctuations remain gapped and damped. They can thus be discarded from an effective long wavelength description. At the phase transition, there is a `condensation' of $\partial_t\theta_i=E$ (i.e., the angular velocity picks up a finite value), while the choice which mode $\theta_i$ starts to rotate is made spontaneously. The equal-time correlator of the phase fluctuations, shown in Tab.~\ref{tab:linear_fluctuations}. displays an enhanced divergence $\sim\vecq^{-4}$, as expected in the vicinity of a \ac{CEP}.

This \ac{CEP} transition does not fall into any known universality class a priori. We thus first discuss the scaling behavior of the linear fluctuations in the vicinity of the \ac{CEP} in more detail. This discussion is exact above the upper critical dimension of the transition, which we determine also to be $d_c=4$ in Sec.~\ref{sec:ExceptionalRG}. There, we will also analyze the problem beyond Gaussian fluctuations.

In the following, $v$ sets the highest momenta, i.e.\ we work at  $q\lesssim v$,  where our effective field theory at low momenta is valid. We are also close to the \ac{CEP} i.e. we work with $\delta^{1/2}\lesssim v,Z $. In the opposite regime, we are deep in one of the ordered phases and the formulas given in Table~\ref{tab:linear_fluctuations}  apply.
In this regime for finite damping $\delta>0$,  the dispersions of the phase fluctuations are
\begin{align}
    \omega_{1,2}(\vecq)=-\frac{i}{2}(\delta+Z\vecq^2)\pm\sqrt{v^2\vecq^2-\frac{\delta^2}{4}}.
\end{align}
There is thus a non-critical EP at a finite momentum scale $q_{\mathrm{EP}}=\frac{\delta}{2v^2}$. It only affects the dynamics, separating overdamped, purely dissipative modes from underdamped, propagating modes. This translates to a length scale
\begin{align}
    \xi_{\text{EP}}\sim v/\delta,
\end{align}
separating both regimes. In contrast to a critical length scale, it does not signal the divergence of a correlation function. The  correlation function displays an enhanced divergence $\sim\vecq^{-4}$ as expected for a \ac{CEP}. The additional divergence as the damping gap $\delta$ is tuned to zero generates is indeed not controlled by $\xi_{\mathrm{EP}}$, but by a divergent length scale 
\begin{align}
    \xi_c=\delta^{-1/2},
\end{align}
 indicating a critical exponent 
\begin{align}
    \nu=\frac{1}{2}
\end{align}
for the mean-field transition. 

\begin{figure}
    \centering
    \begin{tikzpicture}
        \draw[thick,->] (-4,0) -- (4,0);
        \node at (4.2,-0.2) {$q$};

        \draw[thick] (-1.75,-0.1) -- (-1.75,0.1);
        \node at (-1.75,-0.4) {$q_{\mathrm{EP}}=\delta/v$};

          \draw[thick] (0.65,-0.1) -- (0.65,0.1);
          \node at (0.65,-0.4) {$\delta^ {\frac{1}{2}}$};

          \node[align=center] at (2.25,1.25) {Gaussian CEP\\ underdamped};
          \node[align=center] at (-0.75,1.25) {Goldstone\\ underdamped};
          \draw[thick, dashed] (0.65,0)--(0.65,1);
          \node[align=center] at (-3.25,1.25) {Goldstone\\ overdamped};

        \node[align=center] at (2.25,0.5){$G^K(\mathbf{q},t=0)=q^{-4}$};
          \node[align=center] at (-1.5,0.5) {$G^K(\mathbf{q},t=0)=q^{-2}$};
        \draw[thick] (3.6,-0.1) -- (3.6,0.1);
          \node at (3.6,-0.4) {$v$};
        \end{tikzpicture}
    \caption{Mean-field theory scales and regimes. The critical regime, characterized by the $q^{-4}$ divergence of $G^K$, is left for $q < \delta^{1/2}$, while the EP scale $q_{\mathrm{EP}}$ separating over- and underdamped dynamics appears at even smaller momentum scales.}
    \label{fig:Scale1}
\end{figure}
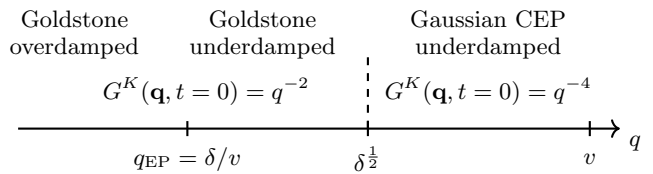

This critical length scale diverges less quickly than the exceptional length scale $\xi_{\text{EP}}$ close to the transition, so that the critical regime is left through the dissipative part before reaching the overdamped regime. The critical regime is therefore found for momenta satisfying 
\begin{eqnarray}
    q\gg\delta^{\frac{1}{2}}\gg q_{\text{EP}}.
\end{eqnarray}
The different scales appearing at mean-field are summarised on Fig.~\ref{fig:Scale1}.

\begin{figure}
\centering
\subfloat[][]{%
  \includegraphics[width=0.23\textwidth]{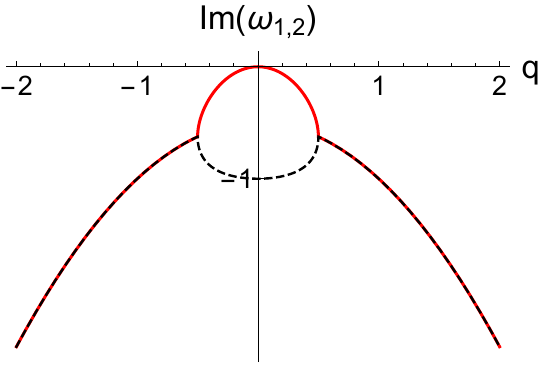}%
}\hfill
\subfloat[][]{%
  \includegraphics[width=0.23\textwidth]{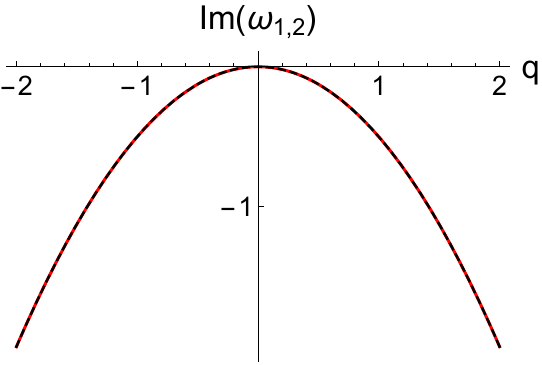}%
}\hfill
\subfloat[][]{%
  \includegraphics[width=0.23\textwidth]{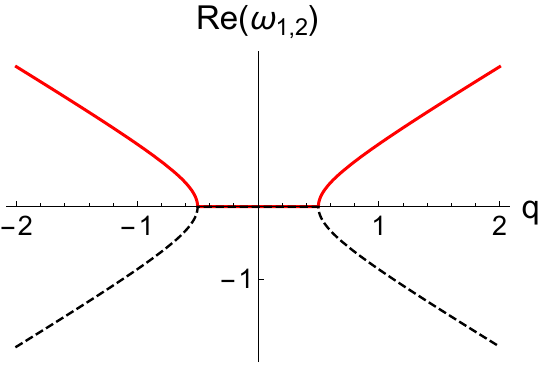}%
}\hfill
\subfloat[][]{%
  \includegraphics[width=0.23\textwidth]{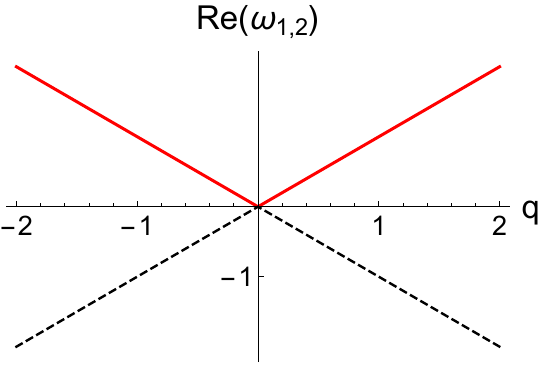}%
}
\caption{The dispersions of the phase fluctuations close to the critical exceptional point. a) and c) show the imaginary (dissipative) and real (propagating) part of the dispersion at a finite damping with $\delta/v^2=1$. The exceptional point separating purely dissipative dynamics from underdamped motion is clearly visible. At vanishing momenta one mode becomes gapless marking its Goldstone nature, whereas the other mode maintains a gap $\delta$. As one approaches the CEP $\delta=0$ shown in b) and d), the egg-shaped structure in the dissipative part shrinks to zero, both modes dissipate as $\sim \vecq^2$ and display a linear scaling in their real parts, indicating propagation at a constant velocity in real space.}
\label{fig:dispersions}
\end{figure}

At the \ac{CEP} $\delta=0$ however, the coexistence of dissipation and propagation persists down to vanishing momentum, where 
\begin{align}
\label{eq:CritDisp}
    \omega_{1,2}=-\frac{i}{2}Z\vecq^2\pm v|\vecq|,
\end{align}
see Fig.~\ref{fig:dispersions}. The linear scaling of the real part of the dispersion in momentum space will manifest as spherical propagation of excitations at  constant velocity $v$, whereas the dissipative part will lead to diffusive decay in real space around the mean position $|\vecx|=vt$. 
This is also seen by inspecting the correlation function in $(\vecq,t)$ space
\begin{equation}
\label{eq:Gkqt}
    G^K(\vecq,t)\propto \frac{D}{v^2 Z \vecq^4}\exp^{-\frac{1}{2}Z\vecq^2 t}\cos(|v \vecq t|).
\end{equation}
Hence, there is no unique dynamical $z$ exponent: the lifetime of a critical fluctuation scales as $\tau_d\sim q^{-2}$ at the \ac{CEP}, and its oscillation period in momentum space as $\tau_c\sim q^{-1}$. These two scaling behaviors coexist, controlling different properties of the dynamics of excitations, and inhibit the existence of a homogeneous scaling solution of the action and a true scale invariance of correlation functions even at the Gaussian fixed point 
\footnote{
One may then be tempted to keep only the lowest order power in momenta in the dispersion~\eqref{eq:CritDisp} arriving at a dynamical exponent $z=1$. This would amount to neglecting the damping term $Z$ in Eq.~\eqref{eq:S0GoldstoneStat}. This is non-physical, because the system would only receive energy from the noise without any dissipation. More formally, Eq.~\eqref{eq:Gkqt} would be infinite for $Z \to 0$. We are thus forced to keep the lowest power in momenta for both imaginary and real part of the dispersions. This has to be contrasted with the quantum case, where there is no dissipation but where also the noise vanishes as $|\omega|\rightarrow0$. 
}.

\section{Beyond mean-field effects and \texorpdfstring{\ac{CEP}}{CEP} fluctuation induced first order phase transition}\label{sec:ExceptionalRG}

In this section, we analyse the phase transition between the ordered and rotating phase below the upper critical dimension $d_c=4$ in detail. We first argue in Sec. \ref{subsec:EP1storder} that the enhanced fluctuations due to the \ac{CEP} tend to restore the symmetry and we will show that it makes a continuous transition between static and rotating phase impossible below $d_c=4$ for all $N>1$.

We then perform a  deeper analysis of the case of $O(2)$. We show how the inclusion of enhanced fluctuations in the vicinity of the \ac{CEP} give rise to a fluctuation induced, weakly first order transition. We show that the \ac{CEP} induces a resonance condition on momenta, linked to the presence of the additional exceptional momentum scale $q_{EP}=\delta/v^2$, signalling the spectral position of the EP as discussed in Sec \ref{subsec:gaussian_fluctuations}, and rendering the standard derivative expansion impossible. In addition, we find that this leads to a subdominant contribution of two-loop corrections compared to their one-loop counterparts, which on the level of the diagrammatics is reminiscent of Brazovskii's seminal work~\cite{Brazovskii75}, and Swift and Hohenberg's later RG analysis~\cite{Hohenberg1995}. The physical origin is a different one, though. This allows for a resummation of the perturbative series in the long wavelength limit, or equivalently, renders the \ac{DSE} one-loop. We then discuss how this generalizes to the $N>2$ case.

\subsection{Exceptional Fluctuations}\label{subsec:EP1storder}

We now provide a simple argument stating that the enhancement of fluctuations in the vicinity of a \ac{CEP} renders it impossible to reach below four dimensions if interactions are not taken into account. We will see that the fluctuations either restore the full symmetry before the \ac{CEP} is reached, or render the transition between statically ordered and rotating phase first order. We use the phase-amplitude decomposition~\eqref{eq:PhaseEqRep}.
As we have seen, the phase fluctuations become critically exceptional at the transition, and the static correlation function is 
\begin{align}\label{eq:thetaFluct}
    \langle\theta_i(\vecq,t_0)\theta_j(-\vecq,t_0)\rangle=G^K_{0,ij}(\vecq,t=0)\sim\frac{D\delta_{ij}}{\rho_0 v^2\vecq^2(Z\vecq^2+\delta)},
\end{align}
see Table~\ref{tab:linear_fluctuations}.
The \ac{CEP} is reached as the damping $\delta\rightarrow0$. This implies in the Gaussian approximation 
\begin{align}
    G^K_{\theta,ii}(\vecq,t=0)\xrightarrow{\delta\rightarrow0}\frac{1}{\vecq^4}.
\end{align}
Thus, the Gaussian correlation function $G^K_{0,ii}(\vecx=0,t=0)$ develops an infrared divergence in $d<4$ spatial dimensions in the vicinity of the \ac{CEP}, which is regularized by the damping:
\begin{align}
    \langle\theta_i(\vecx_0,t_0)\theta_i(\vecx_0,t_0)\rangle=G^K_{0,ii}(\vecx=0,t=0)=C\frac{\delta^{\frac{d-4}{2}}}{\rho_0}.
\end{align}
Here $C>0$ is a non-singular constant that depends on the dimension and the ultraviolet cutoff of the theory. Its exact value is not important for our argument, we only rely on the fact that it is positive and finite. 
We see that when the damping vanishes, the Gaussian fluctuations of the Goldstone modes diverge and would destroy any order.
Indeed, neglecting amplitude fluctuations, 
\begin{align}
\label{eq:SymRestAbsurd}
    \langle\boldsymbol\varphi(\vecx_0, t_0)\rangle&=\sqrt{\rho_0}\langle\exp{\Big(\sum_{i=2}^N\theta_i(\vecx_0, t_0)T_{1,i}\Big)}\rangle\hat{e}_1\\
    \nonumber&=\sqrt{\rho_0}\exp{\Big(2\text{tr}\langle\theta_i(\vecx_0, t_0)\theta_j(\vecx_0, t_0)\rangle T_{1,i}T_{1,j}\Big)}\hat{e}_1\\
    \nonumber 
    &=\sqrt{\rho_0}\exp{\Big(\frac{-2(N-1)C\delta^{\frac{d-4}{2}}}{\rho_0}\Big)}\hat{e}_1\xrightarrow{\delta\rightarrow0}0,
\end{align}
and the enhanced Gaussian fluctuations due to the \ac{CEP} alone destroy the order parameter before one can reach the \ac{CEP} at $\delta=0$ below four dimensions. The order parameter is suppressed when the argument of the exponential in Eq.~\eqref{eq:SymRestAbsurd} is of order one, i.e.\ at a symmetry restoring scale
\begin{equation}
\label{eq:SymRestScale}
    \frac{\delta_{\text{sym}}}{Z} \sim \left(\frac{v^2 Z}{D}\rho_0\right)^{\frac{2}{d-4}},
\end{equation}
restoring all parameters previously absorbed in $C$. 
This argument  is reminiscent of the Mermin-Wagner theorem, which prevents the existence of symmetry breaking in and below two dimensions in the usual case. However, it applies only to the critical point here, not to the entire phase. On the other hand, the rotating phase exists and is not destroyed by fluctuations above two dimensions, as revealed by the fluctuation analysis in~\ref{subsec:gaussian_fluctuations}.

This leaves three scenarios for the transition upon including the effect of fluctuations and interactions :\\
(i) There is no direct transition between static and rotating phases, but a fully symmetric, disordered regime in between. This is the expectation solely based on the exceptional Gaussian fluctuations. \\
(ii)  There is a (weakly) first order transition, induced by interactions. A non-trivial scaling regime close to the transition may still emerge in principle.\\
(iii) The phase transition is second order. This is only possible, if nonlinear effects reduce fluctuations by generating a sufficiently large anomalous dimension. In equilibrium this happens, for example, for the 2d Ising model, where the anomalous dimension shift the naive lower critical dimension from two to one.\\

The third scenario will be ruled out by our analysis.
We will show, that indeed a first order transition occurs for sufficiently large $\rho_0$. For smaller $\rho_0$ the interaction effects do not have room to build, and as one approaches the \ac{CEP} the enhanced fluctuations push the system back in the symmetric phase through the model A transition.

The same mechanism has to arise while approaching the \ac{CEP} line from the rotating phase and the symmetry restoring nature of the enhanced fluctuations will therefore strongly move the phase boundaries as sketched in Fig.~\ref{fig:PhaseDiagram}. 

\subsection{Phase Fluctuations and Potential Picture}

We now show how a first order phase transition into the rotating phase at finite $\delta$ can occur.

As we have seen in Sec.~\ref{sec:MF+Gauss}, the amplitude fluctuations around the stable state in the broken phase remain damped and gapped in the vicinity of the \ac{CEP} at $\delta=0$, and can be integrated out. For $N=2$, this yields the effective Gaussian action for the phase field ~\eqref{eq:S0GoldstoneStat},
\begin{equation}
\label{eq:S0GoldstoneStat2}
    S_0=\int_{t,\vecx}\tilde\theta(\partial_t^2+(\delta-Z\nabla^2)\partial_t-v^2\nabla^2)\theta-D\tilde\theta^2.
\end{equation}
 We rescaled the fields $\theta \to \theta/\sqrt{\rho_0}$ and $\tilde\theta \to \tilde \theta/\sqrt{\rho_0}$.

The symmetry $O(2)\cong SO(2)\ltimes\mathbb{Z}_2$ acts on the phase field as
\begin{align}\label{eq:symmetry_action}
    SO(2):\;\theta\rightarrow\theta+\alpha, \quad \mathbb{Z}_2:\; (\theta,\tilde\theta)\rightarrow -(\theta,\tilde\theta).
\end{align}
This approach assumes that the fluctuations of the amplitude modes are small $\delta \rho \ll \rho_0$ and thus breaks down once the renormalized amplitude becomes small. Approaching the \ac{CEP} below 
four dimensions, this will be the case if we reach the scale $\delta \sim \rho_0^{(d-4)/2}$, signalling that we instead reach a regime where the symmetry gets restored as we have shown above. In the following we work in a regime with sufficiently large $\rho_0$, assuming that the scale at which the symmetry gets restored is not reached. This yields a criterion, whether symmetry restoration occurs or the scenario laid out below is realized.

We first discuss in greater detail how the transition is explained from this action above the upper critical dimension, and develop an effective potential picture which will turn out to be useful in the following. As discussed in Sec.~\ref{subsec:symmetry_breaking}, when crossing  the phase transition by tuning $\delta$ through zero, the order parameter starts to rotate at a finite angular velocity and the $\mathbb{Z}_2$ symmetry is spontaneously broken.  In terms of the phase variable, it corresponds to the `condensation' of $\Pi= \partial_t\theta$, which evolves in an effective Ising like potential 
\begin{equation}
\label{eq:Veff}
V_{\mathrm{eff}}(\Pi)=\frac{\delta}{2} \Pi^2+ \frac{g_{1}}{4!}\Pi^{4},
\end{equation}
where the fourth order term has been added to make the mean-field theory well-defined in the rotating phase ($\delta <0$). 
In that phase, we obtain $\Pi= \sqrt{\rho_0} E = \sqrt{-6\delta/g_1}=\sqrt{-\lambda \delta/u}$. This is in agreement with the calculation done in Sec.~\ref{subsec:gaussian_fluctuations}, obtained by assuming a finite rotation frequency directly. Therefore, this potential picture works despite the out-of-equilibrium nature of the problem. Beyond mean-field, we will get an effective equation of motion for the dressed order parameter using the effective action formalism (see Eq.~\eqref{eq:effEoM}). The potential picture will in turn remain applicable.

Since a $\mathbb{Z}_2$ is broken spontaneously along the transition, it is natural to compare it to the Ising universality class. Indeed, on the mean-field level, the phase transition is reminiscent to some extent to the usual Ising transition, where the role of the Ising field is played by $\partial_t\theta$. This can be rationalized by noting that the Ising model is recovered when $v=0$. However, recall from Tab.~\ref{tab:linear_fluctuations} that $v$ cannot be set to zero in our model without inducing an instability, and that our model is genuinely different from the Ising model.

\subsection{Beyond Gaussian Fluctuations}\label{ssec:BeyondGaussFluct}

We now determine how interactions lead to a fluctuation induced first order scenario below the upper critical dimension for sufficiently large $\rho_0$. To this end, we approach the \ac{CEP} from the statically ordered phase. The broken $SO(2)$ symmetry, Eq.~\eqref{eq:symmetry_action}, ensures that the field $\theta$ can only appear with derivatives, while invariance under $\mathbb{Z}_2$ excludes cubic -- or higher odd powers -- interactions; in particular, it rules out the Kardar-Parisi-Zhang nonlinearity $\tilde \theta ( \nabla \theta)^2$ and the cubic nonlinearity $\tilde \theta (\partial_t \theta)^2$. The lowest order local interaction terms that one can add to the quadratic action within these bounds are
\begin{align}
    \label{eq:Sint}
    S_{\mathrm{int}}=\frac{g_1}{6}\int_{\mathbf{x},t}\tilde\theta(\partial_t\theta)^3+\frac{g_2}{2}\int_{\mathbf{x},t}\tilde\theta\partial_t\theta(\nabla\theta)^2.
\end{align}

These are the most relevant allowed couplings in an \ac{RG} sense. 
The existence of two time scalings in the Gaussian Green functions, as discussed in Sec.~\ref{sssec:PhaseTrans}, renders a simple  power counting analysis at the Gaussian fixed point impossible. 
Therefore, we will instead calculate the diagrams renormalizing the various couplings, and infer their scaling dimensions from the associated infrared divergences. These two time scalings also suggest that, for non-static quantities, not only $\delta$ but also the quantity $\delta/v^2$ shall control the form of correlation functions. We will see that this scale indeed explicitly appears in the renormalization corrections beyond mean-field.

\subsubsection{Perturbative corrections}
\begin{figure}
    \centering 
    \subfloat[\label{fig:2la}]{%
  \includegraphics{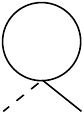}%
}\hfil
    \subfloat[\label{fig:2lb}]{%
  \includegraphics{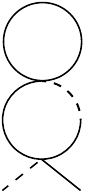}%
}

    \subfloat[\label{fig:2lc}]{%
  \includegraphics{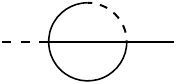}%
}\hfil
    \subfloat[\label{fig:2ld}]{%
  \includegraphics{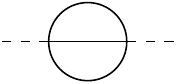}%
}
\caption{Self-energies up to two loop-order. The first three graphs correct the retarded part of the action $\Gamma^{(11)}$, and the last one the noise term $\Gamma^{(20)}$. The solid line denotes the bare Keldysh Green function $G^{K}$, and the solid-to-dashed line the retarded Green function $G^{R}$. The four-point vertices can be either $g_{1}$ or $g_{2}$ defined in~\eqref{eq:Sint}.}
    \label{fig:2loop2pointFunc}
\end{figure}

\begin{figure}
    \centering
    \hspace{-0.3cm}\includegraphics{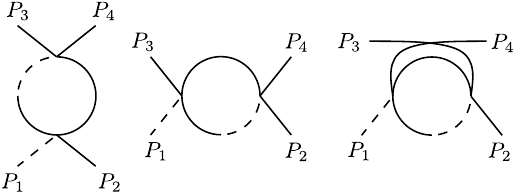}
    \caption{One-loop corrections to $\Gamma^{(13)}$ renormalizing the interactions, $P_{i}=(\mathbf{p}_{i},\omega_{i})$.}
    \label{fig:1loop4P}
\end{figure}

The diagrammatic rules associated to the four-point vertices~\eqref{eq:Sint} and the perturbative corrections to two-point functions (self-energies) up to two loop-order are presented in Fig.~\ref{fig:2loop2pointFunc}. One-loop corrections to the four-point functions are given in Fig.~\ref{fig:1loop4P}.

\begin{figure}
\centering
   \includegraphics{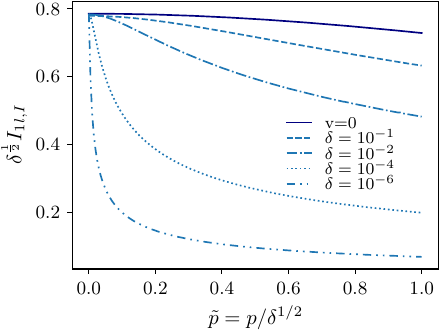}
       \caption{$I_{1l,I}(\tilde{p}=p/\delta^{1/2},\omega_p=0)$ defined by Eq.~\eqref{eq:1loop} in $d=3$ for $v=0$ (solid line) for $v=1$ and different values of $\delta$ (dashed/dotted lines). For $v=0$, the rescaled integral is independent of $\delta$ and diverges as $\delta^{-1/2}$. For $v \neq0$, this divergence is found only for smaller and smaller $\tilde{p}\lesssim \delta^{1/2}/v$ as $\delta \to 0$, and the integral is more and more peaked around zero. For $\delta\to 0$, $I_{1l,I}$ therefore becomes non-analytic and is non negligible only at $\tilde p=0$.}
       \label{fig:1loopPlot}
\end{figure}

\paragraph*{Interactions} --
We now discuss how the parameters of the effective action are renormalized perturbatively, and how they are impacted by the presence of the nonanalyticity of the \ac{CEP} in the spectrum. For this sake we first take a look at the one-loop diagrams renormalizing the four point vertices displayed in Fig.~\ref{fig:1loop4P}, but the phenomenology will go beyond this particular example. First, we consider the case where $g_1$ is renormalizing itself. The first diagram in Fig.~\ref{fig:1loop4P} with $g_{1}$ as vertices is equal to $\omega_2 \omega_3 \omega_4\, I_{1l,I}$ with
\begin{equation}
\label{eq:1loop}
I_{1l,I}(\vecp,\omega_p)  =\int_{\mathbf{q},\omega}^{} i(\omega+\omega_{p})\omega^2G ^{R}(\mathbf{q}+\mathbf{p},\omega+\omega_p)G ^{K}(\mathbf{q},\omega),
\end{equation} 
where $\mathbf{p}=\mathbf{p}_{1}+\mathbf{p}_{2}$, $\omega_p= \omega_1+\omega_2$ and $\int_{\vecq,\omega}\equiv \int d\mathbf{q}d\omega/(2\pi)^{(d+1)}$. The two other diagrams are obtained by permutation of momenta. 
 We now are interested in the infrared behavior of this loop as one tunes $\delta\rightarrow0$, i.e. approaches the \ac{CEP}. When $\delta/v^2$ becomes small, we find that it diverges as
\begin{align}\label{eq:upper_critical}
    I_{1l,I}(\vecp,\omega_p)\sim\delta^{\frac{d-4}{2}},
\end{align}
for small dimensionless momenta $\tilde{p}=\frac{p}{\sqrt{\delta}} \ll \sqrt{\delta}/v$, 
but that this IR divergence is smaller for finite dimensionless momenta
\begin{align}
\label{eq:tranfer_momentum_scaling}
    I_{1l,I}(\vecp,\omega_p)\sim \delta^{\frac{d-4}{2}}(\frac{\delta}{v^2})^\frac{d-1}{4}h(\vecp,\omega_p),
\end{align}
with $h$ some nonsingular scaling function, and therefore become subleading. This implies a very sharp non-analytic behavior as shown in Fig.~\ref{fig:1loopPlot}. This is due to the peculiar form of the dispersions at the \ac{CEP} which induces a resonance condition in the integral to get the highest divergence, as discussed below and in App.~\ref{app:Sunset} where Eqs.~\eqref{eq:upper_critical} and~\eqref{eq:tranfer_momentum_scaling} are also proven.

Similar scaling shapes hold for all combinations of the vertices $g_{1,2}$.
From \eqref{eq:upper_critical}, we can infer that the upper critical dimension is $d_c=4$. Above it, all interactions are irrelevant and the Gaussian theory is exact asymptotically at long wavelengths.
\eqref{eq:tranfer_momentum_scaling} implies, that only loops with transfer momentum $\tilde{p}\ll\frac{\sqrt{\delta}}{v}\xrightarrow{\delta\rightarrow0}0$ contribute to the renormalization of the vertices as we approach the CEP. We can thus regard all finite transfer momenta to lead to subleading contributions.

This means, that divergences of vertex corrections depend on the momentum configuration of the respective vertex in a highly non-analytic way and a derivative expansion around $p=0$ is not possible. In particular, we find that the two limits $\tilde {\vecp} \to 0$ and $\delta \to 0$ cannot be exchanged, see Fig.~\ref{fig:1loopPlot}. This non-analytic structure can be related to the nonanalyticity of the exceptional point. Intuitively, this is indicated by the EP momentum scale $q_{EP}=\frac{\delta}{v}$ already seen in the linear spectrum in Sec.~\ref{subsec:gaussian_fluctuations} above which dimensionful transfer momenta $\vecp$ are cut off (since the critical regime is described by $q/\sqrt{\delta}\sim1$, we are generally interested in momenta $q\gg q_{EP}$).

We now illuminate the origin of these different scalings, which result from a resonance condition on the external momentum. In a nutshell, after frequency integration, rescaling of momenta by introducing $\tilde{\vecq}= {\mathbf{q}}/\delta^{1/2}$, and for small values of $\delta$, the diagram~\ref{fig:2lc}  at zero external frequency reduces to
\begin{widetext}
\begin{equation}
\label{eq:IwithDL}
I_{1l,I} =\frac{\delta^{\frac{d-4}{2}}}{2}\int_{\tilde{\mathbf{q}}}\,\frac{f_1(\tilde{\mathbf{q}},\tilde{\mathbf{p}})+O\left(\delta\right)}{ \frac{v^2 }{\delta } \left(\tilde{\mathbf{p}}^2+2 \tilde{\mathbf{p}} \cdot \tilde{\mathbf{q}}\right)^2  \Delta
   \left(\tilde{\mathbf{q}}^2\right)+ f_1(\tilde{\mathbf{p}},\tilde{\mathbf{q}}) \Delta
   \left(\tilde{\mathbf{q}}^2\right) \left(\Delta
   \left(\tilde{\mathbf{q}}^2\right)+\Delta \left((\tilde{\mathbf{p}}+\tilde{\mathbf{q}})^2\right)\right)+O\left(\delta\right)}.
\end{equation}
 \end{widetext}
In this expression, we use $\Delta(y)= y+1$, and $f_1(\tilde{\mathbf{q}},\tilde{\mathbf{p}})= (\tilde{\mathbf{p}}+\tilde{\mathbf{q}})^2\Delta(\tilde{\mathbf{q}}^2)+ \tilde{\mathbf{q}}^2\Delta((\tilde{\mathbf{p}}+\tilde{\mathbf{q}})^2)$.
In the denominator, we keep a higher order in terms of $\delta$ since it becomes the dominant term in the expansion as soon as $v^2\left(\mathbf{p}^2+2 \mathbf{p} \cdot \mathbf{q}\right)$ is small. This is always true for $\mathbf{p}=0$, but only occurs for special configuration of momenta when $\mathbf{p}\neq0$. When this is fulfilled the integrand behaves as $\delta^{d/2-2}$ and only as $\delta^{d/2-1}$ when it is not: there is a resonance condition to get the highest divergence. Mathematically, the integrand in~\eqref{eq:IwithDL} becomes non-analytic and behaves as a Dirac distribution in the $\delta/v^2\to 0$ limit to still give the stronger divergence. This behavior can in turn be used to compute the integrals, see App.~\ref{app:Sunset}.

This is in sharp contrast with more standard renormalization corrections, where the leading momentum dependent term scales accordingly to the momentum independent part, and where higher order terms in momentum are negligible in the infrared in the spirit of a gradient expansion. This expansion in momentum cannot be used here because of the non-analytic structure. Indeed, we show in App.~\ref{app:scaling} that such an expansion generates spurious divergences with arbitrarily high power in $\delta$. This also illustrates why standard power counting does not work: the presence of the additional scale $v^2/\delta$ allows for a more complex scaling of integrals, which breaks the generic scaling behavior. 

\paragraph*{Self-energies} -- This structure also strongly impacts the perturbative corrections at higher loop orders. We now discuss that matter for the two-point vertex $\Gamma^{(2)}$. We will see that it makes sunset diagrams~\ref{fig:2lc} and~\ref{fig:2ld} less divergent than the tadpoles one~\ref{fig:2la} and~\ref{fig:2lb}. We begin the analysis with the tadpole diagrams. They are linear in the external frequency, and momentum independent. They therefore only renormalize the momentum independent damping coefficient $\delta$. The one-loop tadpole gives
\begin{equation}
    \label{eq:Itadpole}
    I_{1l}=\frac{K_d}{2}(g_1+g_2)\int_{}^{}dq \frac{q^{d-1}}{q^2+\delta},
\end{equation}
where $K_{d}=S_{d}/(2\pi^{d})$ with $S_{d}$ the surface of the $d$-dimensional sphere. In these expressions, new dimensionless quantities have been introduced via the following rescaling: $\delta \to \delta Z$, $g_1 \to g_1 Z^2/D$ and $g_2 \to g_2 Z^2 v^2/D$. Performing the integral over momentum gives
\begin{align}
    \label{eq:Dyson1}
    \overline\delta&=\delta'-K'_{d}\frac{g_{1}+g_{2}}{2}\delta^{\frac{d-2}{2}},
\end{align}
with $K'_{d}=-K_d \pi/(2 \sin(\pi d/2))>0$, and $\delta'= \delta+(g_{1}+g_{2})K_{d}/2\int_{0}^{\Lambda}dq/q^{2}$. Here, $\Lambda$ denotes the UV cutoff used to regularize the loops. This is consistent with $d_{c}=4$, since the perturbative corrections in~\eqref{eq:Dyson1} to the damping become non-negligible below four dimensions. The contribution of the two-loop tadpole diagram~\ref{fig:2la} is simply given by the square of~Eq.~\eqref{eq:Itadpole} and behaves as $\delta^{d-3}$.

We now turn our attention to the loop integrals of the sunset diagrams Fig.~\ref{fig:2lc} (for two $g_1$ vertices going into the loop, the same however holds for all vertex combinations) at vanishing external momenta. It can be written as 
\begin{align}
\label{eq:1boucle=2bloucle}
    I_{2l}=g_1^2\int_{\vecq,\omega'}\omega'^2G^K(\vecq)I_{1l,I}(\vecq,\omega).
\end{align}
I.e. the bubble diagram analysed earlier reappears as a subgraph of the sunset diagrams and their transfer momentum is integrated over. Since the point of vanishing transfer momentum  at which the resonance occurs is a zero measure set, only the subleading scaling of $I_{1l}(\vecq,\omega)$ contributes to the sunset diagram. Thus the whole sunset diagram, even at finite momentum or frequency, is subleading when compared to the other terms in the renormalized two-point function at small damping. Indeed, while $I_{2l}$ scales as $\delta^{d-3}$ for $\delta/v^2 \gg 1$ (like the tadpole diagram \ref{fig:2lb}), it is suppressed in the critical regime $\delta/v^2\ll 1$ where
\begin{align}
\label{eq:scalingSun}
    I_{2l}\sim \delta^{d-3}\Big(\frac{\delta}{v^2}\Big)^{\frac{d-1}{4}} \ll \delta^{d-3}.
\end{align}
Eq.~\eqref{eq:scalingSun} is proven in App.~\ref{app:Sunset}, and a similar result is found for diagram~\ref{fig:2ld} with $\delta^{d-3}$ replaced by $\delta^{d-4}$. At finite external momentum, the sunsets are even less divergent since they display the same non-analytic structure in their $p$ dependencies than the one found for the one-loop diagrams (see Fig.~\ref{fig:1loopPlot}). More details can be found in App.~\ref{app:Sunset}.

This shows that the presence of the EP leads, for $d>1$ and in particular for the dimensions of interest $d\geq2$, to smaller infrared divergences in two-loop sunset diagrams, which in turn indicates that they will contribute only subdominantly in the critical regime and can be neglected. Only one-loop contributions without transfer momentum survive and the corrections to $v$, $K$, and $D$ associated to anomalous dimensions and $z$ exponents all vanish. 

Formally, this is a valid assumption if $I_{2l}(p)$ remains very
small compared to all terms in Eq.~\eqref{eq:Dyson1} i.e. to the renormalized damping $\bar\delta$. Because $I_{2l}$ diverges when the damping becomes small, this necessarily implies a condition on the prefactor of the loop i.e.\ on the interactions $g_1+g_2$, which have to be sufficiently small.  
This condition can only be self-consistently checked once we have computed $\bar\delta$, and we therefore defer its discussion to Sec.~\ref{ssec:SolSelfCons}.

In principle, one has to check that higher loop terms for the self-energies and for interactions follow a similar pattern and are also negligible.
The discussed pattern however extends to all diagrams in the perturbative series that contain loops with more than one momenta. Thus, only graphs with a one-loop structure i.e.\ graphs that are products of one-loop graphs, and without momentum transfer survive when $\delta/v^2$ becomes small.  Alternatively, this is elegantly recovered in the \ac{DSE} framework since the full effective action can be computed solely from (dressed) tadpole and sunset diagrams, see App.~\ref{app:SDE}.

\subsubsection{Self-consistent equations and first-order phase-transition}

Because of the emergent one-loop structure, with negligible higher loop effects, it is possible to resum the entire perturbation series, or equivalently to solve the \ac{DSE}, see App.~\ref{app:SDE}. For the retarded two-point function, the remaining diagrams (the so called cactus diagrams) form a geometric series which leads to the following self-consistent Hartree equation,
\begin{align}
    \label{eq:Dyson2}
    \overline\delta&=\delta'-K'_{d}\frac{g_{1}+g_{2}}{2}\bar\delta^{\frac{d-2}{2}}.
\end{align}
It is also found by using the renormalized damping in the tadpole diagram~\ref{fig:2la}.

For interactions, in order to represent the full renormalized couplings e.g.\ $\overline{g}_{1}(\mathbf{p}_{2},\mathbf{p}_{3},\mathbf{p}_{4})$ (frequency dependencies are implicit), we have to take care of momentum dependencies because of the non-analytic structure discussed above. Note that there is no need to consider such an ansatz in the frequency domain, see App.~\ref{app:Sunset}. We can however truncate the couplings to the most dominant hotspot regions and do not need to consider the full momentum dependencies of the vertices, since the renormalization approximately does not depend on the value of momenta but only on their configuration, when $\delta/v^2$ is sufficiently small. To be specific, we need to introduce three different couplings. The first one is
\begin{equation}
    g_{1,a}= \overline g_{1}(\mathbf{p},-\mathbf{p},\mathbf{p}')
\end{equation}
 with $\mathbf{p} \neq \mathbf{p}'$ for which only the third diagram in Fig.~\ref{fig:1loop4P} contributes,
 \begin{equation}
 g_{1,b}= \overline g_{1}(\mathbf{p},-\mathbf{p},\mathbf{p})
 \end{equation}
  for which the last two diagrams contribute and
  \begin{equation}
  \label{eq:g1c}
  g_{1,c}= \overline g_{1}(0,0,0)    
  \end{equation}
for which all diagrams have to be taken into account. Other generic configurations are not renormalized. This pattern also occurs in Brazovskii's phase-transition scenario as discussed in~\cite{Hohenberg1995}.

These one-loop diagrams also generate a geometric series which can be resummed. Alternatively, this can be seen by using Dyson-Schwinger equations as done in App.~\ref{app:SDE}, where more details and equations for all couplings can be found. It yields, e.g., the following self-consistent equations for the full macroscopic couplings,  
\begin{subequations}
\label{eq:g1SDE}
\begin{align}
\label{eq:g1SDEa}
g_{1,a}&= g_{1}-\frac{g_{1}+g_{2}}{2}g_{1,a}\int_{\mathbf{q}}^{}\frac{1}{(q^{2}+\overline\delta)^{2}} ,\\
\label{eq:g1SDEb}
g_{1,b}&= g_{1}- 2\frac{g_{1}+g_{2}}{2}g_{1,a}\int_{\mathbf{q}}^{}\frac{1}{(q^{2}+\overline\delta)^2},\\
\label{eq:g1SDEc}
g_{1,c}&= g_{1}- 3\frac{g_{1}+g_{2}}{2}g_{1,a}\int_{\mathbf{q}}^{}\frac{1}{(q^{2}+\overline\delta)^2}.
\end{align}
\end{subequations}  
The integral that appears in Eqs.~\eqref{eq:g1SDE} can be calculated and the equations can be inverted to give
\begin{subequations}
\label{eq:g1SDEsolved}
 \begin{align}
\label{eq:g1aSDEsolved}
g_{1,a}&= \frac{g_{1}}{1+ \alpha_{2}\overline\delta^{\frac{d-4}{2}}},  \\
\label{eq:g1bSDEsolved}
g_{1,b}&= g_{1}\frac{1- \alpha_{2}\overline\delta^{\frac{d-4}{2}}}{1+ \alpha_{2}\overline\delta^{\frac{d-4}{2}}}, \\
\label{eq:g1cSDEsolved}
g_{1,c}&= g_{1}\frac{1-2 \alpha_{2}\overline\delta^{\frac{d-4}{2}}}{1+ \alpha_{2}\overline\delta^{\frac{d-4}{2}}}, 
\end{align}  
\end{subequations}
with $\alpha_{2}=(g_{1}+g_{2}) K_{d}'(d-2)/2>0$. 
We see that, while the coupling $g_{1,a}$ at finite momenta is always positive, the couplings $g_{1,c}$ with zero incoming momenta can be negative for sufficiently small $\overline\delta$. This therefore opens the route to a fluctuation induced first-order phase transition since a potential with a negative quartic term typically displays a first-order transition~\cite{Amit2005}. The fourth-order is now momentum dependent, and one has to specify which quartic couplings should enter the effective potential Eq.~\eqref{eq:Veff} for the order parameter $E=\partial_t\theta/\sqrt{\rho_0}$ and check if it is negative. The condensation mechanism occurs at zero momenta, and $E$ is given by minimizing the effective equation of motion $\Gamma^{(10)}$ with a constant order parameter $\partial_t\theta(x,t)=E$. It translates, in momentum space, to
\begin{equation}
  \partial_t\theta(\mathbf{p},\omega)=E\Pi(\mathbf{p},\omega)= \sqrt{\rho_0}E\delta(\mathbf{p})\delta(\omega).
\end{equation}
In the effective equation of motion for $E$, the fourth order term is proportional to $\overline g_1(0,0,0)\partial_t\theta(\mathbf{p}=0,\omega=0)^3=g_{1,c}(\sqrt{\rho_0}E)^3$. The coupling that fixes the limit cycle rotation frequency $E$ is therefore $g_{1,c}$ defined in~Eq.~\eqref{eq:g1c} which can indeed turn negative because of Eq.~\eqref{eq:g1cSDEsolved}. This will drive the first-order phase transition. The coupling $g_{1,b}$ can also turn negative (see Eq.~\eqref{eq:g1bSDEsolved}) which could indicate some instability at finite momentum close to the transition. However, it is larger than $g_{1,c}$ and turns negative for even smaller damping, for which the first-order transition we discussed has already taken place. It therefore does not alter the first-order scenario we describe.

Now, to have a well defined potential, we need to add a sextic term in the potential, i.e.\ $u_{1}\tilde{\theta}(\partial_{t}\theta)^{5}/5!$ in the action. Exactly as for the quartic couplings, there are several hotspot configurations of momenta for which only one-loop diagrams contribute. One has to consider different couplings associated to each of these hotspot regions. To describe the effective potential, we however only need the value of this coupling at zero external momenta,
 \begin{equation}
 u_{1,e}=\overline u_{1}(0,0,0,0,0).
 \end{equation}
\begin{figure}
    \centering
    \includegraphics{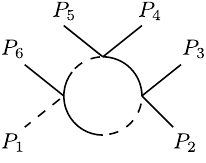}
    \caption{One-loop contribution to the six-point function $\Gamma^{(15)}$. The diagrams obtained by permutation of external and internal lines are not shown.}
    \label{fig:Diag6point}
\end{figure}
Being an irrelevant coupling, its value is entirely set by the quartic couplings at small $\bar \delta$. Its renormalization is then given by the one-loop diagram displayed in Fig.~\ref{fig:Diag6point}. The resummed expression is obtained by using dressed propagators and interactions. This leads to
\begin{equation}
\label{eq:g6e}
u_{1,e}= 15 g_{1}^{2} \frac{\alpha_{3}\overline\delta^{\frac{d-6}{2}}}{(1+\alpha_{2}\overline\delta^{\frac{d-4}{2}})^3}, 
\end{equation} 
where $\alpha_3= (g_1+g_2)K_{d}'(d-2)(4-d)/8$.

We are now in the position to solve the resulting equations, and discuss in greater details how the first-order transition takes place.

\subsection{Solution of self-consistent equations}\label{ssec:SolSelfCons}

The resulting system of equations constituted by Eqs.~\eqref{eq:Dyson2},~\eqref{eq:g1SDEsolved} and~\eqref{eq:g6e} is solved by extracting the damping $\bar\delta$ from the first equation, and inserting it into the others.

\begin{figure}

    \begin{center}
    
        \subfloat[][First-order scenario]{\label{fig:Scale2}
        \begin{tikzpicture}
               \draw[thick,->] (-4,0) -- (4,0);
        \node at (4.2,-0.2) {$q$};

        \draw[thick] (-0.,-0.1) -- (-0.,0.1);
   
        \node at (-0.,-0.4) {$g^{2/(4-d)}$};

          \draw[thick] (3.6,-0.1) -- (3.6,0.1);
          \node at (3.6,-0.4) {$v$};

          \node[align=center] at (2.25,1.25) {Gaussian CEP};
          \draw[thick, dashed] (0,0)--(0,1);
               \node[align=center] at (0,1.5) {1st order\\transition};
          \node[align=center] at (-2.25,1.25) {Goldstone regime\\ rotating};
          \foreach \i in {1,...,8}{
         \draw[xshift=5*\i,thick,red] (-4.2,0.2) --(-4.0,-0.2);}
        \node[align=center] at (2.25,0.5){$G^K(\mathbf{q},t=0)=q^{-4}$};
          \node[align=center] at (-2.25,0.5) {$G^K(\mathbf{q},t=0)=q^{-2}$};

        \draw[thick] (-3.5,-0.1) -- (-3.5,0.1);
        \node at (-3.25,-0.5) {$(\rho_0^{-1})^{2/(4-d)}$};

    \end{tikzpicture}
    }
            \hfil
        \subfloat[][Symmetry restoration scenario]{\label{fig:Scale3}
        \begin{tikzpicture}
               \draw[thick,->] (-4,0) -- (4,0);
        \node at (4.2,-0.2) {$q$};

        \draw[thick] (-0.,-0.1) -- (-0.,0.1);
   
        \node at (-0.,-0.4) {$(\rho_0^{-1})^{2/(4-d)}$};

          \draw[thick] (3.6,-0.1) -- (3.6,0.1);
          \node at (3.6,-0.4) {$v$};

          \node[align=center] at (2.25,1.25) {Gaussian CEP};
          \draw[thick, dashed] (0,0)--(0,1);
               \node[align=center] at (0,1.5) {Escape of CEP};
          \node[align=center] at (-2.25,0.5) {Model A physics};
          \foreach \i in {1,...,8}{
         \draw[xshift=5*\i,thick,red] (-4.2,0.2) --(-4.0,-0.2);}
        \node[align=center] at (2.25,0.5){$G^K(\mathbf{q},t=0)=q^{-4}$};

        \draw[thick] (-3.5,-0.1) -- (-3.5,0.1);
        \node at (-3.25,-0.5) {$g^{2/(4-d)}$};
       
    \end{tikzpicture}
    }
    \end{center}
    \caption{Summary of scales for the two different scenarios obtained while approaching the \ac{CEP}.  (a) The first-order scenario occurs before reaching the point where the symmetry get restored since $\rho_0^{-1}\gg g$, and the system ends up in the rotating phase. (b) The symmetry gets restored at large distances, and the system is  in the disordered phase. In both cases the red dashed area indicates the scale which is never reached because the other scenario takes place first.}
    \label{fig:ScaleSum}
\end{figure}
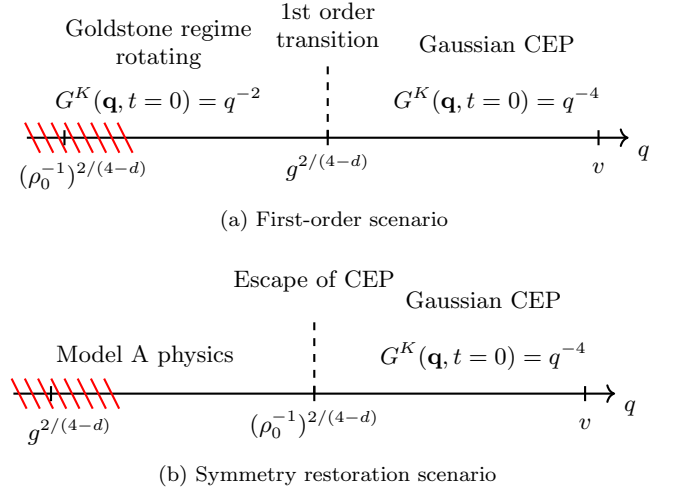

\begin{figure}
    \centering
    \includegraphics{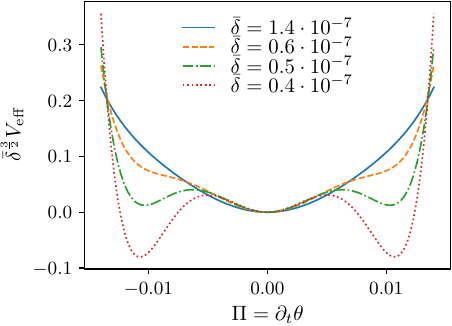}
    \caption{The effective potential as a function of $\Pi=\partial_t\theta$ obtained by solving Eqs.~\eqref{eq:Dyson2},~\eqref{eq:g1SDEsolved} and~\eqref{eq:g6e} becomes characteristic of a first-order phase transition at a finite renormalized damping $\bar\delta$. The results are presented for $d=3$, $g_1=g_2=10^{-1}$ and $v=1$.}
    \label{fig:Veff}
\end{figure}

Asymptotically, the system does not reach any fixed-point, ruling out the second-order phase transition scenario (iii) of Sec.~\ref{ssec:EPFluct}. We find that the $(\partial_{t}\theta)^{4}$ coupling becomes negative and the effective potential describes a first-order  transition for sufficiently small $\delta$ as shown in Fig.~\ref{fig:Veff}: new minima appear for a finite $\partial_t \theta=E$, and the order parameter jumps from zero to a finite value. From Eq.~\eqref{eq:g1cSDEsolved}, the phase transition happens approximately when the quartic term becomes negative, i.e. at a first order transition scale
\begin{equation}
(g_1+g_2)\overline\delta_{\text{fo}}^{(d-4)/2}\sim 1,
\end{equation}
giving
\begin{equation}
\label{eq:delta1stOrder}
    \overline\delta_{\text{fo}}\sim (g_1+g_2)^{2/(4-d)}.
\end{equation}
 We are left to compare this scale for the onset of a first order transition to our previous result on the symmetry restoration $\delta_{\text{sym}}$ via the suppression of $\rho_0$. The system displays a new scale (in terms of the original non rescaled variables)
 \begin{align}
 \label{eq:multicritical_point}
     \frac{\delta_{\text{fo}}}{\delta_{\text{sym}}}=\rho_0\frac{Z(g_1+v^2g_2)}{v^2}\equiv\rho_0 g,
 \end{align}
 which sets whether there is symmetry restoration ($\rho_0g\ll1$) or a fluctuation induced first order transition ($\rho_0 g\gg1$) separated by the multicritical point ($\rho_0 g\sim 1$) where both transition lines meet. The resulting qualitative phase diagram is shown in Fig.~\ref{fig:PhaseDiagram}. Furthermore we remark, that the first order transition scale $\delta_{\text{fo}} $ coincides with the Ginzburg criterion where non-Gaussian fluctuations are expected to play a role. (The Ginzburg criterion can be simply derived by comparing one-loop contributions to the order parameter fluctuations to the bare one). This means that in a situation close to the weakly first-order situation, i.e. $1\gg \delta>\delta_{\text{sym}},\delta_{\text{fo}}$, one can observe the Gaussian scaling behavior described in Sec.~\ref{subsec:gaussian_fluctuations} on length scales $\xi<\sqrt{\frac{Z}{\delta}}$ but there is no intermediate regime where one can observe interaction corrections to that scaling, including anomalous dimensions,  before reaching the regime of either symmetry restoration or first order transition. This is different from e.g. driven-dissipative condensates below the lower critical dimension, where one can observe KPZ scaling at finite length scales smaller than the length scales at which order breaks down~\cite{Sieberer2016,Fontaine2021}.
 The different possible scenario are summarized on Fig.~\ref{fig:ScaleSum}.

 To get a complete picture describing all regimes, one needs a method that can describe both the broken phase within which the first-order transition occurs and the regime in which the amplitude goes to zero. One possible route would be to use the \ac{frg} which is known to describe both the phase transition and the broken phase in equilibrium~\cite{Dupuis_2021}.

\paragraph*{Validity} -- Let us finally assess the validity of the assumptions made, and discuss quantitatively under which conditions the subleading corrections are negligible. The sunset contribution~Eq.~\eqref{eq:scalingSun} (using the renormalized damping in the loop) to $\bar\delta$ can be neglected when it is small compared to all terms in Eq.~\eqref{eq:Dyson2}. The most stringent condition is obtained by demanding it to be negligible with respect to the zero-order term $\bar \delta$. It gives the following condition,
\begin{equation}
\label{eq:ConditionApp}
    (g_{1}+g_{2})^{2}\bar\delta^{d-4}\bar\delta^{(d-1)/4}\ll 1.
\end{equation} 
The one-loop diagrams with momentum transfer can be neglected when~\eqref{eq:tranfer_momentum_scaling} is way smaller than $g_1+g_2$, and the sunset~\ref{fig:2ld} when it is way smaller than $D$, which both lead to the very same condition. Equivalently, the condition~\eqref{eq:ConditionApp} is recovered nonperturbatively using \ac{DSE} as discussed in App.~\ref{app:SDE}.

The condition~\eqref{eq:ConditionApp} becomes, using Eq.~\eqref{eq:delta1stOrder}, $(g_1+g_2)^{(d-1)(4-d)/8} \ll 1$, which is satisfied in $2<d<4$ for sufficiently small values of the bare coupling constants, i.e.\ for a microscopic theory not too far away from the Gaussian fixed-point.
In that case, $\overline\delta$ is generically small close to the transition because of Eq.~\eqref{eq:Dyson2}. The transition is then weakly-first order and the condition $\overline\delta/v^2 \ll 1$ is in turn also not violated, and our calculation is fully justified in this regime. 

One can formally still try to solve the equations for even smaller values of $\bar\delta$ i.e\ deep in the ordered phase where the true damping is instead defined at the nonzero minima. This always gives a solution with $\bar\delta>0$, and the minimum at $\partial_t\theta=0$ does not disappear. Note that this issue also arises in Brazovskii's phase-transitions scenario~\cite{Hohenberg1995}. However, the condition~\eqref{eq:ConditionApp} is not satisfied in this regime, and the solution does not apply anymore. This regime is anyway well-described by the Gaussian theory for the rotating phase done in Sec.~\ref{subsec:gaussian_fluctuations}.

\subsection{\texorpdfstring{$\mathbb{Z}_2$}{Z2} symmetry breaking and \texorpdfstring{$SO(2) \simeq U(1)$}{SO(2)~U(1)} case}
\label{ssec:Sym_break}
\paragraph*{Explicit $\mathbb{Z}_2$ breaking -- }

We now discuss the case where the symmetry is $SO(2)$ or $U(1)$ instead of the $O(2)$ symmetry discussed so far. In that case the $\mathbb{Z}_2$ symmetry is explicitly broken, and a linear term $\partial_t\theta=\mu_0+...$ is allowed in~\eqref{eq:S0GoldstoneStat2}, together with the  cubic non-linearity $\tilde \theta (\partial_t\theta)^2$ and the \ac{KPZ} non-linearity $\tilde \theta \nabla \theta \cdot \nabla \theta$. This induces an explicit rotation of the order parameter, and thus no static phase. Since there is no unbroken internal symmetry left that could by spontaneously broken, no second-order phase transition can occur at the mean-field level, and no CEP is found. This is equivalent to adding a  magnetic field $\mu_0$ in the Ising case: the effective potential for $\partial_t \theta$ generically does not display spontaneous symmetry breaking anymore, but rather describes a first-order phase transition at the mean-field level already between phases with different rotation speeds. There is thus no divergent correlation length occurring, and there is no way to get the enhancement of the fluctuations found at the \ac{CEP}.

The \ac{CEP} transition can still be reached by tuning only one additional parameter: $\mu_0$ can be chosen such that there is an emergent additional $\mathbb{Z}_2$ symmetry at the critical point, where our model is then recovered. There is thus a first-order phase transition line whose end point is exactly the CEP described in this work. This is the transition discussed in~\cite{Hanai2020}. However, their study of fluctuations include the cubic and \ac{KPZ} non-linearity, while their values are zero at the \ac{CEP} because of the additional fine-tuning. It therefore does not describe the \ac{CEP} transition of interest here. All of this is analogous to the second order transition found at the endpoint of the liquid-gas transition that falls in the Ising universality class with upper critical dimension $d_c=4$. It has an emergent $\mathbb{Z}_2$ symmetry at the transition, and one does not consider the cubic non-linearities.

\paragraph*{Spontaneous $\mathbb{Z}_2$ breaking --} Within the rotating phase, the cubic and \ac{KPZ} non-linearities are also effectively present because of the spontaneously broken $\mathbb{Z}_2$ symmetry, or more technically since $\partial_t\theta(\mathbf{q}=0,\omega=0)=E$ has a non-zero value. In particular, we expect the usual \ac{KPZ} physics in the rotating phase. It means that in $d=2$, the rotating phase should instead realize \ac{KPZ} physics at intermediate scale only~\cite{Altman2015,Wachtel2016,Sieberer2016} while the non-rotating ordered phase should correspond to the \ac{BKT} quasi-long range order phase.

\subsection{\texorpdfstring{$O(N>2)$}{O(N)} case}\label{ssec:Nfields}

We now turn to the generic $O(N)$ case. We first discuss the corresponding action and the additional interactions that arise between Goldstone modes for $N>2$ that add some complexity. We then explain how we can generalize the previous results for the first-order scenario even in the presence of these new interactions. 

The model one obtains after phase-amplitude decomposition and integration of the amplitude mode, defines what is often referred to as \ac{Nlsm}.
The Gaussian part of the action in the static phase is given by
\begin{equation}
\label{eq:S0_NL}
    S_0=\int_{\mathbf{x},t}\tilde{\boldsymbol{\pi}}\cdot(\partial_t^2+(-K\Delta+\delta)\partial_t-v^2\Delta)\boldsymbol{\pi} - D \tilde{\boldsymbol{\pi}}\cdot\tilde{\boldsymbol{\pi}},
\end{equation}
where $\boldsymbol\pi= (\theta_2,\dots, \theta_{N})$. Beyond mean-field, we need to consider the generalization of~\eqref{eq:Sint},
\begin{align}
\label{eq:Sint_NL}
     S_{\mathrm{int}}&=\frac{g_1}{6}\int_{\vecx,t}\tilde{\boldsymbol{\pi}}\cdot\partial_t\boldsymbol{\pi}(\partial_t\boldsymbol{\pi})^2+\frac{g_2}{2}\tilde{\boldsymbol{\pi}}\cdot\partial_t\boldsymbol{\pi}(\nabla\boldsymbol{\pi})^2. 
\end{align}

However, contrary to the $O(2)$ case, there are higher order terms that do not only involve derivative terms for $N>2$ as usual for \ac{Nlsm}~\cite{ZinnJustin}. This is due to the fact that the $O(N)$ symmetry does not act anymore as a shift symmetry for the Goldstone modes when $N>2$~\cite{Schwartz2014}.  For example, the term $\tilde{\boldsymbol{\phi}}^T\partial_t \boldsymbol{\phi}$ leads, following the procedure explained in Sec.~\ref{subsec:gaussian_fluctuations}, to 
\begin{align}
\label{eq:NewTerm}
\tilde{\boldsymbol{\phi}}^T\partial_t \boldsymbol{\phi}=\tilde{\boldsymbol{\pi}}\cdot\partial_t\boldsymbol{\pi}+\frac{\rho_0^{-1}}{6}((\boldsymbol{\pi}\cdot\boldsymbol{\pi})\tilde{\boldsymbol{\pi}}-(\tilde{\boldsymbol{\pi}}\cdot\boldsymbol{\pi})\boldsymbol{\pi})\cdot\partial_t\boldsymbol{\pi}+ \dots,
\end{align}
where the neglected terms are irrelevant. A similar pattern arises for every operator present in~\eqref{eq:S0_NL} and \eqref{eq:Sint_NL}.
The coefficients of these new operators are not independent from the one in~\eqref{eq:S0_NL} and \eqref{eq:Sint_NL} because they are generated by the same operator.
This originates from the underlying $O(N)$ symmetry of the model and therefore remains true even beyond mean-field. It is then common to refer to $\rho_0^{-1}$ as a coupling constant in the \ac{Nlsm}. The new operators then lead to a non-trivial (self-)renormalization of the amplitude which is absent in the $O(2)$ case. 
Within the statically ordered phase, the renormalized amplitude is finite, and at sufficiently small scale, the higher order terms can be neglected and the action reduces to its Gaussian part. Indeed, for large $\rho_0$, only fluctuations with $|\boldsymbol{\pi}|\lesssim 1$ contribute to the functional integral and higher order terms become negligible since they come with powers of $\rho_0^{-1}$~\cite{ZinnJustin}~\footnote{
This argument can be made quantitative by the \ac{RG}. The coupling $\rho_0^{-1}$ goes to zero in dimensionless units at the Goldstone fixed point associated to the ordered phase~\cite{ZinnJustin} and the action reduces to its Gaussian part at low-energy.}.

We are now ready to discuss the situation when approaching the \ac{CEP}. 
From previous subsections, we expect  $g_{1}$ and $g_2$ but also $\rho_0^{-1}$ to have dimension $4-d$. This can be checked diagrammatically. Mean-field results can then be used above four dimension sufficiently deep in the ordered phase~\footnote{The irrelevance of $\rho_0^{-1}$ does not preclude the existence of non-trivial corrections above four dimensions (by analogy with the usual equilibrium \ac{Nlsm} above two dimensions).} and we start by discussing it since it will be useful below. It turns out that we can generalize the potential picture developed for the $O(2)$ case.  Omitting the addtional terms coming from~\eqref{eq:NewTerm}, we again have a potential of the form
\begin{equation}
\label{eq:VeffN}
V_{N,\mathrm{eff}}(\rho_{\boldsymbol{\pi}}=(\partial_t \boldsymbol\pi)^2)=\frac{\delta}{2} \rho_{\boldsymbol{\pi}}+ \frac{g_1}{4!}\rho_{\boldsymbol{\pi}}^2.
\end{equation}
This potential is simply the generalization of Eq.~\eqref{eq:Veff}. It would display the typical spontaneous symmetry breaking of $O(N-1)$ down to $O(N-2)$. Here, we have additional terms of the form~\eqref{eq:NewTerm} that prevent our model from reducing to the equilibrium $O(N)$ case for $\partial_t \boldsymbol{\pi}$ even when $v=0$. However, they are present to ensure that the $O(N)$ symmetry is intact, and they fix the value of the amplitude. We can therefore expect that they do not play any role when it comes to the rotational angular velocity. Indeed, our potential picture reproduces the calculation done by expanding directly around the rotating frame: once $E$ is determined by minimizing Eq.~\eqref{eq:VeffN} with $\boldsymbol{\pi}_0= (\sqrt{\rho_0}E t,0,\dots,0)$, these additional terms lead to time-dependent terms that can be eliminated by going into the rotating frame. This then gives back the mean-field expression~\eqref{eq:MF_frequency} for $E$. In addition, one can also study the fluctuations around the rotating order by writing the field as $\boldsymbol{\pi}= (\sqrt{\rho_0}E t+\theta_{\parallel},\boldsymbol{\theta_{\perp}})$, with  $\theta_{\parallel}$ the longitudinal mode and $\boldsymbol{\theta_{\perp}}\in \mathbb{R}^{N-2}$ the transverse modes of the broken $O(N-1)$ symmetry. Their respective action matches~\eqref{eq:S0par} and~\eqref{eq:S0perp} obtained directly from the expansion around the rotating phase. This fully justifies our assertion that the potential picture works also in the $O(N)$ case. We therefore again rely on it also beyond mean-field as we discuss now.

Below four dimensions, $g_1$, $g_2$ and also $\rho_0^{-1}$ become relevant. They have the same dimension around the Gaussian fixed-point and therefore grow at the same rate next to it. We can use the same strategy as in the $O(2)$ case: We expect the $g_1$ and $g_2$ couplings to again favour the first-order phase transition. Sufficiently deep in the ordered phase at the bare scale, we can therefore neglect the restoring effect linked to $\rho_0^{-1}$ since its contribution to loops will become non-negligible only at larger scale. In that case, we get the generalizations to $O(N-1)$ field of the different diagrams discussed in Sec.~\ref{ssec:BeyondGaussFluct}. This only adds $N$ dependent prefactors in front of the loop integrals but leaves the integrals involved unchanged, and therefore their momentum structures and divergences.

The self-consistent equations are therefore similar to the $O(2)$ case. The same mechanism for first-order transition apply again because the structure pointed out for the quartic couplings $g_{1,c}$ which led to this scenario in the $O(2)$ also arises.
Explicitly, we find (see App.~\ref{app:SDE}) that
\begin{equation}
    g_{1,c}=g_1\frac{ (9-4 \alpha_2\bar \delta^{\frac{d-4}{2}} (\alpha_2\bar \delta^{\frac{d-4}{2}} (N'+2)+3))}{(2 \alpha_2+3)
   (\alpha_2\bar \delta^{\frac{d-4}{2}}\ (N'+2)+3)},
\end{equation}
with $N' =N-1$. We can conclude that $g_{1,c}$ turns negative when
\begin{equation}
    \bar\delta=(\frac{2\alpha_2 2(N'+2)}{3(\sqrt{3+N'}-1)})^{\frac{2}{4-d}},
\end{equation}
i.e.\ at the same scale we have in the $O(2)$ case, given by~\eqref{eq:delta1stOrder}, up to a $N$-dependent factor.

In the opposite limit, the $\rho_0^{-1}$ coupling grows first and we approach the point where the symmetry gets restored first. In a generic situation, the \ac{Nlsm} will break down again. We reach the same conclusion as for the $O(2)$ case, and also obtain the qualitative phase diagram Fig.~\ref{fig:PhaseDiagram}. However, contrary to the $O(2)$ case, the \ac{Nlsm} is expected to describe the complete phase diagram whenever the  Goldstone fluctuations dominate the amplitude ones, e.g.\ in a large $N$ expansion. The full phase diagram could be obtained from a renormalization group analysis, which is left for future work.

\section{Possible Realization Schemes}\label{sec:implementation}

We now provide physical schemes that can realize the effective field theory described above. For $N=2$, we put forward a model for a driven easy plane ferrimagnet that realizes our model \eqref{eq:Langevin}. For larger $N$, we construct a Lindblad evolution that realizes our model in a semiclassical approximation.  Furthermore, we demonstrate that the universal phenomena of so-called nonreciprocal phase transitions \cite{Fruchart2021,Hanai2020} are also captured by our model. Altogether this opens up a route to driven dissipative (quantum) systems that realize the phenomenology presented above.

\subsection{Driven magnets close to thermal equilibrium}\label{subsec:drivenMagnets}

Most solid state systems are characterized by fast relaxation rates, which makes it difficult to drive them far out of thermal equilibrium. 
The phase diagram of Fig.~\ref{fig:PhaseDiagram} suggests that to reach the exceptional critical point deep in the ordered phase ($r<0$), one needs a large negative damping $\gamma \sim - |r|$. In contrast, in the following, we will show that it is possible to reach this exceptional critical line in a weakly driven magnet.
\subsubsection{Rotating order and CEP transition in the driven ferrimagnet}
To that end we consider an equilibrium spin system on a cubic lattice. The system is assumed  to have an anisotropy breaking the $SO(3)$ spin symmetry down to $U(1)$ rotations within an easy $xy$ plane and a $\mathbb{Z}_2$ reflection symmetry along the $z$-axis. We assume the system spontaneously orders in the easy plane for equilibrium temperatures $T<T_N$ constituting an $xy$ (anti)ferromagnet. Furthermore it can undergo a ferrimagnetic Ising like transition below a temperature $T_c<T_N$ where an out of plane magnetization along the $z$-axis develops. Instances of systems showing such type of phase transitions are, for example, found in Refs.~\cite{Suergers2014,Omi2021,McConnell2012}. In the vicinity of this phase transition the slow, long wavelength dynamics is captured by the $U(1)$ Goldstone mode of the order in the $xy$ plane, $\theta$, and an Ising variable $m_z$ describing the ferrimagnetic order parameter. We thus construct the effective dynamics for these degrees of freedom. Since there is no additional conserved charge, there is no hydrodynamic modes that need to be considered on top of them. The symmetries act as
\begin{align}\label{eq:symmact}
\begin{split}
    &U(1):\quad \theta\rightarrow \theta+\alpha,\\
    &\mathbb{Z}_2:\quad \theta\rightarrow-\theta,\; m_z\rightarrow-m_z.
    \end{split}
\end{align}
Since the reflection and $U(1)$ rotation do not commute the symmetry group of this system is $U(1)\ltimes\mathbb{Z}_2\cong O(2)$.  We now drive the system out of thermal equilibrium by applying a rapidly oscillating magnetic field with amplitude $B_0$. Since the drive is very fast, the effective dynamics of $\theta$ and $m_z$ is still Markovian and the drive effectively couples the Ising and the Goldstone mode. The $O(2)$ symmetry of the system, the absence of conserved currents and the markovianity of the long time dynamics indicate, that its coarse grained dynamics will be described the $O(N=2)$ model discussed above. We derive the effective dynamics of the spatially averaged collective Goldstone and Ising modes $\langle\theta\rangle,\,\langle m\rangle$ explicitly from a microscopic model in Sec.~\ref{subsec:microscopic_ferri}. Since the correlation length $\xi$ is orders of magnitude larger than the microscopic lattice spacing, $\xi\gg a$, the spatial fluctuations of the dynamics can be treated in a continuum limit with emergent rotational symmetry in space, as usual for the effective dynamics close to a critical point. We thus model the spatial fluctuations beyond the effective single mode with phenomenological constants $K_\theta,\,K_0,\,K_m$ at order $\nabla^2$. This procedure yields the following effective dynamics
\begin{align}\label{eq:drivenGoldstone}
\begin{split}
     \alpha_\theta \partial_t \theta =&  \alpha_\theta \gamma_z m_z + K_\theta\nabla^2\theta-\partial_t m_z+\xi_{\theta},\\  
    \alpha_m \partial_tm_z=&-\frac{\delta V}{\delta m_z}+\partial_t \theta+ K_0 \nabla^2 \theta +\xi_{m}, 
\end{split}
\end{align}
where $\alpha_{\theta,m_z}$ stem from the Gilbert damping of the original spin system. $\xi_\theta,\xi_{m}$ are respective Gaussian white noises which close to equilibrium are set by temperature and $\alpha_{\theta,m_z}$. $V$ is an Ising potential for the ferrimagnetic order parameter, which can be parametrised as
\begin{align}
    V=\int_{\vecx,t}\frac{1}{2}\left(rm_z^2+K_m(\nabla m_z)^2\right)+\frac{\lambda}{4!}m_z^4 \label{VGL}
\end{align}
with $r=T-T_c$ the distance from the equilibrium ferrimagnetic transition. The effect of the drive are nonvanishing values of $K_0$ and $|\gamma|\propto B_0^2$ which do not exist in equilibrium. We give a microscopic derivation of these dynamics below. Evidently, the presence of a finite $\gamma$ indicates that a build up of ferrimagnetic order $\langle m_z\rangle\neq0$ immediately induces a finite angular velocity for the $xy$ order causing it to rotate in the easy plane. For time and length scales above $(\alpha_\theta\gamma)^{-1}$ the effect of $\xi_\theta$ in the first equation get suppressed, and we can use this first equation of motion to eliminate $m_z$ and plug it into the second to indeed reproduce the nonlinear $\sigma$ model for the $N=2$ CEP discussed in Sec.~\ref{sec:ExceptionalRG},
\begin{align}
    \left(\partial_t^2+(\delta-Z\nabla^2)\partial_t-v^2\nabla^2\right)\theta+\frac{g}{6}\left(\partial_t\theta\right)^3+\xi=0. \label{eq:effmodel2}
\end{align}
Since the spin damping coefficients $\alpha_{m,\theta}$ are typically very small, we can restrict ourselves to leading order contributions in these. We then have $\delta=\frac{\gamma_z\alpha_\theta(r-\gamma_z)}{r}$, and thus there is a transition into a rotating phase occuring at $r=\gamma_z$ rendered first order by \ac{CEP} fluctuations as discussed above for all finite drivings $\gamma_z$. In the vicinity of the transition, $r\approx\gamma_z$, the remaining parameters are $Z=\frac{K_\theta}{\alpha_\theta},\,v^2=K_\theta\gamma_z+K_0 \alpha_\theta \gamma_z\approx K_\theta\gamma_z,\,g=\frac{\alpha_\theta\lambda}{\gamma_z^2},\,\xi=\frac{\gamma_z}{\alpha_\theta}\xi_m$. In these units, we have $\rho_0=1$ and thereby, by the criterion found in the field theoretic analysis \eqref{eq:multicritical_point}, there is a first order phase transition between $xy$ order and rotating ferrimagnet if 
\begin{align}
    \frac{\lambda \alpha_\theta^2}{\gamma_z}\gg 1, \label{eq:first}
\end{align}
and the $xy$ order is destroyed in the opposite limit. 

Finally we remark on the connection to the equilibrium case $\gamma_z,K_0\rightarrow 0$, where there is an Ising transition into a static ferrimagnet. At the Ising fixed point however, the nonequilibrium coupling $\gamma_z$ is relevant, so that once it is  allowed via the breaking of equilibrium conditions in terms of the drive, it will flow to a value of $\mathcal{O}(1)$ under the RG for sufficiently large system sizes. In that sense the equilibrium transition constitutes a multicritical point which will not impact the transition phenomenology once one drives the system out of equilibrium. 

Our results are summarized in the phase diagram sketched in Fig.~\ref{fig:phasediagramFerri}, which explores the phases as function of temperature $T$ and driving power $P_D$. Here $T$ is the  equilibrium temperature of the undriven system which in experiments is set by phonon or electron baths and their coupling to a cryostat. 
The  phase diagram is based on the assumption, 
that in equilibrium, $P_D=0$, the system undergoes a sequence of two phase transitions upon lowering $T$, first into an $xy$ ordered phase and then into the ferrimagnetic phase as discussed above.
Driving the system has, first, the effect that the effective temperature and thus the fluctuations grow linearly in the driving power $P_D$ for small $P_D$. Importantly, the coupling $\gamma_z$ linear in $P_D$ emerges in the effective field theory, Eq.~\eqref{eq:drivenGoldstone}, which is highly relevant in the renormalization group sense. Due to $\gamma_z$, the static ferrimagnetic order is transformed into a rotating ferrimagnet for arbitarily small $P_D$ as discussed above. Arbitrarily small, but finite driving also destabilizes the second-order phase transition and one obtains instead a weak fluctuation-induced first order transition characteristic of the CEP as for small $\gamma_z$, the condition of Eq.~\eqref{eq:first} is always obeyed. At larger driving, $\gamma_z \sim \alpha_\theta^2 \lambda$,
the line of first-order transition ends when the xy order is destroyed by the  strong fluctuations arising from the superthermal mode occupation in the vicinity of the CEP. We therefore expect, as sketched in Fig.~\ref{fig:phasediagramFerri}, that the long-ranged xy-order melts most easily just above the transition temperature $T_c$. In the figure, we also took into account that a finite $P_D$ always leads to a net heating of the system proportional to $P_D$, thus all transition lines bend to the left in Fig.~\ref{fig:phasediagramFerri}.

Let us finally compare the symmetry $U(1) \ltimes \mathbb{Z}_2$, Eq.~\eqref{eq:symmact}, to the symmetry $U(1) \times \mathbb{Z}_2$, realized by replacing the second line of that equation by $m_z\to -m_z$, while leaving $\theta$ invariant. In this case, on the right hand side of the first line of~\eqref{eq:drivenGoldstone}, a field-independent constant (as well as a KPZ non-linearity) is also symmetry allowed, and will be generically non-vanishing once the drive is switched on. The system is thus always in a rotating phase for finite drive, and no phase transition of the above type would be realized. In other words, time translation invariance is broken explicitly, as opposed to spontaneously as in our case.

\subsubsection{Microscopic derivation}\label{subsec:microscopic_ferri}
In the following, we provide a  microscopic theory to show how a rotation of Goldstone modes is induced at a ferrimagnetic transition if the system is driven out of thermal equilibrium by an oscillating magnetic field $B_z(t)$. We consider classical spins $\Svec_i$, $|\Svec_i|=1$, on a three-dimensional cubic lattice with 
\begin{align}
H=& J \sum_{\langle i,j\rangle} S^x_i S^x_j+ S^y_i S^y_j- \Delta S^z_i S^z_j \nonumber \\
&+ \sum_i \delta_2 {S_i^z}^2 + \delta_4 {S_i^z}^4 - g_i B_z(t) S_i^z. \label{eq:H}
\end{align}
The model is invariant under global spin rotations around the $z$ axis and we assume $J, \Delta,\delta_4>0$. The sign in front of  $\Delta$ is chosen to obtain a ferrimagnet.  At $T=0$, $B_z=0$, the system orders antiferromagnetically in the xy-plane for  $\delta_2> z J (\Delta-1)$ ($z=6$ is the number of nearest neighbors) but the spins tilt out of the plane for $\delta_2< z J (\Delta-1)$ developing a uniform out-of-plane magnetization. By tuning $\delta_2$, one can thus describe the transition from an xy antiferromagnet to a ferrimagnet. 

The dynamics of the system is obtained from the Langevin (or, equivalently,  Landau-Lifshitz-Gilbert) equation
\begin{align}
    \partial_t \Svec_i=- \Svec_i \times \left(\frac{\partial H}{\partial \Svec_i} + \alpha \partial_t  \Svec_i + \vecxi_i(t) \right),\label{eq:LLG}
\end{align}
 where the Gilbert damping $\alpha$ allows for a relaxation of the magnetization.

\begin{figure}[t]
\center{}
   \includegraphics[width=0.8 \linewidth]{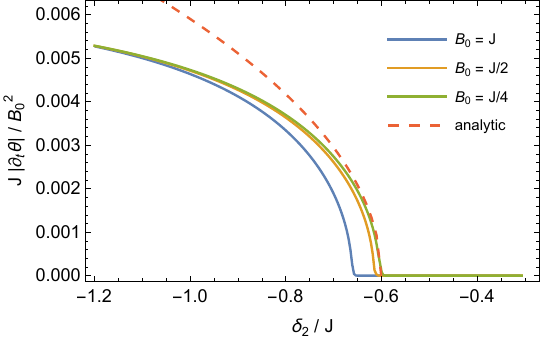}
   \caption{Average precession rate $\partial_t \theta$ as function of the anisotropy $\delta_2$ for the model defined in Eqs.~\eqref{eq:H} and~\eqref{eq:LLG}. For sufficiently large $\delta$, the magnet develops an out-of-plane magnetization and, simultaneously, the spins start to rotate. Solid lines: mean-field numerics (i.e., for a noiseless model) for three different amplitudes $B_0$ of the oscillating field, dashed line: analytical result valid close to the phase transition for small $B_0$ using Eq.~\eqref{eq:drivenGoldstone}, Eq.~\eqref{eq:gammaz} and the mean-field order parameter $|m_z|=\sqrt{\frac{-\delta_2-z J (1-\Delta)}{\delta_4}}$. The parameters are $B_z(t)=B_0 \cos(\omega t)$, $\omega=7.2\,J$, $\alpha=0.2$, $g_A=1$, $g_B=2$, $\Delta=0.9$,  $\delta_4=6\,J$.
  } \label{fig:precessionRate} 
\end{figure} 
To obtain an equation for the time-dependence of the angle $\theta$ parametrizing the Goldstone mode, Eq.~\eqref{eq:drivenGoldstone}, and thus for $\gamma_z$, it is most convenient \cite{Tengdin2022} to compute first the time-dependence of the relevant conservation laws, i.e., of  the total magnetization $M_z=\sum_i S_i^z$. Due to the damping terms, the magnetization $M_z$ is not conserved and one obtains 
\begin{align}
\frac{\partial M_z}{\partial t} = - \alpha  \sum_i \left(\Svec_i \times \partial_t  \Svec_i \right)_z= -\alpha \sum_i \left(1-{S^z_i}^2 \right)\partial_t \theta_i, \label{eq:Mz}
\end{align} 
where $\theta_i$ is the angle describing the in-plane orientation of $\Svec_i$ and we ignored contributions from $\vecxi_i(t)$ which at low temperature will only give rise to small corrections to the value of $\gamma_z$. Next, we average Eq.~\eqref{eq:Mz} over time in the presence of an oscillating magnetic field $B_z(t)$. The time average of $\partial_t M_z$ vanishes, $\overline{\partial_t M_z}=0$, as it is a total derivative of a bounded quantity. In contrast, $\overline{ \partial_t \theta_i}$ can be finite, as the angle is not bounded and can have a net growth in each oscillation period $T$, $\overline{ \partial_t \theta_i}=(\theta_i(t+T)-\theta_i(t))/T$.
Thus, we obtain a remarkably simple equation for the average angle $\theta=\overline{\langle \theta_i \rangle}=\frac{1}{N} \sum_i \overline{\theta_i}$  which is {\em  independent} of the friction constant $\alpha$
\begin{align}
    \partial_t \theta =\overline{\langle {S_i^z}^2 \partial_t \theta_i \rangle},\label{eq:theta2}
\end{align}
where $\langle \dots \rangle$ denotes the average over different sites $i$. Assuming that our system is weakly driven out of thermal equilibrium by a small, time-dependent field $B_z(t)$, we evaluate Eq.~\eqref{eq:theta2} in second-order perturbation theory and {\em linear} in the uniform magnetization $m_z=\overline{\langle S^z_i\rangle}$. Comparing to Eq.~\eqref{eq:drivenGoldstone}, we find 
\begin{align}
\gamma_z = 2 \overline{\langle {S_i^z} \partial_t \theta_i \rangle}_c, \label{eq:gammaz1}
\end{align}
where we omitted corrections from $\overline{\langle {S_i^z}^2 \rangle}\, \overline{ \partial_t \langle \theta_i \rangle}$  as they are of higher order in either $m_z$ or $B_z$. Eq.~\eqref{eq:gammaz1} should be evaluated at the critical point, i.e., for $\delta_2=z J(\Delta-1)$. The contribution to second order in $B_z(t)$ can be obtained by evaluating both $S_i^z$ and $\partial_t \theta_i$ to first order and we find for an oscillating field of the form $B_z(t)=B_0 \cos(\omega t)$
\begin{figure}[t]
\center{}
   \includegraphics[width=0.8 \linewidth]{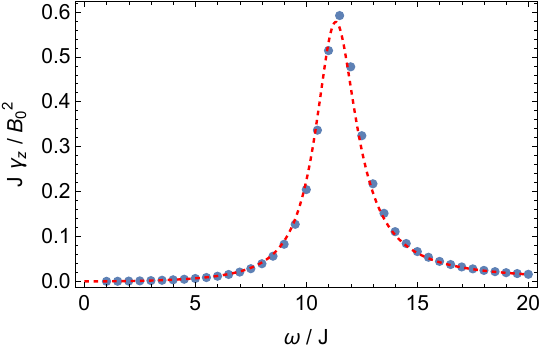}
   \caption{Coupling $\gamma_z$ defined in Eq.~\eqref{eq:drivenGoldstone} as function of the driving frequency $\omega$ (points: numerics for $B_0=0.25\,J$, line: analytical result, Eq.~\eqref{eq:gammaz}). The parameters are $\alpha=0.2$, $g_A=1$, $g_B=2$, $\Delta=0.9$, $\delta_2=-0.66\,J$, $\delta_4=6\,J$. }\label{fig:gammaz} 
\end{figure} 
\begin{align}
\gamma_z
& \approx \frac{B_0^2 (g_A-g_B)^2 z J \omega^2}{2 \left(\omega^2 (1+\alpha^2)-4 (z J)^2 \Delta \right)^2+\alpha^2 8 (z J)^2 \omega^2 (\Delta+1)^2},  \label{eq:gammaz}
\end{align}
 where $g_A$, $g_B$ are the $g$ factors on the two sublattices.

 In Fig.~\ref{fig:precessionRate} the precession rate $\partial_t \theta$ obtained from a noiseless solution, $\vecxi_i(t)=0$, of the equation of motions, Eq.~\eqref{eq:LLG}, is shown in comparison to the analytical results. In the noiseless case (or, equivalently, in a $T=0$ mean-field theory) one can use translational symmetry and simulate only the equation of motion of two spins, one on each sublattice of the antiferromagnet. Close to the phase transition and for small amplitudes of the oscillating fields, excellent agreement of numerics and analytics is obtained. The frequency dependence of $\gamma_z$ is shown in Fig.~\ref{fig:gammaz}. The analytical result fits the numerical result for all frequencies. For $\omega\approx 2 z J \sqrt{\Delta}$ the oscillating field resonantly couples to a magnon mode, giving rise to a pronounced peak in $\gamma_z$ as a function of $\omega$.

Our analytical and numerical results confirm that a rotation of the Goldstone mode (and thus also a critical exceptional point) can be induced by driving the system only weakly away from thermal equilibrium. In our specific (noiseless) model, we used an oscillating magnetic field and different $g$ factors on the two sublattices to induce a non-equilibrium state. General symmetry arguments suggest, however, that the coupling $\gamma_z$ defined in Eq.~\eqref{eq:drivenGoldstone} is always finite when the system is not in thermal equililbrium. For example, one could instead use a laser tuned to an electronic resonance. In this case an absorbed photon will trigger a complex cascade of electronic, spin and phonon excitations which are difficult to describe quantitatively. We expect, however, that their net effect can be absorbed in  an effective parameter $\gamma_z$ which describes that the spins will start to precess in the ferrimagnetic phase.

The combination of the results presented above shows that the microscopic model defined by Eq.~\eqref{eq:H} with the equation of motion, Eq.~\eqref{eq:LLG} provides a direct realization of the effective field theory as given by Eq.~\eqref{eq:symmact}. First, in the absence of noise, there is a second-order phase transitions from an xy-ordered phase into the ferrimagnetic phase with finite magnetization and a rotating order parameter, as shown in Fig.~\ref{fig:precessionRate}. This justifies a Ginzburg-Landau expansion around this transition point, Eq.~\eqref{VGL}. Second, we have shown both numerically and analytically, that the precession of $\theta$ is described by the $\gamma_z$ term in Eq.~\eqref{eq:symmact}, which is calculated analytically in Eq.~\eqref{eq:gammaz1}. Third, the remaining derivative terms follow in a straightforward way from the equation of motion, Eq.~\eqref{eq:LLG} with $\alpha_\theta, \alpha_m\sim \alpha$. Only the $K_0$ term was not explicitly derived by us but its presence follows from symmetry arguments and the discussion below Eq.~\eqref{eq:effmodel2} shows that our results remain valid for $K_0=0$. 
Finally, the presence of noise terms in Eq.~\eqref{eq:symmact} is in equilibrium enforced by fluctuation-dissipation theorems. Thus, only the precise amplitude of the noise may be affected by the non-equilibrium terms. This concludes the microscopic derivation of Eq.~\eqref{eq:symmact}. 

\subsection{Implementation via driven dissipative bosons}\label{ssec:Lindblad_implementation}

The model \eqref{eq:Langevin} can also emerge as a semiclassical limit of driven dissipative bosons subject to Lindbladian time evolution. We consider the dynamics of $N$ spatially extended bosonic fields with creation operators $a^\dagger_i(\vecx,t)$ which is symmetric under $O(N)$ rotations of the bosonic fields. Based on symmetry, it is to be expected that the universal phenomenology of the vector valued expectation value of these bosons after coarse graining is captured by the model \eqref{eq:Langevin}.  We now give an explicit example for Lindblad jump operators together with an $O(N)$ symmetric Hamiltonian time evolution, where the coarse graining procedure to obtain \eqref{eq:Langevin} can be done analytically. This has to be understood as a proof of principle, that given the right symmetries \eqref{eq:Langevin} emerges as an effective theory and is expected to happen for different microscopic setups where the coarse graining is not straightforward, as well. To generate the nonlinear dampings, two-body or higher order loss terms are necessary while the negative damping or pumping required for the rotating phase can be obtained by an effective single particle pump. This is somewhat similar to the single bosonic field case discussed in \cite{Sieberer2013,Sieberer2014,Sieberer2016} which has an additional $U(1)$ symmetry of the complex phase of the mode operators not present in our case.\\
We consider a Lindbladian time evolution for the density matrix $\hat\rho$ 
\begin{equation}
\begin{split}
   \partial_t\hat\rho =&-i[\hat H,\hat\rho]\\
   &+ \sum_\alpha \gamma_\alpha\Big(\int_\vecx \hat L_\alpha(\vecx)\hat\rho \hat L^\dagger_\alpha(\vecx)-\frac{1}{2}\{\hat L^\dagger_\alpha(\vecx) \hat L_\alpha(\vecx),\hat\rho\}\Big). 
\end{split}
\end{equation}
The Hamiltonian is split into a quadratic and an interacting part $\hat H=\hat H_0+\hat H_{\text{int}}$ with 
\begin{align}
    \hat H_0=\sum_{i=1}^N\int_\vecq r_c(\vecq) \hat{a}^\dagger_i(\vecq)\hat{a}_i(\vecq),
\end{align}
and we add a generic $O(N)$ symmetric $\phi^4$ interaction
\begin{align}
    \hat H_{\text{int}}=\lambda_c\sum_{ij}\int_{\vecx}\hat{\phi}_i(\vecx)^2\hat{\phi}_j(\vecx)^2
\end{align}
with canonical field variable $\hat\phi_j=\frac{i}{\sqrt{2}}(\hat a_j-\hat a^\dagger_j)$. We consider local single particle pump and loss $\hat L_1(\vecx)=\hat a^\dagger_i(\vecx)$ and $\hat L_2(\vecx)=\hat a_i(\vecx)$ with respective rates $\gamma_1\equiv\gamma_{in},\; \gamma_2\equiv\gamma_{out}$, where the identical rates forall $i=1,...,N$ ensure a weak $O(N)$ symmetry. Furthermore, we include $O(N)$ symmetric two body pump and loss processes $\hat L_3(\vecx)=\sum_i\hat \phi_i(\vecx)\hat a_i(\vecx)$ and  $\hat L_4(\vecx)=\sum_i\hat \phi_i(\vecx)\hat a^\dagger_i(\vecx)$ with rates $\gamma_3\equiv\lambda_d,\;\gamma_4\equiv\lambda_p$ similar to the case of the single mode quantum van der Pol oscillator \cite{Ben_Arosh_2021}.  We now perform the following steps: pass to the equivalent Keldysh path integral description of the Lindbladian time evolution, introduce the canonical field momentum $\hat \pi_j(t,\vecx)=\frac{1}{\sqrt{2}}(\hat a_j(t,\vecx)+\hat a_j^\dagger(t,\vecx))$ and take the semiclassical limit to obtain an MSRJD action, see \cite{Sieberer2016} for a review. Since the conjugate momentum appears only quadratically, we can perform the Gaussian integration over it, analogously to passing from a Hamiltonian path integral to a Lagrangian in equilibrium (quantum) field theory. Neglecting irrelevant terms that are higher order in field amplitudes or derivatives acting on noise fields, we arrive at the MSRJD action \eqref{eq:action_mic} with the couplings  $\gamma=\gamma_{out}-\gamma_{in},\, r=\gamma^2+r_c^2,\, u=2\lambda_p,\, u'=3\lambda_d-\lambda_p,\, \lambda=\lambda_c+(3\lambda_d+\lambda_p)\gamma,\, D=\frac{1}{2}(\gamma_{in}+\gamma_{out})(r_c^2+2\gamma^2)$.

The choice of interaction and Lindblad operators, involving $\phi$ operators, while giving simple expressions for the parameters of Eq.~\eqref{eq:action_mic}, is somewhat artificial. However, the calculation can be done analogously when including squeezing terms $\sim c \hat a_i\hat a_i+h.c.$ that the bosonic $U(1)$ symmetry to $\mathbb{Z}_2$ on the quadratic level, and again one arrives at our effective model description: By symmetry, we expect the same effective long wavelength model to emerge when breaking a microscopic $U(1)\times O(N) \to \mathbb{Z}_2 \times O(N)$.

\subsection{Two nonreciprocally coupled fields}\label{subsec:nonreciprocal_realization}

Our model also describes the universal phenomenology of two nonreciprocally coupled $N$ component order parameters $\vecphi_{1,2}\in \mathbb{R}^N$ in $d$ spatial dimensions as it may occur in  active matter scenarios. The respective case for $d=N=2$ has been discussed in \cite{Fruchart2021}. The (anti)flocking phase found there corresponds to the ordered phase of our model, the chiral rotating phase to the rotating phase, the so-called swap phase to van der Pol oscillations and the chiral+swap phase to a possible mixture of rotation and van der Pol oscillation.

Suppressing spatial gradients of the fields which are easily restored, the linearized dynamics of the order parameters are given by
\begin{align}
    \partial_t\vecphi_a=K_{ab}\vecphi_b+\vecxi_a,
\end{align}
where $a,b=1,2$ indexes the two order parameters, but not their components in $\mathbb{R}^N$. $\vecxi_a$ is a Gaussian white noise. The linear dynamics is nonreciprocal in the sense that $K_{ab}\neq K_{ba}$. We parametrize it as 
\begin{align}\label{eq:nonreciprocalModel}
    K=\begin{pmatrix}
        -m_1 & g_1\\
        g_2 & -m_2
    \end{pmatrix}.
\end{align}
Nonreciprocity implies that at least one of the off-diagonal couplings is nonzero and without loss of generality we choose it to be $g_1$. One can promote this model to include nonlinearities by letting the parameters $K_{a,b},\,g_{1,2}$ depend on the $O(N)$ invariant amplitudes $\rho_{1,2} = \vecphi^T_{1,2}\cdot\vecphi_{1,2}$. Analogously one can extend to rotationally symmetric dynamics in $d$ spatial dimensions by including $\nabla^2$ dependencies. The model \eqref{eq:nonreciprocalModel} can be brought into the form of the nonconservative $O(N)$ \eqref{eq:Langevin} model discussed above by using $g_1\neq0$ to solve for $\vecphi_2$ :
\begin{align}
    \vecphi_2=g_1^{-1}\Big((\partial_t+m_1)\vecphi_1-\boldsymbol{\xi}_1\Big).
\end{align}
Plugging this into the equation of motion for $\vecphi_2$ yields
\begin{align}
    \Big(\partial_t^2+2\gamma\partial_t+r\Big)\boldsymbol{\phi}(\vecx,t)+\vecxi(\vecx,t)=0
\end{align}
with
\begin{align}
    2\gamma&=m_1+m_2,\\
    r&=m_1m_2-g_1g_2,\\
    \boldsymbol{\xi}&=g_1\boldsymbol{\xi}_1+(g_1m_2+\partial_t)\boldsymbol{\xi}_2.
\end{align}
It is a straightforward calculation to see that the nonlinearities considered in \eqref{eq:Langevin} are generated by respective nonlinearities in \eqref{eq:nonreciprocalModel}. In the presence of nonreciprocities with $\text{sgn}(g_1)\neq\text{sgn}(g_2)$, we can tune the damping $\gamma$ and the mass $r$ to zero independently from each other and therefore reach the rotating phase, the statically ordered one as well as the \ac{CEP} separating them. This would not be possible if either $g_1$ or $g_2$ vanished. 

We briefly remark that the reduction to our model can also be done after for instance adopting an effective description in the ordered phase for the nonreciprocally coupled fields and then passing to a Lagrangian description. In that case one first arrives at the effective theory~\eqref{eq:drivenGoldstone} which can be completely map to our field theory as was demonstrated also in the implementation via a ferrimagnet. In this context, Eq.~\eqref{eq:drivenGoldstone} also corresponds to the theory describing the second-order phase-transition found in driven-dissipative condensates~\cite{Hanai2020} where the $O(2)$ symmetry is emergent. This scenario is also captured by our mechanism, see Sec.~\ref{ssec:Sym_break}.

\section{Conclusions}

We have developed the theory and phenomenology of the phase diagram and the exceptional critical points for $O(N)$ models driven out of equilibrium. Beyond previous approaches on the level of deterministic non-linear dynamics, as a main conceptual and computational step we have systematically incorporated the effects of stochastic fluctuations.  A main insight is that the key properties of a critical exceptional point together imply that the phase transition is rendered weakly first order below the upper critical dimension, once fluctuations and interactions are properly taken into account. 

We have also shown that exceptional critical points can be realized without any further fine-tuning in magnetic systems, by irradiating a system which in thermal equilibrium displays a certain sequence of phase transitions. As an example, we considered the transition from an $xy$ magnet to a ferrimagnet in the presence of an oscillating magnetic field. Such systems are in the `domain of attraction' of our $O(N)$ models, and map to these in the long wavelength limit.

Although our analysis has focused on the concrete case of $O(N)$ models, the phenomenology unravelled here should be universal, and describe more generally the behavior near critical exceptional points: It is driven by the key features of a CEP, non-analytic spectral properties and enhanced infrared occupation, which together with sufficiently strong interactions drive our scenario. 
The $O(N)$ symmetry instead is not prominently involved in this line of reasoning; it rather plays the role of a paradigmatic symmetry class for analyzing critical behavior in a definite setting. One interesting candidate for the realization of our general scenario is given by the separatrix defining the watershed for the domain of attraction of equilibrium and non-equilibrium fixed points in coupled Ising models  in \cite{young2020nonequilibrium} (and, possibly, their generalizations to larger symmetry groups).

From the viewpoint of $O(N)$ models, our non-equilibrium extension fuses the relativistic $O(N)$ model with inertial time derivative term with the dynamical models of Hohenberg and Halperin for equilibrium dynamical criticality. It offers a surprisingly simple phase diagram with an analytically accessible limit-cycle phase for the angular variable, related to spontaneous time translation symmetry breaking -- finding structure and simplicity out of equilibrium might be more valuable than complex phenomenology in the long run. So far we have focused our in depth analysis on the line of  CEPs only, but many interesting aspects still await exploration. Among them, adding and exploring the axis which brings us to $d$-dimensional, $O(N)$ symmetric generalizations of the van der Pol oscillator, hosting a phase transition to a limit cycle for the amplitude variable. Within the angular limit-cycle phase, it will be interesting to study the interplay of gapless modes and fluctuations, for example, determining the connection to the KPZ equation and possible larger symmetry group variants of it. Furthermore, the multicritical point  at the center of the phase diagram, and the direct transition from the disordered into the limit-cycle phase,  host (critical) behavior which manifestly deviates from the one studied here, based on a counting analysis for the leading infrared divergences. In particular, the phase diagram with its multicritical point in the center shares its shape with scenarios of Lifshitz criticality~\cite{Hornreich1975}. There, the leading derivative term $\sim\nabla^2$ is fine-tuned to zero and replaced by a leading $\nabla^4$ term, in some analogy to our fine-tuning of the damping term $\sim \partial_t$ to zero at the CEP, being replaced by a leading inertial term $\sim \partial_t^2$. Both the universal behavior at the multicritical point as well as at the direct transition from the disordered phase into the limit cycle are promising questions for future work.In the context of experimental realizations, it will be interesting to investigate the role of disorder.
Many of these questions crystallized in our concrete model should occur more broadly in fluctuating non-equilibrium systems. The present non-equilibrium $O(N)$ models might serve as a paradigm for the study of such phenomena, as did their equilibrium counterparts.

One additional question that results naturally from our analysis concerns criticality of fermions. At equilibrium, fermion criticality can  occur in quantum phase transitions only, connected to the fact that there is an infrared suppression of fermion occupation at finite temperature instead of an enhancement, opposite to the bosonic case -- there is no classical fermion criticality. However, as we have seen here, \ac{CEP}s provide a way to enhance the infrared behavior, possibly in a way that compensates for a noise-induced suppression of occupation -- and thus give rise to a scenario of non-equilibrium fermion criticality.

\section*{Acknowledgments}
 We thank M. Buchhold, A. Chiocchetta, D. Hardt, J. Lang and  R. Wolters for useful discussions. We acknowledge support by the Deutsche Forschungsgemeinschaft (DFG, German Research Foundation) CRC 1238 project number 277146847. 


\onecolumngrid
\appendix
\section{Amplitude oscillations}\label{app:vdp}
\paragraph*{Van der Pol oscillations and competing order} --
As stated in the main text, the mean-field equation~\eqref{eq:0+1D_Model} can be solved for all $N$ and $\gamma<0$ by 
\begin{equation}
   \boldsymbol{\phi}=(\phi_{\mathrm{VdP}}(t),0 \dots 0),
\end{equation}
where $\phi_{\mathrm{VdP}}(t)$ is the solution of the one-dimensional van der Pol equation. Its closed form is not known, but it has been proven to be a periodic function and thus to describe a limit cycle. In the corresponding phase, the amplitude of the order parameter is oscillating along a given direction. 
The $O(N)$ symmetry is broken to $O(N-1)$, exactly as in the statically ordered phase, but the Goldstone theorem discussed in Sec.~\ref{sssec:goldstone_theorem} applied to this phase leads to a non decaying Goldstone mode associated to the spontaneous breaking of time translation invariance, without it being also associated to the breaking of $O(N)$ .

For $N=1$, this is the only available phase, but for $N>2$, it competes with the rotating phase.
These two phases can only be stable in the yellow region in Fig.~\ref{fig:PhaseDiagram}, and only one of them is stable, at least at the mean-field level. We investigate it analytically below and we also verified this numerically by solving~\eqref{eq:0+1D_Model}. We find that for sufficiently high value of the ratio $u'/u>(u'/u)_c=f(\rho_0,E)$, the rotating phase is stable and the oscillating phase unstable, the reverse being true below the critical ratio.
To gain analytical insight, the linear analysis around this phase is more complicated because there is no analogue of the rotating frame which allowed us to get an autonomous linearized equation as in the rotating phase. One solution could be to do a Floquet analysis of the Van der Pol oscillator. 

However, we can use the stability analysis done in~\ref{app:PhaseAmp} for the rotating phase to get the boundary at which it becomes unstable in a regime of parameter where the static phases is also unstable. The van der Pol oscillating phase shall then be the stable one. As discussed in Sec.~\ref{sec:EPs}, a solution is unstable as soon as one of the modes has a dispersion with a positive imaginary part. The (mean-field) modes can be extracted from the phase-amplitude decomposition in the rotating phase done in App.~\ref{app:PhaseAmp} before integrating out the amplitude. From~\eqref{eq:GoldAmpCoupled}, we can analytically extract the dispersions of the modes involving the amplitude fluctuations. There is indeed a mode that becomes unstable, for values of the parameters which agree with numerical simulations. The exact expression of this dispersion is rather complicated but simplifies in some limits. In particular, deep in the rotating phase, i.e, at large $E$, the threshold is found to be $u'>u/2$. It confirms that $u$ tends to stabilize the swap phase and $u'$ the rotating phase. On the contrary, for $E \to 0$, the rotating phase is stable for any positive value of $u'$. 
\\
\paragraph*{Phase transition} -- The phase transition from the disordered phase to the oscillating phase has the same pattern as the transition from the disordered to the rotating phase: the damping term goes to zero and the dispersions at the critical point have a finite real part, $\omega_{1,2}=\pm\sqrt{r}$.

However, the transition from the statically ordered to the oscillating phase does not occur via a CEP. Indeed, the linearized equation of motions in the static phase are given by, see Sec.~\ref{subsec:gaussian_fluctuations},  
\begin{align}
    \partial_t^2\delta\rho+(\delta+2u'\rho_0-Z\nabla^2)\partial_t\delta\rho+2\lambda\rho_0-v^2\nabla^2)\delta\rho=0,
\end{align}
 for the amplitude and  
\begin{equation}
    \partial_t^2\theta_i+(\delta-Z\nabla^2)\partial_t-v^2\nabla^2)\theta_i=0,
\end{equation}
for the Goldstone modes. When $u'>0$, upon tuning $\delta$ to zero, the damping of the amplitude mode remains positive, while the Goldstone modes become unstable and start to rotate in order to compensate for the negative damping. However, when $u'<0$, the first instability occurs for the amplitude mode, which starts to display van der Pol oscillation. At the critical point, $u'\rho_0+\delta=0$, 
the dispersions of the amplitude modes are
\begin{equation}
    \omega_{1,2}= \pm \sqrt{2\lambda\rho_0},
\end{equation}
indicating that this transition is similar to the direct transition from the disordered phase at the mean-field level. 
Note that we recover the fact that close to transition between the static and time-dependent orders, the sign of $u'$ sets the stable phase.

\section{Critical Exceptional Points of \texorpdfstring{$N$}{N}-component fields}\label{app:Definition_CEP}
We now elaborate on how any CEP occurring in noisy Markovian dynamics of a vector valued field can be mapped to the damped harmonic oscillator case discussed in the main text.

We first note that we can always map a system of $N$ differential equations of second order in time derivatives into a set of $2N$ first order differential equations by introducing $\boldsymbol{\pi}=\partial_t\vecphi$ as an independent variable. In physics terminology we pass from a Lagrangian to a Hamiltonian representation. Using this, the (diagonal) linearized equation of motion or inverse Green function $\Gamma^R(\vecq,t)$ of the $N$-component damped harmonic oscillator discussed in the main text, see Sec.~\ref{ssec:EPFluct}, can always be written as
\begin{align}
    (\partial_t\mathbb{1}+M(\vecq))\delta\boldsymbol{\Phi}=0
\end{align}
where $\boldsymbol{\Phi}$ is a $2N$ component vector and $\mathbb{1}$ and $M(\vecq)$ are $2N\times2N$ matrices. The eigenvalues of $M(\vecq)$ are the dispersions $i\omega_\alpha(\vecq)$ and the corresponding eigenvectors the $2N$ linearly independent modes. In this representation an \ac{EP}, where two modes coalesce, occurs if and only if $M(\vecq)$ is not diagonalizable at $\vecq^*$, and therefore has at least one $2\times 2$ Jordan block 
\begin{align}
    M(\vecq^*)=\begin{pmatrix}
        i\omega_{EP} & 1\\0 & i\omega_{EP}
    \end{pmatrix}.
\end{align}
The dynamics of excitations close to a \ac{CEP} at $\vecq^*=0$ is governed by an inverse Green function that is block diagonal with blocks that are at most of size $2\times 2$ and with at least one block taking the form
\begin{align}
\label{eq:structCEP}
    \Big(\partial_t\mathbb{1}_{2}+\begin{pmatrix}
        i\omega_{1}(\vecq) & 1\\0 & \omega_{2}(\vecq)
    \end{pmatrix}\Big)\delta\boldsymbol{\Phi}_{CEP}=0
\end{align}
where $\delta\boldsymbol{\Phi}_{CEP}$ are the fluctuations contributing to the \ac{CEP} and $\omega_1(\vecq=0)=\omega_2(\vecq=0)=0$. This structure also implies the superthermal mode occupation in the presence of generic Markovian noise, as shown in~\cite{Hanai2020}.

Reciprocally, by reverting this procedure, any system that has a \ac{CEP} arising from the structure~\eqref{eq:structCEP} can generically be brought back to the form of a damped harmonic oscillator with a diagonalizable inverse Green-function even in presence of noise, as done explicitly for the two examples in the implementation section~\ref{sec:implementation} of the main text.

\section{Phase-amplitude representation}\label{app:PhaseAmp}

In this appendix, we give detailed derivations of the action describing the fluctuations around the static and rotating orders obtained by writing the fields in a phase-amplitude decomposition
\begin{align}
   \label{eq:PhaseEqRepApp}
    \phi=\sqrt{\rho_0+\delta\rho}\exp(E T_{1,2} t)\exp(\sum_{i=2}^{N}\theta_i T_{1,i})\hat{e}_1, \quad \tilde \phi=\sqrt{\rho_0}\exp(E T_{1,2} t) \exp(\sum_{i=2}^{N}\theta_i T_{1,i})\tilde\chi,
\end{align}
where $\tilde\chi \in \mathbb{R}^{N}$ is parametrized as $\tilde\chi=(\tilde{\delta\rho},\tilde \theta_2,\dots,\tilde\theta_N)$. The static ordered phase is included as a special case $E=0$.
For simplicity we drop summation indices in the following, and use $\vec{T}= \theta_i T_{1,i}$. The derivative terms like $\tilde{\vecphi}^T\partial_t\vecphi$ generate terms $\exp(-\vec{T})\partial_t \exp(\vec{T})$ and higher orders in derivatives, which have to be evaluated using the infinitesimal form of the Baker Campbell Hausdorff formula. Since we are only interested in the quadratic action at this point, one can however truncate to
\begin{align}
    \exp(-\vec{T})\partial_t^n \exp(\vec{T})=\sum_i(\partial_t^n\theta_i)T_{1,i}+O(\theta^2)
\end{align}
and equivalently for the gradient terms. In the case of the statically ordered phase, we thus arrive at the quadratic action displayed in the main text in Sec \ref{subsec:gaussian_fluctuations}. In the rotating phase, i.e., at finite angular velocity $E$, we immediately arrive at the quadratic action for the perpendicular phase fluctuations $S_\perp^0$ that is given in the main text. The parallel ($\theta$) and amplitude ($\rho=\frac{\delta\rho}{2\rho_0}$) fluctuations however mix on the quadratic level:
\begin{equation}
\label{eq:GoldAmpCoupled}
\begin{split}
    S^0_{\rho\theta}=\rho_0\int_{X}&\tilde{\rho}(\partial_t^2+(2u'\rho_0-Z\nabla^2)\partial_t-v^2\nabla^2)\rho+\tilde{\theta}(\partial_t^2-Z\nabla^2\partial_t-(v^2+ZE)\nabla^2)\theta\\
    &-2E\tilde{\rho}\partial_t\theta+\tilde{\theta}(2u\rho_0E+2E\partial_t)\rho-2D(\tilde\theta^2+\tilde\rho)^2
\end{split}
\end{equation}
This action clearly violates the thermal symmetry conditions, since the coupling between phase and amplitude are not symmetric. We can access the dynamics of the slow phase fluctuations alone by performing the Gaussian integration over the gapped amplitude field on the level of the path integral. This yields the following effective Gaussian action for the phase field (after proper rescaling of the field)
\begin{align}
    S_\parallel^0=\rho_0\int_X\tilde\theta(\partial_t^2+(\bar\delta-\bar Z\nabla^2)\partial_t-\bar v^2\nabla^2)\theta-2D\tilde\theta^2
\end{align}
where the shifted couplings close to the transition, i.e., for small angular velocities $E$ are
\begin{align}
    \bar\delta=|\delta|+O(E^2),\quad \bar Z=Z+O(E),\quad \bar v^2=v^2+O(E). 
\end{align}

This procedure of integrating out the amplitude mode (i.e., passing to a \ac{Nlsm}) can be carried beyond the quadratic level to derive the nonlinearities for the interacting theory. For the case $N=2$, this can be done without any truncation in $\theta$, while for $N>2$, a truncation in $\theta_i$ leads to terms of the form~\eqref{eq:NewTerm} discussed in the main text.
The precise coefficients obtained through this procedure do not really matter, as they will not remain intact in the \ac{RG} flow once one starts coarse graining the dynamics.
The important point is that with the rescaled fields used in~\eqref{eq:Sint}, there is always a contribution of order one in a $\rho_0^{-1}$ expansion that therefore does not vanish in the large $\rho_0$ limit.

\section{Explicit loop calculations}\label{app:Sunset}

In this appendix, we compute the integrals arising from loop corrections, which are given in the main text. We start by the two-loop sunset diagram because there is no momentum running through the loop which simplifies the analysis and then discuss the one-loop integral with momentum transfer.

\subsection{Two-loop sunset}
\begin{figure}
\center{}
\subfloat[\label{fig:integrandI2l} ]{
   \includegraphics{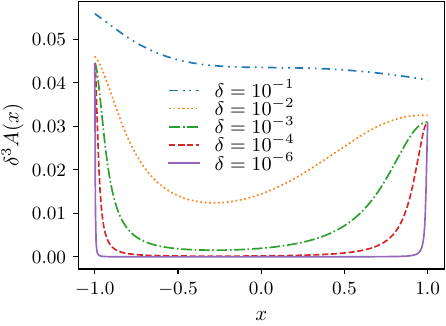}
  }
  \hfil 
 \subfloat[\label{fig:2loopSunset2d}]{
   \includegraphics{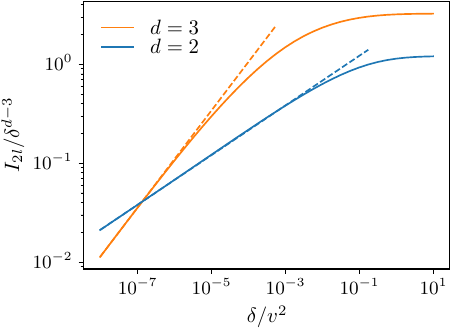}
  }
    \caption{(a) The integrand of~\eqref{eq:IwithDLapp}, denoted $A(x)$, is plotted as a function of $x=\cos(\theta)$ where $\theta$ is the angle between $p_1$ and $p_2$, $\tilde p_{1}=1/2$, $\tilde p_{2}=1$, $v=1$ and different values of $\delta$. At small $\delta$, it becomes sharply peaked around $x=\pm 1$, reflecting the resonance condition. (b) $I_{2l}/\delta^{d-3}$, plotted in $d=2,3$ is a function of $\delta/v^{2}$ only. At large $\delta/v^2\ll 1$ the integral behaves as $\delta^{d-3}$ but gets suppressed by an additional power law when the ratio $\delta/v^2$ is small. The power law behaviors (indicated by the dashed lines) agree quantitatively with~\eqref{eq:NSnd}.}
\end{figure}

Let us prove Eq.~\eqref{eq:scalingSun} which gives the correction to $\delta$ induced by $g_{1}$ and comes from the diagram~\ref{fig:2lc}. In the following, all integrals are considered to be suitably regularized in the UV when divergent. It reads 
\begin{align}
\label{eq:I2lqw}
I_{2l} =& \int_{Q_{1},Q_{2}}^{}\omega_{1}^{2}G ^{K}(Q_{1})\omega_{2}^{2}G ^{K}(Q_{2})i(\omega_{1}+\omega_{2})G ^{R}(Q_{1}+Q_{2}).
\end{align}
  The frequency integrals can be performed, and after rescaling of momenta $q_{1,2}\to q_{1,2}\delta^{1/2}$, we obtain

\begin{align}
\label{eq:IwithDLapp}
I_{2l} =\frac{\delta^{d-3}}{4}\int_{\mathbf{q_{1}},\mathbf{q}_{2}}\frac{f_1(\mathbf{q_{1}},\mathbf{q_{2}})+O(\delta)}{  \frac{v^{2}}{\delta} \left[(\mathbf{q_{1}}\cdot \mathbf{q_{2}})^2-(\mathbf{q_{1}}\mathbf{q_{2}})^{2}\right]^{2}+  f_2(\mathbf{q_{1}},\mathbf{q}_{2})\left[\Delta(\mathbf{q_{1}}^2)+\Delta(\mathbf{q_{2}}^{2})+\Delta((\mathbf{q_{1}}+\mathbf{q_{2}})^{2})\right]+O\left(\delta\right)}\frac{1}{\Delta(\mathbf{q_{1}}^2)\Delta(\mathbf{q_{2}}^2)}
\end{align}
In this expression, we defined $f_1$ and $f_2$, two functions independent of $v$ whose precise forms are not important for the argument. The only property we will use is that $f_1(\mathbf{q_{1}},\mathbf{q_{2}})=f_2(\mathbf{q_{1}},\mathbf{q_{2}})$ when $\mathbf{q_{1}}$ and $\mathbf{q_{2}}$ are aligned.
We neglected the subleading terms in $\delta$ which do not contribute to the leading infrared divergence. However, as discussed in the main text for the one-loop integral~\eqref{eq:IwithDL}, we keep a higher order term in the denominator because the leading term can in fact become small under certain conditions. This is the case for every momentum if $\delta/v^2$ is large. In this regime, the integral reduces to
\begin{equation}
\label{eq:IsingSunset}
\begin{split}
I_{2l}= \frac{\delta^{d-3}}{2}\int_{\mathbf{q_{1}},\vecq_2}\frac{1}{\Delta(\mathbf{q_{1}}^2)+\Delta(\mathbf{q_{2}}^{2})+\Delta((\mathbf{q_{1}}+\mathbf{q_{2}})^{2})} \frac{1}{\Delta(\mathbf{q_{1}}^2)\Delta(\mathbf{q_{2}}^2)},
\end{split}
\end{equation} 
and behaves as $\delta^{d-3}$, i.e., exactly as the tadpole diagram~\ref{fig:2lb}. Now, approaching the \ac{CEP} where $\delta/v^2$ becomes small, the first term in the denominator of Eq.~\eqref{eq:IwithDLapp} dominates for generic momenta, but still vanishes when $\vecq_1$ and $\vecq_2$ are aligned. The integrand thus behaves as $\delta^{d-3}$ when the momenta are almost aligned, But only as $\delta^{d-2}$ when they are not, giving a subleading contribution to the integral. We thus have a resonance condition to get the highest divergence. This is shown in figure~\ref{fig:integrandI2l}. Formally, the denominator of the integrand in~\eqref{eq:subI} behaves as a Dirac distribution. 
To make it apparent, we rewrite the integral over $\vecq_2$ using hyperspherical coordinates around $\mathbf{q_{1}}$,
\begin{align}
\label{eq:I2sphere}
I_{2l} \underset{\delta \to 0}{\sim}\frac{\delta^{d-3}}{4}\int_{\mathbf{q_{1}}}\int q_{2}^{d-1}dq_{2} d\Omega_{d-1}^{q_{2}}\frac{1}{\Delta(q_{1}^2)\Delta(q_{2}^2)} I_{\mathrm{sub}},
  \end{align}
 where 
 \begin{equation}
\label{eq:subI}
I_{\mathrm{sub}}=\int_{}^{}d\theta\frac{\sin(\theta)^{d-2}f_1(q_{1},q_{2},\cos(\theta))}{  \frac{v^{2}}{\delta} (q_{1} q_{2})^2\left[\cos(\theta)^{2}-1\right]^{2}+  f_2(q_{1},q_{2},\cos(\theta))\left[\Delta(q_{1}^2)+\Delta(q_{2}^{2})+\Delta((\mathbf{q_{1}}+\mathbf{q_{2}})^{2})\right]},
\end{equation} 
 where $\theta$ is the angle between $\vecq_{1}$ and $\vecq_{2}$.

This last integral~\eqref{eq:subI} can be brought to the following form,
\begin{equation}
\label{eq:subI2}
I_{\mathrm{sub}}=\eta^{2}\int_{}^{}d\theta \sin(\theta)^{d-2}\frac{a(\cos(\theta))}{\left[\cos(\theta)^{2}-1\right]^{2}+ \eta^{2}b(\cos(\theta))^2 }, 
\end{equation} 
with  $\eta=\sqrt{\delta/v^{2}}$ and $a$, $b$ smooth non-vanishing functions around $\pm 1$. We dropped all the dependencies in momentum for the sake of clarity.
Using $x=\cos(\theta)$, the fraction in~\eqref{eq:subI2} is of the form
\begin{align}
    \label{eq:apart}
    F(x)=\frac{1}{ (x^2-1)^{2}+\eta^2 b(x)^2}=& \mathrm{Im}\left(\frac{1}{((x^2-1)- i \eta b(x)}\right)=\frac{1}{2} \mathrm{Im}\left(\frac{1}{x-1-\frac{i\eta }{2}b(x)}-\frac{1}{x+1+\frac{i
   \eta}{2}b(x)}\right),\\
    \xrightarrow[\eta \to 0]{}  &\frac{\pi}{2} \left(\delta(x-1)+ \delta(x+1)\right).
\end{align}
It indeed behaves as a Dirac distribution, reflecting the resonance condition. At small but finite $\eta$, the Dirac distributions are slightly extended, and while performing the integral in ~\eqref{eq:subI}, we can use that $\eta^2 F(x)$ is essentially equal to one for $|\cos(\theta)\pm 1|< \eta$ and zero otherwise. It means that only small deviations of $\cos(\theta)$ around $\pm 1$ of order $\eta$ contribute to the integral, i.e., only small deviations of $\theta$ of order $\sqrt{\eta}$ around zero and $\pi$, and we can only keep the leading terms in a series expansion around these points for the other terms.
Integration then yields
\begin{align}
    \label{eq:NSnd}
    I_{\mathrm{sub}} &\sim \int_{0}^{\sqrt{\eta}}d\theta \theta^{d-2}a(1)+\int_{\pi-\sqrt{\eta}}^{\pi}d\theta (\pi-\theta)^{d-2} a(-1),\\
   &\sim C \eta^{\frac{d-1}{2}}=C(\frac{\delta}{v^{2}} )^{\frac{d-1}{4} },  
\end{align}
where C is a multiplicative constant, proving Eq.~\eqref{eq:scalingSun}, $I_{2l}\sim \delta^{d-3}(\frac{\delta}{v^{2}} )^{\frac{d-1}{4} }$ since the remaining integrals are free of any parameters. We emphasize that the additional variable $\delta/v^2$ enters the corrections, reflecting the absence of a full scaling solution at the \ac{CEP}. 
This is confirmed by numerical integration over the momenta presented in Fig.~\ref{fig:2loopSunset2d} in two and three dimensions: $I_{2l}/\delta^{d-3}$ is indeed a function of $\delta/v^{2}$ only, and the power law behaviors found at small $\delta$ agree quantitatively with the analytical results.

The same structure arises for the other sunset integrals (e.g., with $g_2$ instead of $g_1$ as vertices or the sunset obtained from diagram~\ref{fig:2ld}), and they can be computed using the same procedure. 

It is also instructive to rephrase this discussion using $(\mathbf{q}, t)$ variables in the loops. The corresponding expressions are obtained by Fourier transform with respect to time to their $(\mathbf{q}, \omega)$ counterparts.
The integral $I_{2l}$ becomes
\begin{align}
\label{eq:2lqt}
I_{2l}= \int_{t}^{}&\int_{\mathbf{q}_{1}}^{}G_D ^{K}(\mathbf{q_{1}},t) \int_{\mathbf{q}_{2}}^{}G_{D} ^{K}(\mathbf{q_{2}},t)G_{D} ^{R}(\mathbf{q_{1}}+\mathbf{q_{2}},t) \cos(v |\mathbf{q}_{1}| t)\cos(v |\mathbf{q}_{2}| t)\cos(v |\mathbf{q}_{1}+\mathbf{q}_{2}| t),
\end{align} 
where we denote $G_{D}^{R}(\mathbf{q},t)= \Theta(t)\exp(-\Delta(\mathbf{q}^{2})t )$ and $G_{D}^{K}(\mathbf{q},t)= \exp(-\Delta(\mathbf{q}^{2})|t| )/\Delta(\mathbf{q}^{2})$, purely dissipative Green's functions which are also those that appear in model A. Absent the oscillating terms associated to $v$, this integral scales as $\delta^{d-3}$ and displays a typical $z=2$ behavior. Now, the oscillating terms (seen as perturbation of this scenario) oscillate faster and faster at finite $t$ in the scaling regime where we choose $t \sim q^{-2}$. In the spirit of a Rotating Wave Approximation (RWA), we can keep only the non-oscillating terms. Since the oscillating terms are of the form $\exp(i v t (\pm|\mathbf{q_{1}}+\mathbf{q_{2}}| \pm |\mathbf{q_{1}}|\pm |\mathbf{q_{2}}|))$, we recover our resonance condition, $\vecq_1$ and $\vecq_2$ have to be aligned to yield a significant contribution.

\subsection{One-loop integral}\label{app:1loop_diagrams}

We now discuss Eqs.~\eqref{eq:upper_critical} and~\eqref{eq:tranfer_momentum_scaling}, involved in the corrections of the quartic interactions coming from the diagrams displayed in Fig.~\ref{fig:1loop4P}. The calculation is closely related to the one done for the sunset integral above, and a similar resonance conditions arises. The diagrams read, with $\omega_{p}$ the frequency and $\vecp$ the momentum entering the loop,  
\begin{equation}
\label{eq:I1lI_app}
I_{1l,I}(\vecp,\omega_p)  =\int_{\mathbf{q},\omega}^{} i(\omega+\omega_{p})\omega^2G ^{R}(\mathbf{q}+\mathbf{p},\omega+\omega_p)G ^{K}(\mathbf{q},\omega) = \int_{\mathbf{q},t}^{}G ^{K}_D(\mathbf{q},t)G_{D} ^{R}(\mathbf{q}+\mathbf{p},t) \cos(v |\mathbf{q}| t)\cos(v |\mathbf{q}+\mathbf{p}| t) \exp(i \omega_p t),
\end{equation} 
respectively in $\vecq,\omega$ and $\vecq,t$ variables.

Before discussing the finite momentum case, we note that for $\vecp=0$, after performing the integral over the frequency and rescaling of momentum $q \to q \delta^{1/2}$, the integral reduces to
\begin{equation}
\label{eq:IzeroMomentum}
    I_{1l,I}(\vecp=0) \sim \delta^{\frac{d-4}{2}} \int_{\vecq} \frac{1}{\Delta(\vecq^2)(-i \tilde\omega_e + \Delta(\vecq^2))},
\end{equation}
in the scaling regime where $\tilde \omega_p = \omega_p/\delta\sim 1$. We also again use $\Delta(y)= y+1$. This is in line with formula~\eqref{eq:upper_critical}.

Now, at finite momentum, there are several resonances that can arise depending on the precise value of the external frequency and momentum. The first resonance is found for frequencies close to $\omega_p\sim \pm v |\vecp|$, and exactly corresponds to the one we have in the sunset integral: $\vecp$ and $\vecq$ have to be aligned. Intuitively, this can be understood from the RWA argument we developed above: the oscillating terms in~\eqref{eq:I1lI_app} are, when $\omega_p= \pm v |\vecp|$,  of the form
\begin{equation}
    \exp(i v t (\pm|\mathbf{q}+\mathbf{p}| \pm |\mathbf{q}|\pm \omega_e)) = \exp(i v t (\pm|\mathbf{q}+\mathbf{p}| \pm |\mathbf{q}|\pm \pm v |\mathbf{p}|),
\end{equation}
which is the same form we got for the sunset diagram. They are not oscillating again exactly when $\vecp$ and $\vecq$ are aligned. This peculiar resonance directly originates from the non-vanishing real part of the relation dispersions at the \ac{CEP}, and it can be checked explicitly that the highest divergences of the Gaussian Green's functions in $(\vecq,\omega)$ coming from~\eqref{eq:S0GoldstoneStat} are obtained exactly for $\omega= \pm v |\vecq|$. In a sense, while the divergence occurs through the imaginary part of the dispersion relations, the real part acts to some extent like a finite frequency scale (because it goes infinitely slower to zero for small momenta) around which the divergences occur. This is technically very reminiscent of the role of a finite momentum scale in the Brazvoskii's scenario~\cite{Brazovskii75} where it is also the reason why loops with momentum transfer are negligible.  

To make the link with the sunset diagram explicit, the integral with $\omega_e= \pm v |\vecp|$ can be written as
\begin{equation}
\label{eq:bbbb}
I_{1l,I}(\vecp, \omega_p= \pm v |\vecp|) \sim  \delta^{\frac{d-4}{2}}\int_{\tilde{\vecq}} \frac{ \pi|\tilde{\mathbf{p}}|\left(\tilde{\mathbf{p}} \cdot \tilde{\mathbf{q}}+\tilde{\mathbf{q}}^2\right)+O(\delta)}{i \frac{v}{\sqrt{\delta}}\left[(\tilde{\mathbf{p}} \cdot \tilde{\mathbf{q}})^2-\tilde{\mathbf{p}}^2 \tilde{\mathbf{q}}^2\right]+ |\tilde{\mathbf{p}}|\left(\tilde{\mathbf{p}} \cdot \tilde{\mathbf{q}}+\tilde{\mathbf{q}}^2\right)\left[\Delta\left(\tilde{\mathbf{q}}^2\right)+\Delta\left((\tilde{\mathbf{p}}+\tilde{\mathbf{q}})^2\right)\right]+O\left(\delta\right)} \frac{1}{\Delta\left(\tilde{\mathbf{q}}^2\right)}.
\end{equation} 
Its real part is exactly the integral over $\vecq_{2}$ in Eq.~\eqref{eq:IwithDLapp}, with $\vecp$ playing the role of $\vecq_{1}$. This renders the intuition developed around Eq.~\eqref{eq:1boucle=2bloucle} rigorous, and we can thus use the analysis done in the previous section. Again, we have to keep a higher order term in the denominator because its contribution can dominate the integral in some cases. 
When approaching the \ac{CEP}, $\delta$ is small with respect to $v^2$. In that case, the second term dominates only when $v^2/\delta[(\tilde{\mathbf{p}} \cdot \tilde{\mathbf{q}})^2-\tilde{\mathbf{p}}^2 \tilde{\mathbf{q}}^2]^2$ is sufficiently small. It is again true for every momenta $\vecq$ when $\tilde p^2v^2/\delta$ is small i.e., for small dimensionless momentum $\tilde p \ll \delta^{1/2}/v$ and the integral behaves as $\delta^{(d-4)/2}$, proving Eq.~\eqref{eq:upper_critical}. But for a finite dimensionless momentum $\tilde p \sim 1$, the resonance condition appears, and the integrand in~\eqref{eq:bbbb} behaves again as a Dirac distribution: the highest divergence is found when $\vecq$ and $\vecp$ are aligned, as in the sunset integral. This allows us to perform the integral, and using the result of the previous section, it leads to~\eqref{eq:tranfer_momentum_scaling}.

Interestingly, there is another type of resonance. In particular, at zero frequency and finite dimensionless momentum, see Eq.~\eqref{eq:IwithDL} where we got a different resonance condition:  the highest divergence is found when $v^2\left(\mathbf{p}^2+2 \mathbf{p} \cdot \mathbf{q}\right)$ is small. This resonance condition also shows up as very sharp and non-analytic behavior, which allows us to get the scaling of the integral with $\delta$ in a similar fashion. We find, for some $\tilde{p}$ not too large,  a slightly different behavior, $I_{1l,I}\sim \delta^{(d-4)/2}(\delta/v^2)^{1/2}$ when $\delta$ is small. We can expect the frequency and momentum contributing the most to the loops to sit on the highest divergences and thus to involve the first scaling~\eqref{eq:tranfer_momentum_scaling}. Anyway, this cannot change the conclusions of the main text because both behaviors lead to smaller divergences at finite momentum for any frequency, no matter the precise power law we get. 

To conclude, we found that for a finite frequency but at zero momentum, the loop is given by~\eqref{eq:IzeroMomentum}, and thus assume a usual scaling form with frequency $\tilde \omega= \omega/\delta$. Therefore, we do not need to use specific frequency dependencies of the different couplings as we do for the momentum dependencies. 

\section{Dyson-Schwinger equations}\label{app:SDE}
\subsection{\texorpdfstring{$N=2$}{N=2} case}
The \ac{DSE} constitute an exact hierarchy of equations between the 1PI vertices. It emerges as consequence of the shift invariance of the effective action. It is discussed for the usual $\phi^4$ case, e.g., in~\cite{Horak2020} but the method can be applied directly within the \ac{MSRJD} framework. The effective action can be written using the shift invariance with respect to both fields, see~\eqref{eq:effective_action},
\begin{equation}
    \Gamma[\vartheta, \tilde \vartheta]= \int \mathcal{D}i\tilde \theta  \mathcal{D} \theta e^{-S[\vartheta+\theta,\tilde  \vartheta+\tilde  \theta]+\frac{\delta\Gamma}{\delta\vartheta}\theta+\frac{\delta\Gamma}{\delta\tilde\vartheta}\tilde\theta}.
\end{equation}

The \ac{DSE} can be obtained by taking functional derivatives with respect to the fields of this equation. To simplify notation in the following, we use the Nambu fields $\Theta= (\vartheta,\tilde{\vartheta})$ and introduce
\begin{align}
    \Gamma^{(n)}_{I_1 \dots I_n}[\Theta] = \frac{\delta^n \Gamma}{\delta \Theta_{i_n}(X_n)\dots\delta \Theta_{i_1}(X_1)}[\Theta],
\end{align}
with $I_j$ a super-index including internal and external indices, $I_j=\{i_j,X_j\}$. Functional derivatives with brackets, e.g., $\Gamma^{(n)}_{I_1 \dots I_n}[\Theta]$ denote functional derivatives before any evaluation on the equation of motion. The formalism can be readily extended to an $N$-component field by including the resulting indices within the internal indices.

Since the action does not contain any term independent of $\tilde\theta$ which would break conservation of probability, we define the master \ac{DSE} by taking a derivative with respect to the response field which yields
\begin{equation}
    \label{eq:DSE10}
     \Gamma^{(10)}[\tilde\vartheta,\vartheta; X_1]=\langle S^{(10)}[\tilde\vartheta+\tilde\theta,\vartheta+\theta;X_1]\rangle.
\end{equation}
This equation tells us that the dressed equation of motion is given by the expectation value of the bare equation of motion.
In the following, we work with the bare action given by~\eqref{eq:S0GoldstoneStat2} and~\eqref{eq:Sint},
\begin{align}\label{eq:GaussianAction_RealSpaceApp}
    S=\int_{\mathbf{x},t}\tilde\theta(\partial_t^2+(-K\nabla^2+\delta)\partial_t-v^2\nabla^2)\theta - D \tilde\theta^{2}+\frac{g_1}{6}\tilde\theta(\partial_t\theta)^3+\frac{g_2}{2}\tilde\theta\partial_t\theta(\nabla\theta)^2,
\end{align}
which describes the fluctuations of the Goldstone mode for $N=2$.
It is a fourth order polynomial, so a series expansion in the fluctuating fields $\theta$ and $\tilde\theta$ stops at finite order and gives
\begin{equation}
   \Gamma^{(1)}_{I_1}= S^{(1)}_{I_1}+\frac{1}{2} S^{(3)}_{I_1}G-\frac{1}{6} S^{(4)}_{I_1} G G G\Gamma^{(3)},
\end{equation}
where sums over indices and $\Theta$ dependencies are implicit.
 
\begin{figure}
    \centering
    \subfloat[][\ac{DSE} for $\Gamma^{(11)}$]{\label{fig:DSE_2}
\(\Gamma^{(11)}=S^{(11)}+ \frac{1}{2}\quad \includegraphics[valign=c] {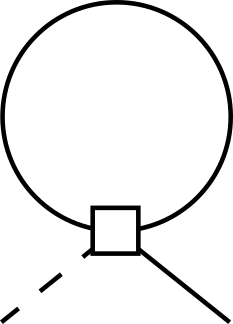} \quad -\frac{1}{2} \quad \includegraphics[valign=c]{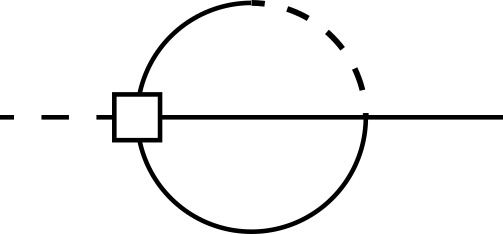}\)
}
\hfil
\subfloat[][One-loop \ac{DSE} for $\Gamma^{(13)}$]{\label{fig:DSE_4}
    $\Gamma^{(13)}=S^{(13)}- \quad \includegraphics[valign=c] {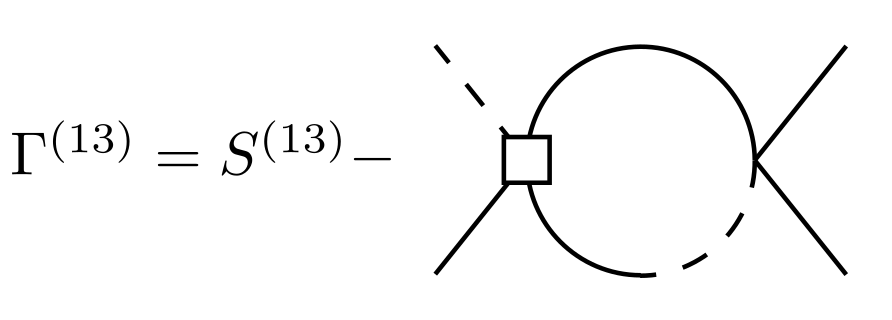}$
    }
\hfil
\subfloat[][One-loop \ac{DSE} for $\Gamma^{(15)}$]{\label{fig:DSE_6}
$  \Gamma^{(15)}=S^{(15)}+ 2\quad \includegraphics[valign=c,raise=8pt] {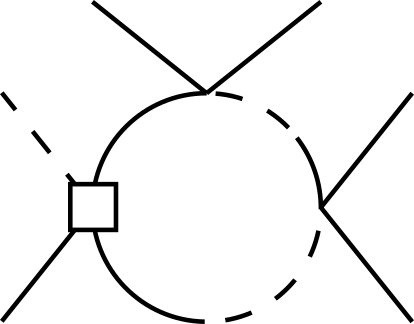} \quad +\quad \includegraphics[valign=c,raise=8pt]{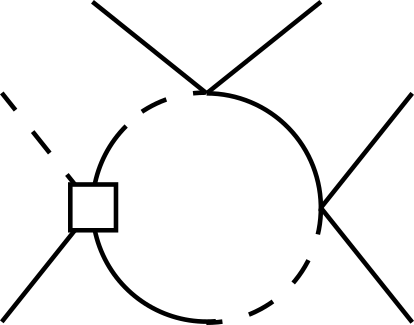}- \quad \includegraphics[valign=c]{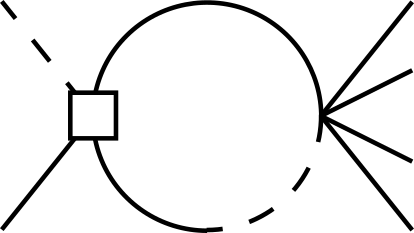}$
}
    \caption{Diagrammatic representation of the \ac{DSE}. The solid and solid-to-dashed lines correspond respectively to the \textit{full} Keldysh and retarded Green functions. The vertices correspond also to the \textit{full} vertices $\Gamma^{(13)}$ and $\Gamma^{(15)}$, except for those that are represented with a square box, which correspond to the bare vertices $S^{(13)}$. Diagrams obtained by permutation of external legs attached to a $\theta$ fields (solid line legs) are not shown. }
    \label{fig:DSE_Diag}
\end{figure}

This master \ac{DSE} gives relations between the full 1PI vertices and involves the renormalized propagator, making it a non-perturbative method. By taking additional functional derivatives with respect to the fields, one can generate equations for higher order vertices. 
Approaching the transition from the phase where $\langle\partial_t\theta\rangle=0$ i.e\ the static ordered phase in our case, we can concentrate on the 1PI vertices evaluated at $\Theta=0$. If a series expansion around $\tilde\vartheta=0$ and $\vartheta=0$ is valid, this also describes the broken phase. We expect it to be the case close to the first-order phase transition. Note that this is also the case in Brazovskii scenario where the calculations done around zero and around a finite order parameter  agree well~\cite{Hohenberg1995}. For the retarded inverse Green-function, we get
\begin{equation}
\label{eq:DSE11}
\begin{split}
    \Gamma^{(1,1)}(P,\Theta=0)=& S^{(1,1)}(P) +\frac{1}{2} \int_Q S^{(12)}(P,-P,Q)G^R(Q)-\frac{1}{2}\int_{Q_1,Q_2} S^{(13)}(P,Q_1,Q_2,-(P+Q_1+Q_2)) \\
    &\times G^K(Q_1)G^K(Q_2)G^R(Q_1+Q_2+P)\Gamma^{(13)}(-Q_1,-Q_2,(P+Q_1+Q_2),-P),
\end{split}
\end{equation} 
where we neglected contribution coming from four point vertices that are higher order in $\tilde\vartheta$. They are irrelevant at the Gaussian CEP for all dimensions of interest $2<d<4$ and only lead to subleading divergences in the following. More generally, all couplings that are irrelevant at the Gaussian \ac{CEP} fixed point induce smaller loop divergences (it is even the way we define and operator to be irrelevant at the Gaussian fixed-point because of the absence of a full scaling solution). Since the \ac{DSE} will become one-loop exact in the regime we are interested in, any irrelevant operator brings subleading divergences in the \ac{DSE}.
Diagrammatically, Eq.~\eqref{eq:DSE11} can be represented by Fig.~\ref{fig:DSE_2}.

In the \ac{DSE} framework, the renormalized effective action is obtained only from dressed tadpole and sunset diagrams. The main difference with the two-loop expression obtained from Fig.~\ref{fig:2loop2pointFunc} is that the sunset involves the full fourth-point vertex $\Gamma^{(13)}$ instead of $S^{(13)}$, and we have to specify the form of the vertices used to solve the \ac{DSE}. 
As discussed in the main text, the effective action can be parameterized using effective couplings $g_1$ and $g_2$ that become momentum dependent upon including interactions because of the structure of the loops as discussed in the main text. To be specific, the renormalization depends on the number of pair of momenta that sum to zero, but not on the precise values of these momenta. Based on these considerations, the four-point vertices can be 
parameterized as 
\begin{equation}
\label{eq:DSE_ansatz}
\begin{split}
    \Gamma^{(13)}(\mathbf{-p_2},\mathbf{p_1},-\mathbf{p_1},\mathbf{p_2},\omega_4,\omega_1,\omega_2,\omega_3) =& i\omega_1 \omega_2 \omega_3 g_{1,a} -i\omega_3 p_1^2 g_{2,a}-i(\omega_1-\omega_2) p_1 \cdot p_2 g_{2,b},\\
    \Gamma^{(13)}(\mathbf{-p_2},0,0,\mathbf{p_2},\omega_4,\omega_1,\omega_2,\omega_3) =& i\omega_1 \omega_2 \omega_3 g_{1,a}\\
    \Gamma^{(13)}(-\mathbf{p_1},\mathbf{p_1},-\mathbf{p_1},\mathbf{p_1},\omega_4,\omega_1,\omega_2,\omega_3) =& i\omega_1 \omega_2 \omega_3 g_{1,b} -i(\omega_1-\omega_2+\omega_3) p_1^2 g_{2,c}, \\
    \Gamma^{(13)}(0,0,0,0,\omega_4,\omega_1,\omega_2,\omega_3) =& i\omega_1 \omega_2 \omega_3 g_{1,c},
    \end{split}
\end{equation}
where $\mathbf{p_1}$ and $\mathbf{p_2}$ are different finite momenta $p_{1,2} \sim \sqrt{\delta}$ and $\omega_4=-(\omega_1+\omega_2+\omega_3)$. All other configurations do not get renormalized.

Anticipating that all couplings will only decrease or stay constant (which can be checked a posteriori), the condition for neglecting the two-loop contributions is therefore given by replacing $\Gamma^{(13)}$ by $S^{(13)}$ in~\eqref{eq:DSE11}. The loop diagrams are then computed exactly as in Sec.~\ref{ssec:BeyondGaussFluct} and App.~\ref{app:Sunset}. The condition Eq.~\eqref{eq:ConditionApp} is thus recovered non-perturbatively, only by considering the sunset topology. When it is fulfilled, the \ac{DSE} have only one-loop contributions which simplify them considerably, and they can be solved. In particular, from~\eqref{eq:DSE11}, the renormalized damping $\bar{\delta}$ satisfies the self-consistent equation~\eqref{eq:Dyson2}.

The \ac{DSE} equation for the four-point vertex $\Gamma^{(13)}$ is represented diagrammatically in Fig.~\ref{fig:DSE_4}. We neglected two-loop contributions and the effect of all irrelevant vertices. The corresponding equation is 
\begin{equation}
\label{eq:Dyson41l}
    \Gamma^{(13)}(P_4,P_1,P_2,P_3)= S^{(13)}(P_4,P_1,P_2,P_3)- \int_{Q}G^K(Q)G^R(Q+P_1+P_2)\Gamma^{(13)}(-(Q+P_1+P_2),Q,P_1,P_2) + \text{perm.}, 
\end{equation}
where $P_4=-(P_1+P_2+P_3)$ and where the permutations apply on the set $P_1, P_2$ and $P_3$.  
When the condition~\eqref{eq:ConditionApp} is met, the loop in the right-hand side is subleading and negligible whenever there is a running momentum going into the loop.
Injecting the forms~\eqref{eq:DSE_ansatz} into Eq.~\eqref{eq:Dyson41l} then allows us to get equations for the different couplings. In particular, we get back Eqs.~\eqref{eq:g1SDEa} and~\eqref{eq:g1SDEc}.
The resulting equations form a linear system that can be inverted. 
The complete solution reads
\begin{align}
\label{eq:gDSEsolvedApp}
g_{1,a}&= \frac{g_{1}}{1+ \alpha_{2}\overline\delta^{\frac{d-4}{2}}}, \quad g_{1,b}= g_{1}\frac{1- \alpha_{2}\overline\delta^{\frac{d-4}{2}}}{1+ \alpha_{2}\overline\delta^{\frac{d-4}{2}}},\quad g_{1,c}= g_{1}\frac{1-2 \alpha_{2}\overline\delta^{\frac{d-4}{2}}}{1+ \alpha_{2}\overline\delta^{\frac{d-4}{2}}}, \\
g_{2,a}&= \frac{g_{2}}{1+ \alpha_{2}\overline\delta^{\frac{d-4}{2}}}, \quad 
g_{2,b}= \frac{g_{2}}{1+ \alpha_{d}\overline\delta^{\frac{d-4}{2}}}, \quad g_{2,c}= \frac{4}{3}\frac{g_{2}}{1+ \alpha_{d}\overline\delta^{\frac{d-4}{2}}}+\frac{2}{3}\frac{g_{2}}{1+ \alpha_{2}\overline\delta^{\frac{d-4}{2}}} -g_2, 
\end{align}
with $\alpha_{2}=(g_{1}+g_{2}) K_{d}'(2-d)/2$ and $\alpha_{d}=g_{2} K_{d}'(2-d)/d$. 

The possibility of a negative quartic coupling (which induces the first-order transition) is cured by taking into account the renormalized six-point vertex $\Gamma^{(15)}$. Being irrelevant, its value is set by the four-point vertex $\Gamma^{(13)}$. Again, its renormalization depends on the configuration of incoming momenta. There are five different configurations which get renormalized differently. We, however, only need this vertex for vanishing momenta,
\begin{align}
   \Gamma^{(15)}(\vecp_1=0,\dots,\vecp_5=0,-\sum_i \omega_i,\omega_1,\dots,\omega_5)=i\omega_1 \omega_2\omega_3\omega_4\omega_5 u_{1,e},
   \end{align}
with 
   \begin{align}
   \label{eq:u1e_app}
   u_{1,e }= 15 (g_1+g_2) g_{1,a}^2\int_q \frac{1}{(q^2+\bar\delta)^3} - \frac{5}{2} u_{1,e}(g_1+g_2)\int_q \frac{1}{(q^2+\bar\delta)^2},
\end{align}
found using the \ac{DSE} diagrammatically represented in Fig.~\ref{fig:DSE_6}. It leads immediately to Eq.~\eqref{eq:g6e}.
\subsection{\texorpdfstring{$N>2$}{N>2} case}

For $N>2$, the \ac{DSE} equations discussed above can be directly used in the $N$-component case by adding the $O(N)$ indices in the internal indices. As discussed in Sec.~\ref{ssec:Nfields}, in the regime where the bare condensate is large, the action is simply the generalization of~\eqref{eq:GaussianAction_RealSpaceApp} to a vector field $\pi \in \mathbb{R}^{N-1} $ with an $O(N-1)$ symmetry. It is given by
\begin{equation}
\label{eq:S_NLapp}
    S_0=\int_{\mathbf{x},t}\tilde{\boldsymbol{\pi}}\cdot(\partial_t^2+(-K\Delta+\delta)\partial_t-v^2\Delta)\boldsymbol{\pi} - D \tilde{\boldsymbol{\pi}}\cdot\tilde{\boldsymbol{\pi}}+\frac{g_1}{6}\int_{\vecx,t}\tilde{\boldsymbol{\pi}}\cdot\partial_t\boldsymbol{\pi}(\partial_t\boldsymbol{\pi})^2+\frac{g_2}{2}\tilde{\boldsymbol{\pi}}\cdot\partial_t\boldsymbol{\pi}(\nabla\boldsymbol{\pi})^2. 
\end{equation}
The loops that appear in the perturbative expansion or in the \ac{DSE} equations (Fig.~\ref{fig:DSE_Diag}) and their scaling properties are therefore the same. The additional $O(N)$ structure only changes the prefactors (sometimes called symmetry factors) of the loops, which become $N$-dependent. The $O(N)$ symmetry factors are the standard $O(N)$ ones and can be found in e.g.,~\cite{Kleinert2001}.

We now show that the quartic coupling controlling the value of the order parameter i.e., the generalization of $g_{1,c}$ also becomes negative and that there is again a first order phase transition. We will therefore concentrate on the renormalization of $g_1$, which at the bare level generates the following vertex:
\begin{align}
   \frac{\delta\Gamma^{(4)}}{\delta\tilde \pi_a(P_4)\delta\pi_b (P_1)\delta\pi_c (P_2)\delta\pi_d (P_3) }=i \omega_1 \omega_2 \omega_3 \Gamma'^{(13)}_{abcd}(P_4,P_1,P_2,P_3)=i\omega_1 \omega_2 \omega_3 g_1 T_{abcd},
\end{align}
where we define
\begin{align}
\quad T_{abcd}=\frac{1}{3}(\delta_{ab}\delta_{cd}+ \delta_{ac}\delta_{bd}+ \delta_{ad}\delta_{bc}),
\end{align}
and $\Gamma'^{(13)}$ which denotes the part of $\Gamma^{(13)}$ encoding the renormalization of $g_1$.

We first look at the set of momenta that corresponds to what we called the $g_{1,a}$ coupling above. In perturbation theory, the diagrams~\ref{fig:1loop4P} lead to
\begin{align}
\Gamma'^{(13)}_{abcd}(\mathbf{-p_1},\mathbf{p_1},-\mathbf{p_2},\mathbf{p_2},\omega_4,\omega_1,\omega_2,\omega_3)= g_1 T_{abcd} - \alpha_2\bar \delta^{\frac{d-4}{2}}\frac{g_1^2}{9}((N' +4) \delta_{ab}\delta_{cd}+ 2\delta_{ac}\delta_{bd}+ 2\delta_{ad}\delta_{bc}),
\end{align}
where $N'=N-1$.  However, the different Kronecker delta functions get different coefficient because only one diagram among the three of Fig.~\ref{fig:1loop4P} contributes. 
We thus see that the ansatz done for the vertex using $g_{1,a}$ in~\eqref{eq:DSE_ansatz} is not sufficient to self-consistently solve the~\ac{DSE}. We need to parametrize it as 
\begin{align}
    \Gamma'^{(13)}_{abcd}(\mathbf{-p_1},\mathbf{p_1},-\mathbf{p_2},\mathbf{p_2},\omega_4,\omega_1,\omega_2,\omega_3) =&  (g^s_{1,a}\delta_{ab}\delta_{cd}+ g^t_{1,a}\delta_{ac}\delta_{bd}+ g^u_{1,a}\delta_{ad}\delta_{bc}).
\end{align}
The \ac{DSE} equation then gives 
\begin{align}
\label{eq:g1a_Ncase}
    g_{1,a}^s= g_{1}\frac{1+ \alpha_{2}\frac{2(N'+2)}{9}\overline\delta^{\frac{d-4}{2}}}{(1+ \alpha_{2}\frac{2}{3}\overline\delta^{\frac{d-4}{2}})(1+ \alpha_{2}\frac{N'+2}{3}\overline\delta^{\frac{d-4}{2}})},\quad g_{1,a}^t=  g_{1,a}^u =\frac{g_{1}}{1+ \alpha_{2}\frac{2}{3}\overline\delta^{\frac{d-4}{2}}}.
\end{align}
Using again the \ac{DSE} for the four point function at zero momenta we find
\begin{align}
\Gamma^{(13)}(0,0,0,0,\omega_4,\omega_1,\omega_2,\omega_3) \equiv g_{1,c}T_{abcd}=\left(g_1-  \alpha_2\bar \delta^{\frac{d-4}{2}}\frac{(N+4)g_{1,a}^s+4 g_{1,a}^t}{9}\right)T_{abcd},
\end{align}
where we define $g_{1,c}$, underlying the fact that we do not need an extra parameter in that case. This gives, using~\eqref{eq:g1a_Ncase},
\begin{equation}
    g_{1,c}=g_1\frac{ (9-4 \alpha_2\bar \delta^{\frac{d-4}{2}} (\alpha_2\bar \delta^{\frac{d-4}{2}} (N'+2)+3))}{(2 \alpha_2+3)
   (\alpha_2\bar \delta^{\frac{d-4}{2}}\ (N'+2)+3)}.
\end{equation}
For $N=2$, we recover the same result we got for $g_{1,c}$ in~\eqref{eq:gDSEsolvedApp}. We indeed see that the couplings $g_{1,c}$ turns negative for sufficiently small damping for all $N$. By inspecting Eq.~\eqref{eq:u1e_app}, we see that the sextic coupling $u_{1,e}$ is clearly positive no matter the precise $N$-factor of the loops, and we therefore find the same first-order scenario.

\section{Scaling and the breakdown of the gradient expansion}\label{app:scaling}

We now elaborate in a bit more detail how the scaling of various operators is inferred in the vicinity of the \ac{CEP} given that canonical power counting does not work due to the breakdown of gradient expansions and the different scaling of coherent period and lifetime of excitations
\begin{align}
    \omega_{\text{CEP}}(\vecq)=-iK\vecq^2\pm v|\vecq|.
\end{align}
The scaling is then fixed by finding an ansatz that renders the dimensionfull RG $\beta$-functions dimensionless. It is on first sight possible to make a homogeneous scaling ansatz for the effective action of the phase fluctuations by choosing a dynamical critical exponent $z_s=1$ and thus being forced to have $[K]=-1$, i.e., (dangerously) irrelevant. This however yields Green functions that diverge for all momenta at the Gaussian fixed point.

One therefore has to analyze the divergences of loop contributions to infer the scaling dimensions of various couplings. Due to the breakdown of the derivative expansion, it is not possible to infer the scaling of momentum dependent operators by taking  momentum derivatives of loops renormalizing for instance the self energy.

We demonstrate this explicitly for the case where one allows a cubic interaction $\sim \lambda\tilde\theta(\partial_t\theta)^2$ breaking $O(2)$ to $SO(2)$ as in \cite{Hanai2020} for comparison. Note, that this coupling is absent at a fixed point with (emergent) $O(2)$ symmetry. This is analogous to the $\mathbb{Z}_2$ symmetric endpoint of the liquid gas transition described by the Ising universality class. Regardless, the loop integral through which such a coupling renormalizes the damping (i.e., the part of the self energy linear in frequencies) is exactly the loop analyzed in the main text which renormalizes interactions, cf Fig.~\ref{fig:1loop4P}. The following result about the breakdown of the gradient expansion thus also immediately applies to the couplings discussed in the main text. At
vanishing momenta, it implies a scaling dimension $6-d$ for this cubic coupling, like for the $\phi^ 3$ coupling in the
Ising case. However, as discussed, the infrared divergence of the loop is lowered at finite transfer momenta and there is no convergent derivative expansion of this loop close to the fixed point. \\ 
If one takes momentum derivatives of this loop, as one would in a derivative expansion of the self energy corrections, one generates spurious singularities as the dependence on dimensionless momenta becomes nonanalytic at $\tilde p =0$, see App.~\ref{app:1loop_diagrams}. Now trying to enforce a scaling form for such an expansion of the self energy, as it usually emerges in equilibrium critical phenomena, leads to operators with apparently larger and larger upper critical dimensions. This is an artifact of the breakdown of the derivative expansion due to the nonanalyticity of the \ac{CEP}.\\
Explicitly, such an expansion to fourth order in dimensionless momenta $\tilde p$ yields
\begin{equation}
    \partial_\omega\Gamma^{(2)}= k^2(Z\tilde p^2+\frac{\delta}{k^2}+ \alpha'_1\lambda^2 k^{d-6}+ \alpha'_2\lambda^2 k^{d-8}\tilde p^2+ \alpha'_3\lambda^2 k^{d-10}\tilde p^4+ O(\tilde p^4)).
\end{equation}
Cutting this expansion at order $\tilde p^2$ would imply that there is an operator $\sqrt{\alpha'_2}\lambda$ with dimension $[\sqrt{\alpha'_2}\lambda]=\frac{8-d}{2}$ inferring an upper critical dimension $d_c=8$ \cite{Hanai2020}. Going to order $\tilde p^4$ one would then however diagnose $d_c=10$ from the operator $\sqrt{\alpha'_3}\lambda$. Clearly, arbitrarily large upper critical dimensions are generated within such an expansion, demonstrating again that a standard derivative expansion is inapplicable in this case.

\twocolumngrid

%

\end{document}